%% file: merged_manuscriptandSM.tex
\documentclass[11pt,a4paper,twoside,listof=totoc,usenames,dvipsnames]{article}

\input{preamble}

\usepackage{refcount}
\usepackage{comment}
\title{Using systemic modeling and Bayesian calibration to investigate the
role of the tumor microenvironment on chemoresistance}
\author[1]{Sabrina Sch\"onfeld}
\author[2]{Laura Scarabosio}
\author[3]{Alican Ozkan}
\author[4]{Marissa Nichole Rylander}
\author[5]{Christina Kuttler}
 
\affil[1,5]{\small Center of Mathematics, Technical University of Munich, Garching, Germany}
\affil[2]{\small Institute for Mathematics, Astrophysics and Particle Physics, Radboud University, Nijmegen, The Netherlands}
\affil[3]{\small Wyss Institute for Biologically Inspired Engineering, Harvard University, Boston, MA, United
States}
\affil[4]{\small Department of Mechanical Engineering, The University of Texas at Austin, Austin, TX, United States}

\begin{document}
\maketitle

\input{only_manuscript}

\pagebreak
\begin{center}
\textbf{\Large Supplemental Material to Using systemic modeling and Bayesian calibration to investigate the
role of the tumor microenvironement on chemoresistance}
\end{center}
\setcounter{equation}{0}
\setcounter{figure}{0}
\setcounter{table}{0}
\setcounter{section}{0}

\makeatletter

\newcommand\eqrefWithArgument[1]{%
\def\myref{\getrefnumber{#1}}
(SM.\myref)
}

\renewcommand{\thefigure}{SM.\arabic{figure}}
\renewcommand{\thetable}{SM.\arabic{table}}
\renewcommand{\thesection}{SM.\arabic{section}}

\input{only_supplementary}

\end{document}

%% file: preamble.tex
\usepackage[reqno]{amsmath}
\usepackage{amsthm}
\usepackage{framed}
\usepackage{multirow}
\usepackage{amssymb}
\usepackage[utf8]{inputenc}
\usepackage{url}
\usepackage{authblk}
\usepackage{sectsty}
\allsectionsfont{\sffamily}
\usepackage{diagbox}
\usepackage[hidelinks]{hyperref}
\usepackage{hvfloat}
\usepackage[bottom]{footmisc}

\usepackage[titles]{tocloft}
\usepackage{mathtools}
\usepackage{etoolbox}
\usepackage{float}
\usepackage{xcolor}
\usepackage{relsize}
\usepackage{makecell}
\usepackage[cmtip,all]{xy}
\newcommand{\longsquiggly}{\xymatrix{{}\ar@{~>}[r]&{}}}

\usepackage[style=numeric-comp,sorting=none,giveninits=true,maxbibnames=99,backend=bibtex]{biblatex}
\renewbibmacro{in:}{%
		\ifentrytype{article}{}{\printtext{\bibstring{in}\intitlepunct}}}
\AtBeginBibliography{}

\usepackage{calc}

\usepackage{tikz}
\usepackage{pgfplots}
\usetikzlibrary{pgfplots.groupplots}

\usepgfplotslibrary{external}
\tikzsetexternalprefix{./Plot-PGF/}
\pgfplotsset{compat=newest, every tick label/.append style={scale=0.8}, every axis/.append style={xlabel style={scale=0.9}, ylabel style={scale=0.9}}, legend image with text/.style={
				legend image code/.code={%
						\node[anchor=center] at (0.3cm,0cm) {#1};
					}
			}
	}

\usepackage[
per-mode=fraction,
locale=US,
]{siunitx}
\usepackage[utf8]{inputenc}
\usepackage[T1]{fontenc}

\usepackage{algpseudocode} 
\usepackage{algorithm} 

\DeclareSIUnit[number-unit-product = \,]\cells{\text{cells}}
\DeclareSIUnit[number-unit-product = \,]\Cells{\num{e5}\,\text{cells}}
\DeclareSIUnit[number-unit-product = \,]\Cellss{\num{e6}\,\text{cells}}
\DeclareSIUnit[number-unit-product = {}]\fbs{\%\,\text{FBS}}
\DeclareSIUnit[number-unit-product = \,]\FBS{\num{10}\%\,\text{FBS}}
\DeclareSIUnit[number-unit-product = \,]\day{\text{day}}
\DeclareSIUnit[number-unit-product = \,]{\molar}{M}

\usepackage{graphicx}
\usepackage[font=small]{subcaption}
\usepackage{bbm}
\usepackage{enumitem}

\usepackage{tabularx}

\usepackage{extarrows}
\usepackage{bm}
\usepackage{nicefrac}
\usepackage{cancel}
\usepackage[normalem]{ulem}

\newcolumntype{C}[1]{>{\centering\arraybackslash}p{#1}}

\usepackage{booktabs}

\usepackage{blindtext}
\usepackage{caption}

\usepackage[figuresright]{rotating}

\usepackage[toc,page,title]{appendix}

\usepackage[nottoc]{tocbibind}

\usepackage{stackengine}
\stackMath

\numberwithin{equation}{section}
\newtagform{brackets}{[}{]}

\newtagform{smallbraces}{\raisebox{1pt}{\smaller\smaller (}}{\raisebox{1pt}{\smaller\smaller )}}
\usetagform{smallbraces}

\def\qm#1{``#1''}

\colorlet{RED}{red}

\usepackage{soul}

\newcommand*{\whiten}[1]{\llap{\textcolor{white}{{\the\SOUL@token}}\hspace{#1pt}}}
\DeclareRobustCommand*\myul{%
	\def\SOUL@everyspace{\underline{\space}\kern\z@}%
	\def\SOUL@everytoken{%
		\setbox0=\hbox{\the\SOUL@token}%
		\ifdim\dp0>\z@
		\raisebox{\dp0}{\underline{\phantom{\the\SOUL@token}}}%
		\whiten{1}\whiten{0}%
		\whiten{-1}\whiten{-2}%
		\llap{\the\SOUL@token}%
		\else
		\underline{\the\SOUL@token}%
		\fi}%
	\SOUL@}

\newcommand*{\mypar}[1]{\vspace*{6pt}\noindent \textsf{\myul{#1}.}~}

\input{mathMacros}

\numberwithin{equation}{section}

\newcommand*\chem[1]{\ensuremath{\mathrm{#1}}}

\makeatletter
\newcommand{\leqnomode}{\tagsleft@true}
\newcommand{\reqnomode}{\tagsleft@false}
\makeatother

%% file: mathMacros.tex
\def\NEW#1{{\color{black} #1}}

\def\txtSub#1#2{{#1}_{\text{#2}}}
\def\txtTop#1#2{{#1}^{\text{#2}}}
\def\Ldots{,\,\ldots\,,}

\def\Triang{\mbox{Triang}}

\def\init#1{#1_0}

\def\stress{\eta}
\def\initH{\init{H}}
\def\initC{\init{C}}
\def\initS{\init{S}}
\def\initD{\init{D}}
\def\initSt{\init{\stress}}
\def\initStCH{\stress_{0,HC}}
\def\initV{\init{V}}
\def\Vtreat{\txtTop{V}{treat}}
\def\Vctrl{\txtTop{V}{ctrl}}

\def\initt{\init{t}}
\def\ttreat{\txtSub{t}{treat}}
\def\tend{\txtSub{t}{end}}

\def\Dthr{\txtSub{D}{thr}}
\def\DthrCH{D_{\text{thr},HC}}
\def\Ddamp{\txtSub{D}{damp}}
\def\cdamp{\txtSub{c}{damp}}
\def\Dnorm{\txtSub{D}{norm}}
\def\DnormCH{D_{\text{norm},HC}}
\def\DnormCHlog{\widehat{D}_{\text{norm},HC}}
\def\Sthr{\txtSub{S}{thr}}
\def\SthrCH{S_{\text{thr},HC}}

\def\adSCH{\sensRate{S}^-\delta^-_{S,HC}}
\def\adDCH{\sensRate{D}^-\delta^-_{D,HC}}

\def\dV{\dot{V}}
\def\dStress{\dot{\stress}}
\def\dD{\dot{D}}
\def\dS{\dot{S}}

\def\CYP{\varrho_{HC}}

\def\Ssupp{\txtSub{S}{supp}}

\def\ESS{\txtSub{P}{eff}}
\def\calIk#1{\mathcal{I}_{:#1}}
\def\calIIk#1#2{\mathcal{I}_{#1:#2}}

\def\pval{\txtSub{p}{sig}}

\def\nat{\lambda}

\def\metD{\gamma_D}
\def\metS{\gamma_S}
\def\metDCH{\gamma_{D,HC}}
\def\metSCH{\gamma_{S,HC}}
\def\met{\gamma}
\def\MetD{\omega_{D}}
\def\MetDCH{\omega_{D,HC}}
\def\MetS{\omega_{S}}
\def\MetSCH{\omega_{S,HC}}

\def\dCH{d_{HC}}

\def\indDeath{\nat_{\text{ind}}}

\def\prol{\beta}

\def\Kv{\txtSub{V}{cap}}

\def\sensRate#1{\alpha_{#1}}

\def\deactFct#1#2{\delta_{#1}^-(#2)}
\def\actFct#1#2{\delta_{#1}^+(#2)}
\def\deact{\delta^-}
\def\act{\delta^+}

\def\underarrow#1#2{\raisebox{-2.4pt}{\footnotesize$\boldsymbol{|}$}\hspace*{-3.7pt}\underbracket[0.14ex]{\hspace*{#1}}_{#2}\hspace*{-7.2pt}\raisebox{-2.8pt}{$\boldsymbol{\Uparrow}$}}

\newcommand{\defas}{\mathrel{\mathop{\raisebox{1.1pt}{\scriptsize$:$}}}=}

\def\smallunderbrace#1{\mathop{\vtop{\m@th\ialign{##\crcr
				$\hfil\displaystyle{#1}\hfil$\crcr
				\noalign{\kern3\p@\nointerlineskip}%
				\tiny\upbracefill\crcr\noalign{\kern3\p@}}}}\limits}
			
\def\ModChemo0{\mathcal{M}^0_{DS}}

\def\mean{\mathbb{E}}
\def\Var{\text{Var}}
\def\unif{\mathcal{U}}

\def\supp{\text{supp}}

\def\AAAst{\raisebox{1pt}{${\scriptstyle ***}$}}
\def\AAst{\raisebox{1pt}{${\scriptstyle **}$}}
\def\Ast{\raisebox{1pt}{${\scriptstyle *}$}}

%% file: only_manuscript.tex
\begin{abstract}
Using a novel modeling approach based on the so-called environmental stress
level (ESL), we develop a mathematical model to describe systematically the collective influence of oxygen concentration and stiffness of the extracellular matrix on the response of tumor cells to a combined chemotherapeutic treatment. We perform Bayesian calibrations of the resulting model using particle filters, with \textsl{in vitro} experimental data for different hepatocellular carcinoma cell lines. The calibration results support the validity of our mathematical model. Furthermore, they shed light on individual as well as synergistic effects of hypoxia and tissue stiffness on tumor cell dynamics under chemotherapy.
\end{abstract}

\section{Introduction}

Cancer, one of the most harmful diseases worldwide, remains a challenging frontier for effective therapeutic interventions. To address this critical issue, researchers are continually striving to develop advanced treatment strategies. In this context, the utilization of cell culture experiments proves invaluable, providing crucial insights into the intricacies of cell growth and how tumor cells respond to treatment. \textsl{In vitro} experiments offer a less restrictive alternative to obtaining data when compared to collecting patient information, for cell cultures and even beyond for single cells~\cite{Szot.2011}.
Mathematical modeling, as introduced e.g. in~\mbox{\cite{Byrne.2006, Preziosi.2003}}, even goes further and provides a general understanding as well as, after a sucessful calibration of the model parameters on experimental data and observations, a possibility to develop an \textsl{in silico} setting. This allows even for some quantitative predictions and by that may go beyond experimental possibilities. 

Our research focuses on \textsl{in vitro} experimental data involving various hepatocellular carcinoma cell lines. Specifically, we investigate the combined effects of hypoxia (oxygen levels) and tissue stiffness on the chemoresistance of these cells. The chemotherapeutic agents employed in this study consist of doxorubicin, the principal drug, supplemented by sorafenib for supportive purposes. The experiment unfolds in three distinct phases: the adaptation phase involves the cultivation of cells within an untreated environment, where a gel substitutes the natural extracellular matrix (ECM), varying in stiffness. This is followed by the treatment phase, and subsequently, another growth phase. Finally, the viability of the cells is assessed.

Our modeling approach, as introduced in~\cite{Schonfeld.2022}, relies on ordinary differential equations (ODEs), with a central variable introduced as the environmental stress level (ESL). The ESL is contingent on various environmental factors affecting the cells, including oxygen saturation, drug concentration, and ECM stiffness. Ultimately, the ESL exerts an influence on the growth of viable tumor cells through what we refer to as \qm{influence functions}, based on the generalized logistic growth equation. In this context, having relatively large cell numbers, we want to apply a deterministic growth model without too many parameters. The generalized logistic growth model is one of the standard approaches for homogeneously distributed cells and conditions, and is closely related to the Gompertz growth model, which is another standard model for tumor cell growth. An overview about such classical models is given e.g. in~\cite{Tabassum_2019}. 
For the equations and terms for the chemotherapeutic drugs and their impact on the cells, classical approaches from pharmacokinetics with exponential decays and Hill functions are applied~\cite{Goutelle2008}.  

Next, we turn our attention to the crucial task of addressing model parameter values. The estimation of these unknown parameters is a critical endeavor, and it can be approached using either deterministic or probabilistic methods. Deterministic approaches seek to provide point estimates of the parameters, whereas probabilistic methods not only estimate parameter values but also provide valuable insights into the associated uncertainties and correlations among them. This probabilistic approach treats parameters as random variables, although it does come at a higher computational cost. In our study, we opt for the latter approach – Bayesian calibration – using special particle filters to speed up the algorithms.

Given the nature of our mathematical model, which describes time dynamics through ordinary differential equations, there arises a unique challenge for calibration from the available experiments. Viability measurements were only available at a single time-point, specifically three days after treatment. Despite this inherent complexity, our approach yields invaluable insights into the impact of hypoxia and tissue stiffness on tumor cell dynamics when subjected to chemotherapeutic treatments, ultimately affirming the model's validity.

The probabilistic method that we use to calibrate our models falls in the category of Sequential Monte Carlo (SMC) methods~\mbox{\cite{Chopin.2002,DelMoral.2006}}. Using these within a Bayesian setting, a population of particles is initiated according to the prior distribution, that is a distribution on the parameters before having seen any data, and then evolved in sequential steps until it approximates the posterior, that is an update of the prior using the information provided by the data. Other methods within a similar spirit are annealed importance sampling~\cite{R.M.Neal.2001} and population Markov Chain Monte Carlo~\mbox{\cite{Liang.2001,Jasra.2007}}. An alternative to these would be more classical Markov Chain Monte Carlo (MCMC) methods, see for instance~\mbox{\cite[Ch. 6-7]{Robert.2004}}. The reason we opt for SMC in our setting is that, despite not having time series data, we can cast our problem in the framework of data assimilation. Then, intermediate calibration steps use only part of the data and require, accordingly, the evaluation of the forward model for the settings of that data, and this leads to an overall lower computational cost compared to MCMC. Since our parameter space is relatively high dimensional, in order to obtain robust results with a feasible number of particles, we use a combination of filtering (or data assimilation) and tempering~\cite{Kantas.2014} when defining the intermediate steps in SMC, and we improve on the adaptive strategies from~\cite{Beskos.2015}. The methods mentioned so far are fully Bayesian, in the sense that they aim at sampling from the full posterior. Examples of their use for biological and medical applications are~\mbox{\cite{Cho.2021,Falco.2023,lima2016selection,paun2021markov}}. An alternative would be to aim at approximate posteriors, especially if we know some characteristics a priori, for example unimodality. Within this strategy, a possibility is ensemble Kalman filtering (EnKF)~\cite{Evensen.2003}, which samples from a Gaussian approximation for the posterior, and needs a lot of care when quantifying uncertainty with nonlinear forward models~\mbox{\cite{ernst2015analysis,weissmann2022gradient}}. Another option when looking for approximate posteriors are variational methods, see for instance~\cite{jaakkola2000bayesian} and~\mbox{\cite{opper2009variational,el2012bayesian}}, which, with an approach closer to optimization than sampling, look for a posterior approximation within a family of probability distributions, and fit to the data the parameters characterizing that family.

The structure of this article is as follows. In Section 2, we provide a comprehensive overview of our methodology, delving into the intricate details of the experimental setup, the resulting data, their organization, the modeling framework, and the specific approach taken, which includes a reduced model version. Additionally, we elucidate the methods used for uncertainty quantification and parameter calibration. The results stemming from these approaches are presented in Section 3, followed by a comprehensive discussion in Section 4, which encapsulates our findings and ends with an outlook for the future. This paper is accompanied by supplementary material, and all labels that refer to the latter can be recognized by the prefix SM.

\section{Materials and Methods}
We first describe, in Section~\ref{ssec:expdata}, the \textsl{in vitro} experiments that have been used to validate and calibrate our mathematical model. Sections~\ref{ssec:model} resp.~\ref{sub:MOD-App2-sol} provide information on the model construction resp. its solution. The statistical procedure used for calibration, therefore bridging model and data, is described in Section~\ref{ssec:bayesian}. Finally, Section~\ref{sec:PAR-SMC} presents the implemented algorithm. 

\subsection{Experimental data}\label{ssec:expdata}

Our model calibrations use the raw data from the \textsl{in vitro} experiments described in \cite{Ozkan.2021}, of which we now recall the main features.

\paragraph*{Experimental setting.} Thanks to the \textsl{in vitro} setting, the cell population is placed in a particular, controlled environment, which allows for investigating the influence of ambient features, in our case the (combined) effect of tissue stiffness and oxygen level on the cells' chemoresistance. Namely, in order to mimic specific mechanical properties of the ECM~\cite{Antoine.2015}, the cell populations are seeded directly into a collagen hydrogel with a given stiffness, under an adjustable temperature, humidity and oxygen supply, which is later supplemented with chemotherapeutic drugs. The latter consist of a main drug, doxorubicin (DOX), and possibly a supportive drug, sorafenib (SOR).

\begin{figure}[H]
	\centering
		\includegraphics[width=\linewidth]{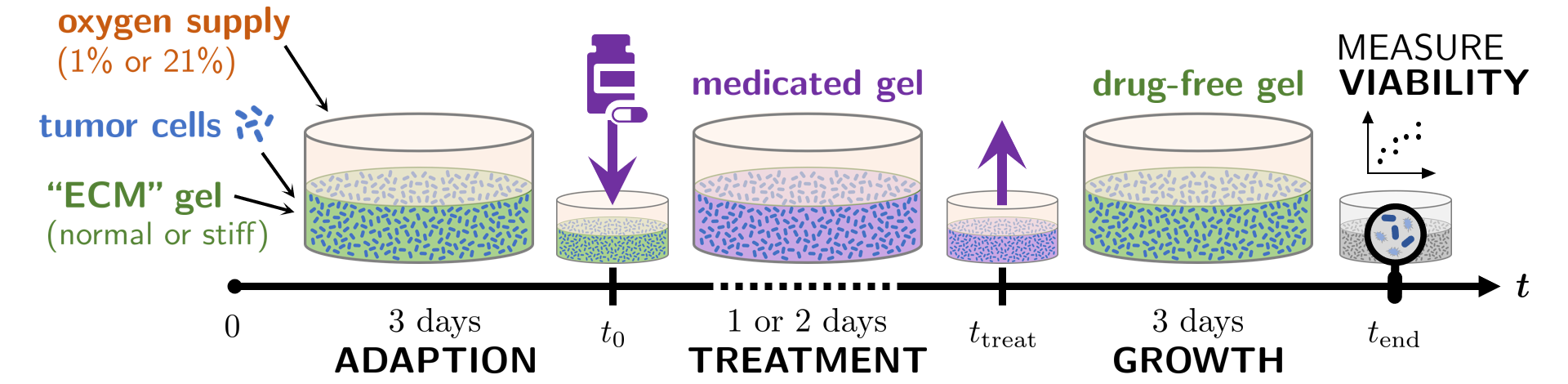}
		\caption[Phases of the underlying experimental setting of the chemoresistance models]{Schematic overview of the phases of the underlying experimental setting. Here $t$ denotes the time, measured in days.}
		\label{fig:chemoPhases}
\end{figure}

\begin{figure}[h]
	\centering
		\includegraphics[width=0.9\textwidth]{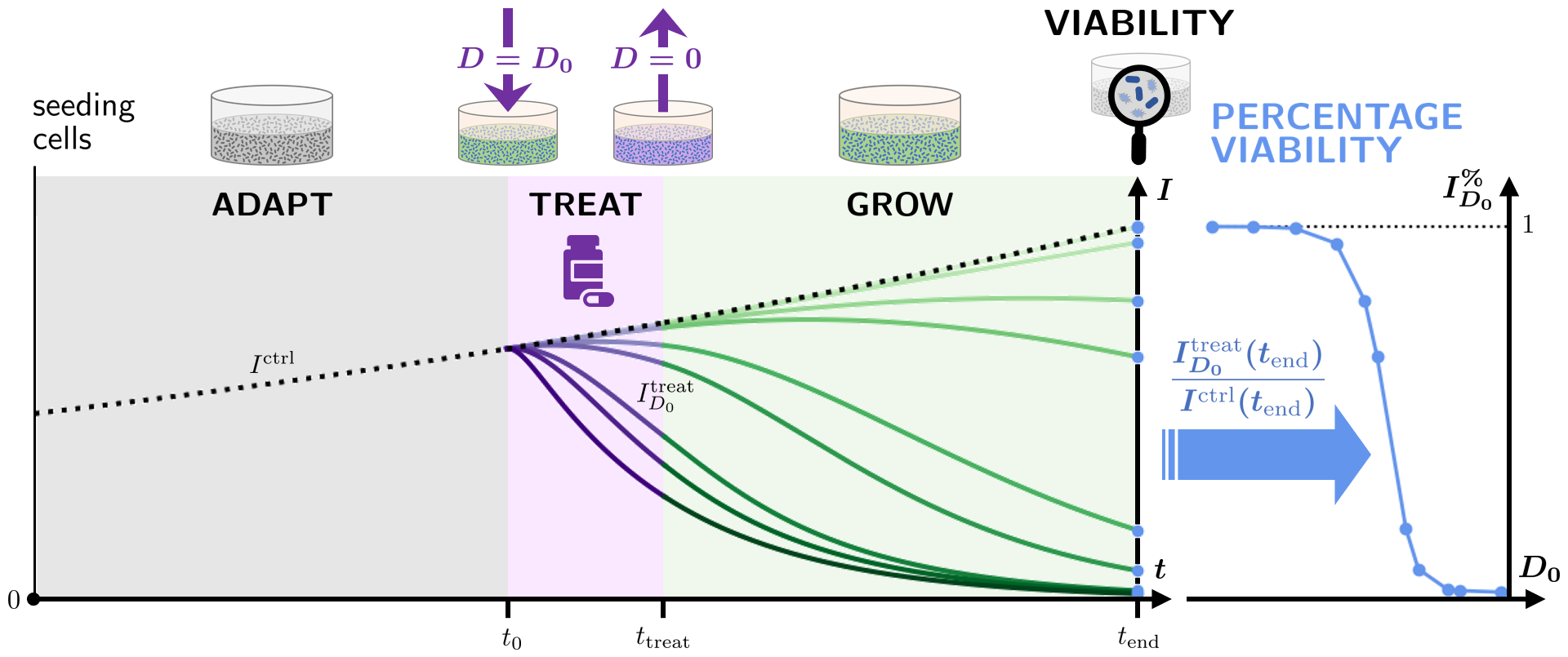}
		\caption[Experimental design yielding the model calibration data for models~\ref{eq:M-chemo-0} and~\ref{eq:M-chemo}]{Outline of the experimental design yielding the model calibration data
		. \\
		\textbf{Left plot:} Exemplary time evolution of the cell density over the course of the experiment (adaption/treatment/growth phase), showing the untreated growth of a control population (black dotted curve) compared to populations exposed to different DOX dosages~$\initD$ in the treatment phase~${T=[\initt,\ttreat]}$ (colored solid curves, where a darker shade indicates a higher dosage). Viability is measured once at~$\tend$. \\
		\textbf{Right plot:} Resulting percentage viability data in form of a dose-response relationship.}
		\label{Fig:chemoData}
\end{figure}

The experiments start from an initial cell density of \SI{}{\Cellss/\milli\litre} and can be divided into three phases, illustrated in Figure~\ref{fig:chemoPhases}:
\begin{enumerate}
	\item \emph{Adaption phase}: The tumor cells grow under specific oxygen (normoxic/hypoxic) and stiffness (normal/cirrhotic) conditions for three days to reach native morphology before starting the drug treatment.
	\item \emph{Treatment phase}: Chemotherapy is applied for 24 or 48 hours with a certain dosage of DOX, potentially in combination with a normal/high dosage of SOR.
	\item \emph{Growth phase}: After treatment, remaining drugs are washed from the cells and the population is grown for further three days. Eventually, viability is measured to investigate the treatment effect. 
\end{enumerate}
This experiment is done for several cell lines, from which we use the data of three particular ones (Hep3B2, HepG2, C3Asub28). These especially differ in their expression of the drug metabolizing enzyme CYP3A4 (CYP in short): Hep3B2 shows no considerable CYP expression, while the other two do to a variable extent~\cite{Ozkan.2021}.
\paragraph*{Structure of the experimental data.} 
Viability is monitored via a CellTiter-Blue\textsuperscript{\textregistered} assay, resulting in fluorescence intensity measurements, denoted by~$I$, which are assumed to be proportional to the number of viable cells (for more details, see Subsection~\ref{sec:PAR-Unc}). The data used for calibration is given as \emph{percentage viabilities} at a specific time point~$\tend$, i.e. the ratio between the measured intensities/viabilities of a treated population (\qm{treat}) and of a control population (\qm{ctrl}) without any treatment, neither DOX nor SOR:
\begin{equation}
	I^\%_{\initD}=\dfrac{\txtTop{I}{treat}_{\initD}(t_\text{end})}{\txtTop{I}{ctrl}(t_\text{end})}\in[0,1]
	\,.\label{eq:percI}
\end{equation}
By performing the experiment described above for various DOX dosages, denoted by~$\initD$\,, we get a discrete representation of a reversed dose-response relationship (usually, a \qm{classical} dose-response is given by a term like~${1-I^\%_{\initD}}$), where~${I^\%_{\initD}\equiv0}$ represents the maximal response to DOX, while~${I^\%_{\initD}\equiv1}$ means no response. Figure~\ref{Fig:chemoData} illustrates the experimental design and the resulting percentage viability data, which is used for model calibrations. 
\noindent Repeating this procedure for different conditions regarding the supportive drug dosage~$\initS$, treatment duration~$\ttreat$, oxygen supply~$\initH$ and ECM stiffness~$\initC$, allows to investigate the environment's influence on the treatment efficacy. Per cell line, this results in measurements~$I^\%_{\initD}$ for each combination of \enlargethispage{\baselineskip}
\begin{itemize}\setlength\itemsep{0pt}
	\renewcommand{\labelitemi}{\tiny$\blacksquare$}
	\item~$\initD \in\{0.0001,0.001,0.01,0.1,0.5,1,5,10,50,100,1000\}$ (\SI{}{\micro\molar}) DOX dosage,
	\item~$\initS \,\in\{0,0.5,1\}\propto$ 0/11/22~\SI{}{\micro\molar} SOR dosage (supportive drug),
	\item~$\ttreat\in\{1,2\}\propto$ 24/48 hours treatment duration,
	\item~$\initH \in\{0,1\}\propto$ normoxic/hypoxic oxygen supply,
	\item~$\initC \,\in\{0,1\}\propto$ normal/cirrhotic ECM stiffness,
\end{itemize}
i.e. in total we have data for~${11\cdot 3 \cdot 2 \cdot 2\cdot 2=264}$ different environmental settings. The details on the derivation of the notations~${\initD, \initS, \ttreat, \initH, \initC}$ and their interpretation will be provided in Subsection~\ref{sec:MOD-App2}.

%
\paragraph*{Utilization of the data.} For each combination of the above quantities, there are three biological replicates. Due to the complexity of the chemoresistance models, we want to minimize the effect of biological variation and potential outliers on the model calibrations. Hence, we consider the median over the corresponding replicates instead of the separate measurements~$\txtTop{I}{treat}_{\initD,i}$ and~$\txtTop{I}{crtl}_i$~(\mbox{$i$-th} replicate) to construct the percentage viability~$I^\%_{\initD}$ of~\eqref{eq:percI}. Due to biological variation and measurement inaccuracies some (median) data points appear outside of the range of~${0-100\%}$. However, our subsequent models can only give percentage viabilities within this range by construction, i.e. we clip these data points to the interval~${[0.001,1]}$\,. 
Note that we cannot choose the lower bound of the data arbitrarily small without excessively increasing the noise variance, which could lead to numerical issues in the calibration algorithm. This is due to the multiplicative noise modeling, see \eqref{eq:noisemodel}, which is explained in more detail in later Subsection~\ref{sec:PAR-Unc}.

\subsection{Modeling the influence of hypoxia and tissue stiffness on chemoresistance}\label{ssec:model}

In order to systematically model the impact of various factors in the tumor microenvironment (TME) on cell growth, we have adopted the comprehensive framework presented in \cite{Schonfeld.2022}. This framework centers around the concept of environmental stress level (Subsection~\ref{sec:MOD-general}) and forms the basis for our research. We subsequently apply this framework to the scenario previously outlined, with a specific focus on factors such as ECM stiffness, oxygen levels, and drug concentration (Subsection~\ref{sec:MOD-App2}).
Furthermore, we explore the implications of this model when cells lack the expression of the drug metabolizing enzyme CYP3A4. In our calibration process, we utilize the full model for one of the cell lines, HepG2, while employing the simplified model for the other, Hep3B2, lacking CYP enzyme expression.

\subsubsection{General approach using the environmental stress level}\label{sec:MOD-general}
Let us shortly summarize the modeling framework introduced in the work \cite{Schonfeld.2022}, to which we refer for further explanations. We use the term \emph{environmental factors} to denote the features of the TME which potentially affect the tumor cells' survival in a harmful or beneficial way. Here, these will correspond to the oxygen saturation, drug concentrations and ECM stiffness. Each environmental factor, which should be considered in the model, is mathematically represented by a time-dependent system variable:~${E_1(t)\Ldots E_n(t)}$, where~${n\in\mathbb{N}}$ is the number of factors/variables and~${t\geq 0}$ is the time. These quantities might influence each other as well as be influenced by present tumor cells. Such dynamics can be captured by a respective reaction function~$g_j$~(${1\leq j \leq n}$), leading to an initial value problem for each variable of the form
\begin{equation}
	\dot{E}_j = g_j(E_1\Ldots E_n,V,t)\,,\quad E_j(\initt)=E_{j,0}\,, \label{eq:reactEj}
\end{equation}
with~${V=V(t)\geq 0}$ being the density of viable (i.e. alive) tumor cells and~${\initt}$ an initial time point.

The collective influence of all environmental variables on the tumor cells is modeled by an auxiliary time-dependent variable~${\stress=\stress(t)}$, the  \emph{environmental stress level} or \emph{ESL} for short. It is an immeasurable quantity and describes how stressful the conditions, generated by the present environmental factors, are for viable cells. By definition, the ESL is 
\NEW{bounded by~${0\leq \stress(t)\leq 1~\forall t}$}, where the bounds~${\stress=0}$ resp.~${\stress=1}$ mean that cells find optimal resp. most inexpedient survival conditions. Note that \qm{most inexpedient} always has to be seen in the context of the given environmental factors, since the ESL is normalized. This means that there is no point in comparing the ESL for two settings which respectively consider different sets of environmental factors. Given the ESL $\eta$, we model the growth of viable cells over time as
\begin{equation}
	\dV =(1-\stress)\cdot\prol\, V\left(1-\left(\frac{V}{\Kv}\right)^b \right)-(\nat+\stress\cdot\indDeath)\, V\,.\label{eq:ODE-V}
\end{equation}
The right hand side of this ODE is a combination of a generalized logistic growth and an 
exponential death term. 
In the logistic growth, ${\prol}$ is the maximal possible growth rate, $\Kv$ the corresponding carrying capacity of the
biological system and $1/b$, with ${b>0}$, is a dimensionless measure for the strength of contact inhibition. 
In the exponential death term, $\nat$ denotes the natural death rate and~$\indDeath$ the maximal possible death rate due to the stressful environment. As a higher level of stress inhibits proliferation and potentially promotes death, the ESL~$\stress$ scales the rates~$\prol$ and~$\indDeath$ accordingly. 

To describe how exactly the environmental factors influence the ESL, we use \emph{influence functions}. They are defined, for each~$E_j$ (${1\leq j\leq n}$), by
\begin{align*}
	\act_j: (E_1\Ldots E_n) &\mapsto \act_j(E_1\Ldots E_n)\in[0,1]\\
	\mbox{and}\quad\deact_j: (E_1\Ldots E_n) &\mapsto \deact_j(E_1\Ldots E_n)\in[0,1]\,,
\end{align*}
where the superscript \qm{$+$} resp. \qm{$-$} of~$\delta_j$ indicates whether the function describes a positive or negative influence of the variable~$E_j$ on the cells' survival. Hence, we call them \emph{positive/negative influence function} accordingly. An influence function~$\delta_j^\pm$ can depend on the other environmental variables~$E_i$~(${i\neq j}$), as well: for instance, the influence of a chemotherapeutic drug on the cells' viability might be affected by the presence of another drug. We use~$E_{1:n}$ as a short notation for~${E_1\Ldots E_n}$ in the following.

\begin{figure}[H]
	\centering
	\includegraphics[width=0.4\linewidth]{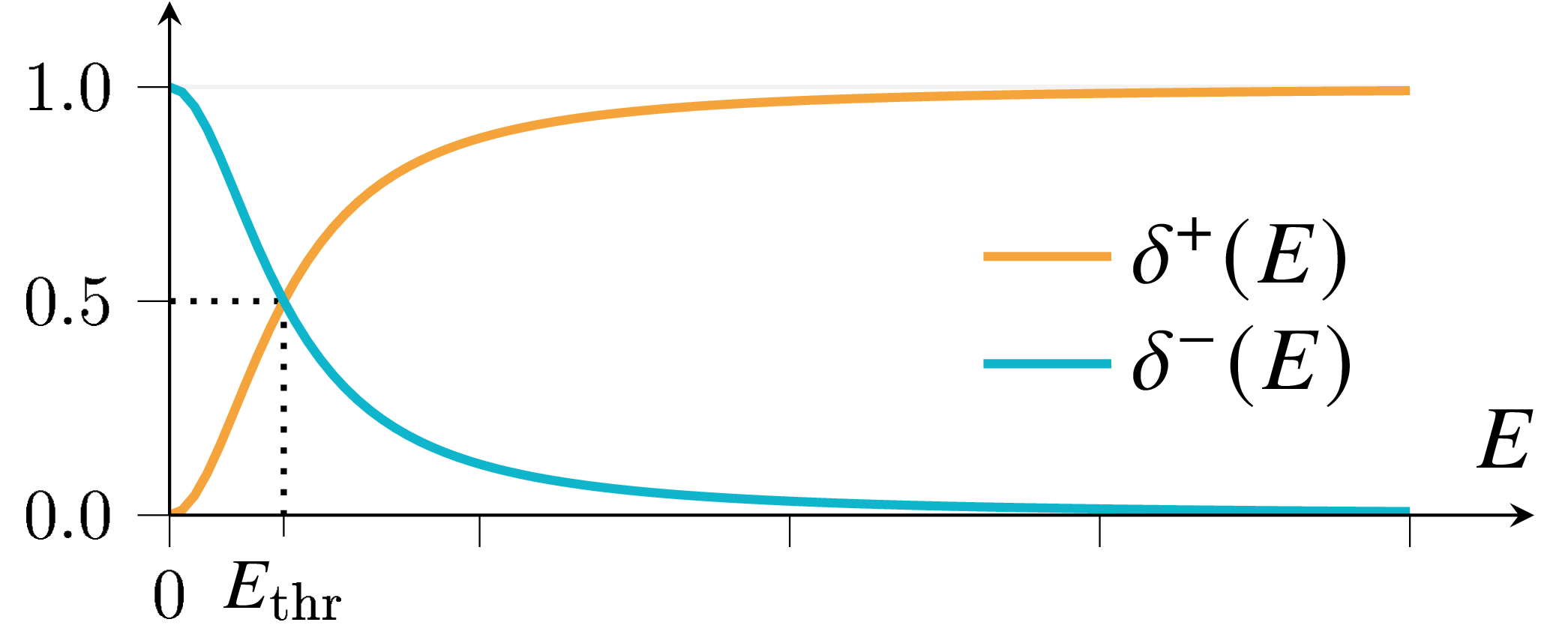}
	\caption[Exemplary influence functions~${\act(E)=\frac{E^2}{\txtSub{E}{thr}^2+E^2}}$ and~${\deact(E)=1-\act(E)}$]{Plots of exemplary influence functions~$\act,\deact$ for a system with only one (i.e.~${n=1}$) beneficial environmental factor~${E_1=E}$, here using~${\act(E)=\frac{E^2}{\txtSub{E}{thr}^2+E^2}}$ and~${\deact(E)=1-\act(E)}$.}
	\label{fig:beneficialE}
\end{figure}

We call an environmental factor \emph{beneficial} if increasing its corresponding variable~$E_j$ (with fixed~$E_i$,~${i\neq j}$) increases the value of~$\act_j$ and decreases the value of~$\deact_j$ (see Figure~\ref{fig:beneficialE}). We call it \emph{harmful} if it is the other way round. Examples for a beneficial and harmful environmental factor can be nutrient saturation and anti-cancer drug concentration, respectively.
\noindent To model the influence functions, 
we assume that a response is triggered by a stimulus in a switch-like manner. Therefore, using ideas from so-called ultrasensitive responses~\cite{Ferrell.1998}, we use Hill-type functions to model this behavior. Namely, if~$E$ is the variable of a beneficial resp. harmful environmental factor, we define~${\act(E)=\mathcal{H}(E)}$ resp.~${\deact(E)=\mathcal{H}(E)}$, where
\begin{equation*}
	\mathcal{H}:~\NEW{\mathbb{R}_+\to[0,1]}\,,~~ E \mapsto \frac{E^m}{\txtSub{E}{thr}^m+E^m}\,,\quad \mbox{with }m> 1\,.
\end{equation*}
The (switch)~threshold~$\txtSub{E}{thr}$ can be interpreted as the critical value of~$E$, where the cells get influenced by this environmental factor.
Here, we model potential interactions of $E$ with other environmental factors by making $\txtSub{E}{thr}$ dependent on all these factors. The Hill coefficient~$m$ governs the cells' sensitivity to changes in~$E$. 
To capture the desired sigmoidal, switch-like behavior, we impose $m>1$.

Using the influence functions, we model the collective of positive and negative influences of all environmental factors on the cells' survival as the dynamics of the ESL:
\begin{equation}
	\dStress=\underbracket[0.187ex]{\left(\sum_{j=1}^{n}\alpha^-_j\cdot \delta_j^-(E_{1:n})\right)(1-\stress)}_{\substack{\text{increasing stress level} \\ \text{(stressful conditions)}}}\underbracket[0.187ex]{-\left(\sum_{j=1}^{n}\alpha^+_j\cdot \delta_j^+(E_{1:n})\right)\stress}_{\substack{\text{recovery from stress} \\ \text{(beneficial conditions)}}}\,,\quad\mbox{with }\stress(\initt)=\initSt \in[0,1]\,. \label{eq:ODE-eta}
\end{equation}
We call the parameters~$\alpha^-_j$ resp.~$\alpha^+_j$ \emph{impact rates}. These variable-specific rates describe how fast the stress level increases ($\alpha^-_j$) or decreases ($\alpha^+_j$) if the cells' viability is impacted negatively resp. beneficially by the environmental factor of the associated variable~$E_j$. A distinction between~$\alpha^-$ and~$\alpha^+$ per variable 
accounts for varying response rates of cells to increasingly stressful or beneficial environments.
For our application, it has been sufficient to assume these rates to be constant. However, for other biological processes, it might make sense to consider time-dependent impact rates, for instance, for tumor cells potentially building a resistance to a chemotherapeutic drug, see e.g.~\cite{Claret.2009}. The model \eqref{eq:ODE-eta} for the ESL dynamics ensures its positivity and boundedness via~${0\leq\eta(t)\leq 1}$, for all $t\geq t_0$, \cite{Schonfeld.2022}.

The final model consists of a combination of the ODEs~\eqref{eq:reactEj}, \eqref{eq:ODE-V} and~\eqref{eq:ODE-eta}, with respective initial conditions. In  \cite{Schonfeld.2022}, this model has been validated against experimental data for the case where nutrient saturation is the only environmental factor. There, it has also been compared to a standard model, not using the ESL approach, for that specific case. We refer to that previous work for a discussion of potential advantages of using an ESL-based model.

\subsubsection{Application to effect of ECM stiffness and hypoxia on response to chemotherapy}\label{sec:MOD-App2}
We now specify the general model introduced in the previous subsection to study the effect of oxygen level and ECM stiffness on the response of tumor cells to chemotherapeutic treatments. In the corresponding experimental setting from~\cite{Ozkan.2021}, for which we design a suitable mathematical model, the cell population size stays sufficiently smaller than the system's capacity~$(V\ll\Kv)$ during the whole experiment. As a feature of the logistic part of the ODE~\eqref{eq:ODE-V}, this allows to simplify~\eqref{eq:ODE-V} to an initial value problem with a purely exponential, stress- and hence time-dependent rate~$\prol_\stress(t)$:
\begin{equation}
	\dV \overset{V\ll\Kv}{=} \Big(\big(1-\stress(t)\big)\cdot\prol-\big(\nat+\stress(t)\cdot\indDeath\big)\Big) \,V=\Big(\underbrace{\prol-\nat-\big(\prol+\indDeath\big)\stress(t)}_{\prol_\stress(t)}\Big)\, V\,. \label{eq:ODEVchem}
\end{equation}
Depending on its sign,~$\prol_\stress(t)$ is the actually observable net growth/death rate. In a stress-free environment (i.e.~${\stress(t)\equiv0}$), the population grows with constant rate~${\prol_\stress(t)\equiv\prol-\nat>0}$. The population size starts to decline once the stress level exceeds a certain threshold:
\begin{equation}
	\stress(t)>\frac{\prol-\nat}{\prol+\indDeath}\quad\Rightarrow\quad \prol_\stress(t)<0\,. \label{eq:maxStressCH}
\end{equation}
\paragraph*{Environmental factors/variables.} 
The model aims at describing the influence of oxygen concentration, ECM stiffness and a combination therapy with two chemotherapeutic drugs on viable tumor cells. In this setting, the nutrient supply is maintained optimal for cell growth. Therefore, it does not influence the stress level and hence it is not necessary as an environmental variable in the model. In total, we need four environmental variables~${E_1\Ldots E_4}$. The first two variables~$E_1$ resp.~$E_2$ represent the dosage of the drugs doxorubicin (DOX) and sorafenib (SOR), respectively. The oxygen supply varies between normoxic (${21\%\,\chem{O_2}}$) and hypoxic (${1\%\,\chem{O_2}}$) conditions. For an easier mathematical description, the corresponding variable~${E_3(t)\in[0,1]}$ describes the present \emph{hypoxia level}, i.e.~${E_3\equiv0}$ resp.~${E_3\equiv1}$ represent a normoxic resp. hypoxic environment. Similarly, the last variable~${E_4(t)\in[0,1]}$ models the \emph{cirrhosis level}, as the ECM stiffness ranges between normal (${E_4\equiv0}$) and cirrhotic (${E_4\equiv1}$) conditions. For better readability, instead of~${E_1\Ldots E_4}$ we use more intuitive notations for the environmental variables: 
\begin{alignat*}{3}
	\text{\underline{d}oxorubicin concentration: } D(t)&=E_1(t)\,, &\qquad \text{\underline{s}orafenib concentration: } S(t)&=E_2(t)\,, \\
	\text{{\underline{h}ypoxia level}: } H(t)&=E_3(t)\,, & \text{{\underline{c}irrhosis level}: } C(t)&=E_4(t)\,.
\end{alignat*}
During all experiments, the cells grow under constant oxygen (either normoxic or hypoxic) and stiffness (either normal or cirrhotic) conditions, i.e.
\begin{equation*}
	H(t)=H_0\in\{0,1\} \quad\text{and}\quad C(t)=C_0\in\{0,1\}\quad \forall\,t\geq 0\,.
\end{equation*}
\paragraph*{Influence on the tumor cells.} Following the definition of the general stress equation~\eqref{eq:ODE-eta}, we now consider an ODE of the form
\begin{equation*}
	\dStress=\Big(\sensRate{D}^-\deact_D+\sensRate{D}^-\deact_S+\sensRate{H}^-\deact_H+\sensRate{C}^-\deact_C\Big)\cdot(1-\stress)-\Big(\alpha^+_D\act_D+\alpha^+_S\act_S+\sensRate{H}^+\act_H+\sensRate{C}^+\act_C\Big)\cdot\stress\,,
\end{equation*} 
with~${\stress(0)=0}$. In general, each influence function can depend on all environmental variables, which means~${\deact_*=\deact_*(D,S,H,C)}$ and~${\act_*=\act_*(D,S,H,C)}$, with~${*\in\{D,S,H,C\}}$.
To specify these functions, we have to consider the three phases (adaptation, treatment and growth) of the underlying experiment as explained in previous Subsection~\ref{ssec:expdata}, which also yield different reaction terms for each environmental variable.

\smallskip
\textsl{Adaption phase:~${t\in[0,\initt]}$.} There are no drugs involved in this phase, hence the only environmental variables influencing the ESL are the hypoxia level~${H\equiv\initH}$ and cirrhosis level~${C\equiv\initC}$, and the reaction terms of the environmental variables are all zero:
\begin{equation*}
	\dD\equiv0\,,\qquad\dS\equiv0\,,\qquad\dot{H}\equiv0\,,\qquad\dot{C}\equiv0\,.
\end{equation*}
The cells were grown under optimal conditions before starting the experiment, i.e. there is no initial stress at the beginning of the adaption phase. Overall, this leads to the initial value problem
\begin{equation*}
	\dot\stress= \Big(\sensRate{H}^-\cdot\deact_H+\sensRate{C}^-\cdot\deact_C\Big)\cdot(1-\stress)-\Big(\sensRate{H}^+\cdot\act_H+\sensRate{C}^+\cdot\act_C\Big)\cdot\stress\,,\quad\stress(0)=0\quad\text{for }t\in[0,\initt]\,.
\end{equation*}
We assume that \NEW{the cells' reaction to hypoxia resp. cirrhosis is not influenced by the stiffness resp. oxygen supply, i.e. that }the influence functions of~$H$ do not depend on~$C$ and vice versa:
\begin{equation*}
	\deact_H=\deact_H(H)\,,~	\act_H=\act_H(H)\,,~\deact_C=\deact_C(C)\text{ and }\act_C=\act_C(C)\,.
\end{equation*}
The adaption phase allows the tumor cells to reach native morphology, i.e. to completely adapt to normoxic/hypoxic and normal/cirrhotic conditions before the drug treatment is started. Therefore, it is reasonable to assume that the stress level~$\stress$ has reached its steady state not later than the end of this phase, i.e. with~${H\equiv\initH}$ and~${C\equiv\initC}$ we have
\begin{equation*}
	\dot\stress\big|_{\stress=\stress(\initt)}=0\quad\Rightarrow\quad\stress(\initt)=\frac{\sensRate{H}^-\cdot\deactFct{H}{\initH }+\sensRate{C}^-\cdot\deactFct{C}{\initC }}{\sensRate{H}^-\cdot\deactFct{H}{\initH }+\sensRate{C}^-\cdot\deactFct{C}{\initC }+\sensRate{H}^+\cdot\actFct{H}{\initH }+\sensRate{C}^+\cdot\actFct{C}{\initC }}\,.
\end{equation*}
If we assume that the cells react with the same sensitivity to stress inducing and stress reducing oxygen resp. stiffness conditions, i.e.~${\alpha_*^-=\alpha_*^+}$ (abbreviated~$\alpha_*$) and~${\deact_*=1-\act_*}$ for~${*\in\{H,C\}}$, this steady state simplifies further, leading to the stress level~$\stress_{HC}$ at the end of the adaption phase:
\begin{equation}
	\stress_{HC}(\initH ,\initC )=\stress(\initt)=\frac{\sensRate{H}\cdot\deactFct{H}{\initH }+\sensRate{C}\cdot\deactFct{C}{\initC }}{\alpha_H+\alpha_C}\in\left[0,\frac{\prol-\nat}{\prol+\indDeath}\right] \label{eq:initStress}
\end{equation}
under the constant environmental conditions~${H\equiv\initH}$ and~${C\equiv\initC}$. The upper bound of~$\stress_{HC}$ follows from relation~\eqref{eq:maxStressCH}, since it is reasonable to assume that stress induced solely by hypoxic and/or stiff environment will not lead to a declining population size. We do not have any particular information on how the influence functions~$\deact_H$ and~$\deact_C$ can be defined. Therefore, we will proceed with the more general notation~$\stress_{HC}$\,. Note that in the following we omit the arguments of~$\stress_{HC}$ for better readability. Analogously, all upcoming quantities depending on the constant environmental factors~$\initH$ and~$\initC$, are indicated by the subscript~\qm{$HC\,$}.

\begin{figure}[H]
	\centering
		\includegraphics[width=0.75\linewidth]{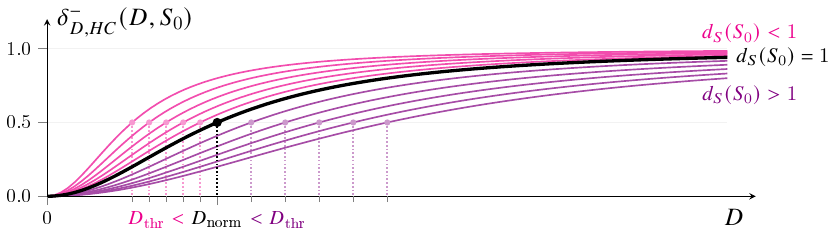}
		\caption[Illustration of how~${\DthrCH(\initS)}$ is shifted compared to~$\Dnorm$ by~$d_S(\initS)$.]{Illustration of how the threshold~${\Dthr=\DthrCH(\initS)}$ of the influence function~$\deact_{D,HC}$ is shifted compared to~$\Dnorm$ by~$d_S(\initS)$ for fixed~${d_H(H),d_C(C)=1}$\,.}
		\label{fig:deactD}
\end{figure}

\smallskip
\textsl{Treatment phase:~${t\in[\initt,\ttreat]}$.} After the cells have adapted to the oxygen and stiffness conditions (${H\equiv\initH }$ and~${C\equiv\initC }$), chemotherapeutic drugs are added to the biological system:
\begin{equation*}
	D(\initt)=\initD >0\qquad \text{and}\qquad S(\initt)=\initS\geq 0\,.
\end{equation*}
Both drugs might degrade over time with constant, \NEW{drug-specific} rates. Additionally, \NEW{liver cells, as they are used in the underlying experiments,} can express enzyme CYP3A4, which is able to metabolize drugs. In the human body, the half-life of DOX and SOR is approximately 1-2 days, where the metabolization in the liver via CYP3A4 plays a major role~\mbox{\cite{Gerson.2018, EMA.2022}}. Assuming constant exponential decay of the drugs, this half-life translates to total decay rates in the range of approximately~$\SIrange{0.35}{0.70}{\per\day}$. Since this decay includes both degradation and metabolization, it is reasonable to assume that the degradation rates are significantly smaller than these total decay rates. Hence, we choose to omit the degradation rates in our model and only focus on the metabolization rates. These are expected to be proportional to the cell-specific CYP expression. We see in~\cite{Ozkan.2021} that the latter can depend on the environmental factors~$\initH$ and~$\initC$, \NEW{which is modeled by} the following reaction terms for~$D$ and~$S$ in the treatment phase:
\begin{equation*}
	\dot{D}=-\metDCH\cdot D\quad \text{and}\quad \dot{S}=-\metSCH\cdot S\qquad \text{for }t\in[\initt,\ttreat],
\end{equation*}
with~$\metDCH$ and~$\metSCH$ being the environment-dependent metabolization rates of DOX and SOR, respectively. For~${t\in[\initt,\ttreat]}$, these ODEs have the analytical solutions
\begin{equation}
		D(t) = \initD \,\exp\big(-\metDCH\cdot t\big)\qquad\mbox{and}\qquad
		S(t) = \initS \,\exp\big(-\metSCH\cdot t\big)\,. 
	\label{eq:chemo-solDS}
\end{equation}
The stress level is now influenced only by the present drugs, since the cells have already adapted to the oxygen and stiffness conditions. Hence, we use the steady state~\eqref{eq:initStress} from the previous phase as initial condition for the ESL. It is reasonable to assume that the cells do not significantly recover from the drug treatment during the period of the experiment, i.e.~${\sensRate{D}^+,\sensRate{S}^+\approx 0}$\,. Overall, for~${t\in[\initt,\ttreat]}$ we arrive at
\begin{equation*}
	\dot\stress= \Big(\sensRate{D}^-\cdot\deact_D+\sensRate{D}^-\cdot\deact_S\Big)\cdot(1-\stress)\quad\text{with}\quad\stress(\initt)=\initStCH=\stress_{HC}\,.
\end{equation*}
Both, DOX and SOR can directly induce stress to the cells. As described in Subsection~\ref{sec:MOD-general}, Hill-type functions are a suitable choices for~$\deact_D$ and~$\deact_S$. In this context, the respective threshold values can be interpreted as a measure for the cells' chemoresistance to the corresponding drug. A large threshold means that the cells tolerate higher drug dosages before their survival is impaired. We know that the chemoresistance can be influenced by other environmental factors. On the one hand, we know that varying SOR dosage, oxygen supply and ECM stiffness can influence the chemoresistance of the tumor cells to DOX. On the other hand, the cytotoxic effect of SOR might be influenced by the present oxygen and stiffness conditions~\cite{Ozkan.2021}. Overall, this translates to
\begin{equation*}
	\deact_{D,HC}(D,S)=\frac{D^{m_1}}{\big(\DthrCH(S)\big)^{m_1}+D^{m_1}}\qquad\mbox{and}\qquad \deact_{S,HC}(S)=\frac{S^{m_2}}{\SthrCH^{m_2}+S^{m_2}}\,,
\end{equation*}
where the thresholds~${\DthrCH(S)=\DthrCH(S,\initH ,\initC )}$ and~${\SthrCH=\SthrCH(\initH ,\initC )}$ are functions depending on the environmental factors, which influence chemoresistance to the respective drug. In particular, we assume that each variable of the threshold's argument can shift the critical (switch) value of the corresponding Hill function. For DOX, this can be written mathematically as
\begin{equation}
	\DthrCH(S)=\txtSub{D}{norm}\cdot d_S(S)\cdot d_H(\initH )\cdot d_C(\initC )\,, \label{eq:Dthr}
\end{equation}
where each~$d_*$ is a positive function, describing how the respective variable~${*\in\{S,H,C\}}$ scales the \emph{unaffected threshold}~$\txtSub{D}{norm}$ (i.e. the chemoresistance under normal conditions, unaffected by other environmental influences):
\begin{align*}
	d_* <1 \qquad&\Rightarrow \mbox{ reduces chemoresistance,}\\
	d_* =1 \qquad&\Rightarrow \mbox{ does not influence chemoresistance,}\\
	d_* >1 \qquad&\Rightarrow \mbox{ increases chemoresistance.}
\end{align*}
Therefore,~$d_*(0)=1$ should hold by definition. Figure~\ref{fig:deactD} visualizes the change of chemoresistance to DOX for an example case of only SOR influencing the threshold.
\noindent For defining~$d_S$, we know that a sufficient dosage of SOR can decrease the cells' chemoresistance to DOX, i.e.~$d_S$ is a monotonically decreasing function on~$[0,1]$. To implement an appropriate dose-response relation, we can again use a Hill-type function:
\begin{equation*}
	d_S(S)=1-\frac{\NEW{\txtSub{a}{max}} S^{m_3}}{\Ssupp^{m_3}+S^{m_3}}\,,\qquad(m_3>0)\,,
\end{equation*} 
where~$\Ssupp$ is the critical dosage of SOR, after which the chemoresistance to DOX is significantly influenced. \NEW{The parameter~$\txtSub{a}{max}\in(0,1)$ ensures that~$d_S$ does get arbitrarily small for very large~$S$.} For the remaining functions~$d_H$ and~$d_C$ in~\eqref{eq:Dthr}, we do not have enough understanding of how hypoxia and/or high ECM stiffness influence chemoresistance to explicitly define them. Therefore, for better readability, we summarize them in a shorter notation:
\begin{equation*}\label{eq:dHC}
	\dCH=d_H(\initH)\cdot d_C(\initC)\,.
\end{equation*}
For SOR, we could define~$\SthrCH$ similarly to~\eqref{eq:Dthr}, but since we do not have further information to explicitly define appropriate scaling functions to shift the respective threshold, we proceed with the more general notation~$\SthrCH$. Altogether, in the treatment phase, i.e. for~${t\in[\initt,\ttreat]}$, we consider the ODE system
\begin{align}
	&\left\{\begin{aligned}
			\dV &= \Big(\prol-\nat-\big(\prol+\indDeath\big)\stress\Big)\,V\,, \\[4pt]
			\dStress &= \left(\frac{\sensRate{D}^- D^{m_1}}{\big(\DthrCH(S)\big)^{m_1}+D^{m_1}}+\frac{\sensRate{S}^-S^{m_2}}{\SthrCH^{m_2}+S^{m_2}}\right)(1-\stress)\,,
		\end{aligned} \right. & \label{eq:ODEtreat}
\end{align}
with~$D$ and~$S$ as given in~\eqref{eq:chemo-solDS} and the initial conditions~${V(\initt) =\initV}$ and~${\stress(\initt)=\initStCH=\stress_{HC}}$\,.

\smallskip
\textsl{Growth phase:~${t\in[\ttreat,\tend]}$.}
At the beginning of the experiment's last phase, all drugs are removed from the cells' environment, motivating the initial conditions~${D(\ttreat),S(\ttreat)=0}$\,. The cell population is left to grow under drug-free conditions for the remaining time of the experiment, meaning the reaction terms of the environmental variables are again all zero:
\begin{equation*}
	\dD\equiv0\,,\qquad\dS\equiv0\,,\qquad\dot{H}\equiv0\,,\qquad\dot{C}\equiv0\,.
\end{equation*}
For the other variables, the corresponding solutions~$V(\ttreat)$ and~$\stress(\ttreat)$ from system~\eqref{eq:ODEtreat} are used as initial conditions. Since the stress level is not influenced by treatment anymore and the cells are still fully adapted to oxygen and stiffness conditions, the ESL stays constant. Hence, we can use a single ODE to model the dynamics of the growth phase, i.e. for~${t\in[\ttreat,\tend]}$:
\begin{equation}\label{eq:ODEgrow}
	\dV = \Big(\prol-\nat-\big(\prol+\indDeath\big)\,\stress(\ttreat)\Big)\,V\,.
\end{equation}
After numerically solving ODE system~\eqref{eq:ODEtreat} from the treatment phase to obtain~$V(\ttreat)$ and~$\stress(\ttreat)$, the initial value problem above has the analytical solution
\begin{equation*}
	V(t)=V(\ttreat)\cdot\exp\bigg(\Big(\prol-\nat-\stress(\ttreat)\cdot\big(\prol+\indDeath\big)\Big)\cdot (t-\ttreat)\bigg),\quad\mbox{for }t\in[\ttreat,\tend]\,.
\end{equation*}
%

\paragraph*{Cell line-specific chemoresistance models.} We collect all results from the previous paragraphs into one model~\ref{eq:M-chemo}. Since the dynamics of the adaption phase are covered by the value of~${\stress(\initt)=\initStCH}$, we set the initial time of this model to~${\initt=0}$. By denoting the treatment phase with~${T=[\initt=0,\ttreat]}$ \NEW{and~${\mathbbm{1}_T(t)}$ the indicator function on~$T$}, we achieve the \emph{complete chemoresistance model} for~${t\geq 0}$:
\leqnomode
\begin{align}\label{eq:M-chemo}\tag{$\mathcal{M}_{DS}^\text{CYP}$}
	&\hspace*{28pt}\left\{\begin{aligned}
		\dV &= \Big(\prol-\nat-\stress\cdot \big(\prol+\indDeath\big)\Big)\, V\,, \\[4pt]
		\dStress &= 
		\left(\frac{\sensRate{D}^- D^{m_1}}{\left(\DnormCH\left(1-\dfrac{\txtSub{a}{max} S^{m_3}}{\Ssupp^{m_3}+S^{m_3}}\right)\right)^{m_1}\hspace*{-6pt}+D^{m_1}}+\frac{\sensRate{S}^- S^{m_2}}{\SthrCH^{m_2}+S^{m_2}}\right)(1-\stress)\,,\\[4pt]
		D(t)&= \initD \,\exp\big(-\metDCH\cdot t\,\big)\cdot\mathbbm{1}_T(t)\,,\\[2pt]
		S(t)&= \initS \,\,\exp\big(-\metSCH\cdot t\,\big)\cdot\mathbbm{1}_T(t)\,,
	\end{aligned} \right. \hspace*{-1.5cm}&
\end{align}
\reqnomode
with short notation~${\DnormCH=\Dnorm\cdot \dCH}$ and the initial conditions
\begin{equation*}
	V(0) =\initV \qquad \mbox{and} \qquad
	\stress(0)=\stress_{HC} 
	\overset{\eqref{eq:initStress}}{\leq} \frac{\prol-\nat}{\prol+\indDeath}\,.
\end{equation*}
The model variables and parameters are summarized in the supplemented Tables~\ref{Tab:model2Vars} resp.~\ref{Tab:model2Pars}. We will refer to~$\DthrCH$ resp.~$\SthrCH$ as \emph{DOX/SOR susceptibility threshold} and to~$\deact_{D,HC}$ resp.~$\deact_{S,HC}$ as \emph{DOX/SOR dose-response function} to clarify their role in the underlying biological setting.
%
In general, model~\ref{eq:M-chemo} is suitable for investigating all cell lines of interest. However, further cell line-specific adaptions of this model might be useful or even necessary, which will be explained in the following paragraphs.

\smallskip
\textsl{Cell line Hep3B2.} Measurements quantifying the CYP expression from~\cite{Ozkan.2021} show that cell line Hep3B2 does not express CYP for any combination of oxygen and stiffness conditions. Hence, for this cell line the drug concentrations stay constant during the whole treatment phase:~${D(t)=\initD}$ and~${S(t)=\initS}$ for~${t\in[0,\ttreat]=T}$. Omitting the ODEs for~$D$ and~$S$ in~\ref{eq:M-chemo} yields the following \emph{reduced chemoresistance model} for~${t\geq 0}$:
\leqnomode
\begin{align}\label{eq:M-chemo-0}\tag{$\mathcal{M}^0_{DS}$}
	&\hspace*{20pt}\left\{~\begin{aligned}
		\dV &= \Big(\prol-\nat-\big(\prol+\indDeath\big)\stress\Big)\, V\,, \\[4pt]
		\dStress &= \left(\frac{\sensRate{{D}}^- \initD^{m_1}}{\left(\DnormCH\left(1-\dfrac{\txtSub{a}{max} \initS^{m_3}}{\Ssupp^{m_3}+\initS^{m_3}}\right)\right)^{m_1}\hspace*{-6pt}+\initD^{m_1}}+\frac{\sensRate{S}^- \initS^{m_2}}{\SthrCH^{m_2}+\initS^{m_2}}\right) (1-\stress)\,\mathbbm{1}_T(t)\,,\\[4pt]
		&\mbox{with }V(0)=\initV \,,~\stress(0)=\initStCH  \,.
	\end{aligned} \right. \hspace*{-2cm}&
\end{align}
\reqnomode

\smallskip
\textsl{Cell line HepG2.} In contrast to Hep3B2, HepG2 show significant CYP expression~\cite{Ozkan.2021}, i.e. for this cell line it is reasonable to use the complete model~\ref{eq:M-chemo}. Attempts to estimate the parameters of that model for HepG2 
exposed that the available data is not informative enough to do so. In particular, the calibration algorithm shows difficulties to properly estimate the drug impact rates~$\sensRate{D}^-$ and~$\sensRate{S}^-$ as well as the metabolization rates~$\metD$ and~$\metS$ (possible reasons are discussed in Section~\ref{sec:DIS-model}). Instead of explicitly estimating the drug metabolization rates, we aim for investigating potential correlations between model parameters and the available measurements of CYP expression from~\cite{Ozkan.2021} by using the reduced model~\ref{eq:M-chemo-0} for HepG2 as well. 
However, the corresponding calibrations repeatedly failed for certain measurements.

\begin{figure}[H]
\centering
\includegraphics[width=\linewidth]{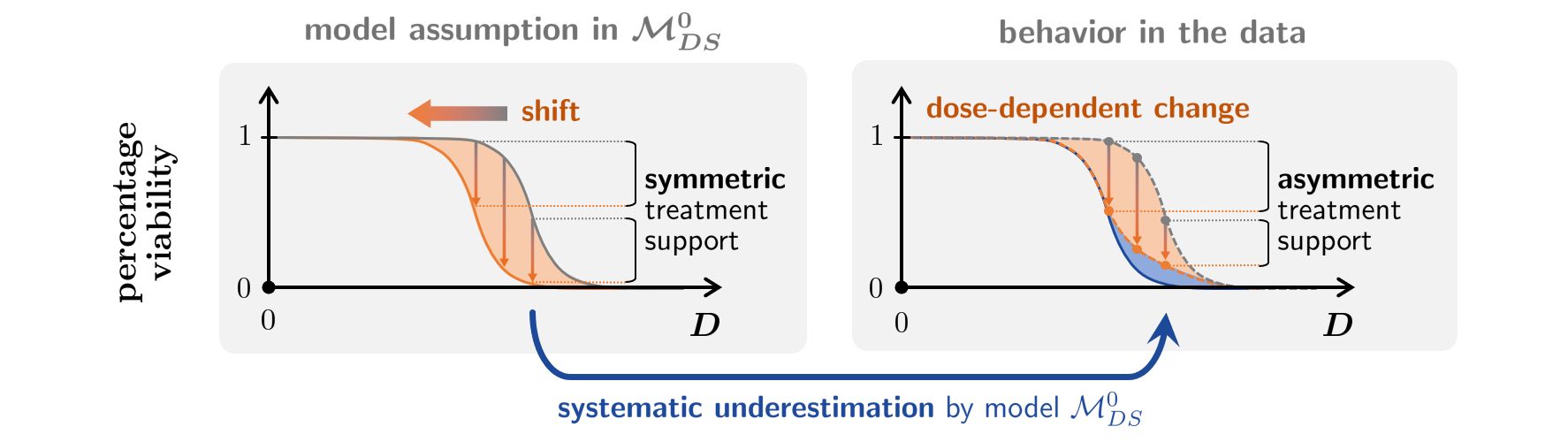}
\caption[How model~\ref{eq:M-chemo-0} fails to appropriately describe the trend of the HepG2 data]{Schematic description of how model~\ref{eq:M-chemo-0} fails to appropriately describe the trend of the HepG2 data in terms of the supportive effect of SOR. The approach of shifting the DOX susceptibility threshold by~${d_S(S)}$ leads to a systematic underestimation of the measurements by the model solution for high DOX dosages.} 
\label{Fig:CL2-supp}
\end{figure}

A closer investigation of the respective data points suggests that the modeling approach for the supportive effect of SOR in model~\ref{eq:M-chemo-0} might be too restrictive to appropriately replicate the measured cell behavior. The observed discrepancy is illustrated in Figure~\ref{Fig:CL2-supp}.
\noindent In particular, by shifting the DOX susceptibility threshold~$\DnormCH$ of the dose-response function~${\deact_{D,HC}(D,\initS)}$ with~${d_S(\initS)\in(0,1]}$, the treatment support shows itself by an analogous shift of the percentage viability. This means, except for the horizontal shift, the shape of the unsupported (gray graph) and supported (orange graph) percentage viability function is the same, which results in a \qm{symmetric} supportive effect (left side of Figure~\ref{Fig:CL2-supp}). By contrast, the measurements indicate an asymmetric supportive influence which can be weaker for larger DOX dosages (right side of Figure~\ref{Fig:CL2-supp}).

\begin{figure}[H]
	\centering
	\includegraphics[width=0.96\linewidth]{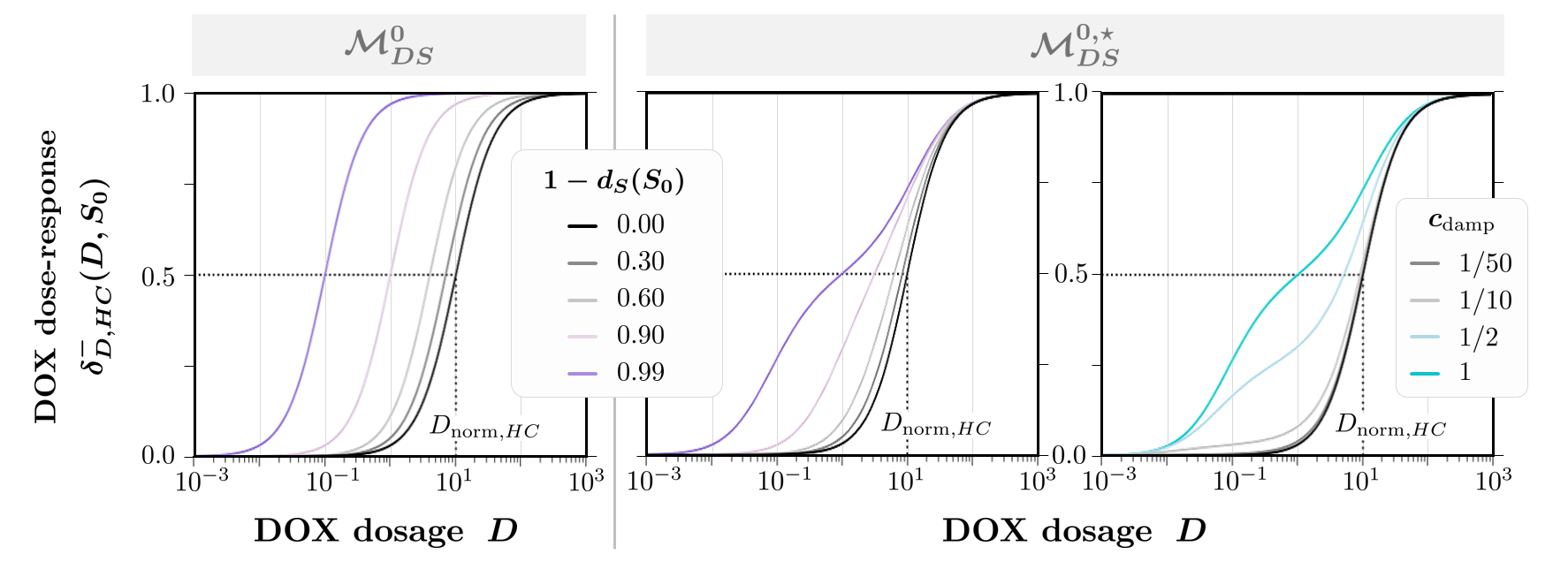}
	\caption[Different approaches of~\ref{eq:M-chemo-0} and~\ref{eq:M-chemo-0-inf} describing the supported DOX dose-response]{Comparison of the different approaches of models~\ref{eq:M-chemo-0} (left) and~\ref{eq:M-chemo-0-inf} (middle and right) to describe the supported DOX dose-response. Each plot illustrates the influence of the terms~${1-d_S(\initS)}$ resp.~$\cdamp$ on the dose-response curve (for all plots:~${\DnormCH=1}$\,, middle plot:~${\cdamp=1}$\,, right plot:~${1-d_S(\initS)=0.99}$).}
	\label{Fig:CL2-damp}
\end{figure}

To accomodate this, we start from the reduced model \eqref{eq:M-chemo-0} and change the modeling of the supportive effect of SOR by introducing a DOX dose-dependent damping factor by \enlargethispage{\baselineskip}
\begin{align*}
	d_S(S)&=1-\,\dfrac{\txtSub{a}{max}\cdot{S}^{m_3}}{\Ssupp^{m_3}+S^{m_3}}\\[2pt]
	\overset{~\text{adapt}~}{\longsquiggly}~~ d_S^\star(S,D)&= 1-\underbracket[0.187ex]{\vphantom{\left(1-\dfrac{D}{\Ddamp+D}\right)}\,\dfrac{\txtSub{a}{max}\cdot{S}^{m_3}}{\Ssupp^{m_3}+S^{m_3}}\,}_{\substack{\text{supportive effect}\\ 1-d_S(S)}}\cdot\underbracket[0.187ex]{\left(1-\dfrac{D}{\Ddamp+D}\right)}_{\substack{\text{damping factor}\\\in(0,1]}}\,,\quad\mbox{with }\Ddamp\in(0,\DnormCH]\,,
\end{align*}
and propose the adjusted model
\leqnomode
\begin{align}\label{eq:M-chemo-0-inf}\tag{$\mathcal{M}^{0,\star}_{DS}$}
	&\hspace*{10pt}\left\{~\begin{aligned}
		\dV &= \Big(\prol-\nat-\stress\cdot \big(\prol+\indDeath\big)\Big)\cdot V\,, \\[4pt]
		\dStress &= \left(\frac{\sensRate{D}^-\cdot \initD^{m_1}}{\big(\DnormCH \cdot d^\star_S(\initS,\initD)\big)^{m_1}+\initD^{m_1}}+\frac{\sensRate{S}^-\cdot \initS^{m_2}}{\SthrCH^{m_2}+\initS^{m_2}}\right)\cdot (1-\stress)\cdot\mathbbm{1}_T(t)\,,\\[4pt]
		&\mbox{with }V(0)=\initV \,,~\stress(0)=\initStCH  \,.
	\end{aligned} \right. \hspace*{-2cm}&
\end{align}
\reqnomode
Figure~\ref{Fig:CL2-damp} visually compares the supportive effect as described by both models~\ref{eq:M-chemo-0} and~\ref{eq:M-chemo-0-inf} as well as shows the role of the term~${1-d_S(\initS)}$ and the damping threshold reparameterized by\enlargethispage{\baselineskip}
\begin{equation}
	\Ddamp=\DnormCH\cdot\cdamp\,,\qquad\mbox{with }\cdamp\in(0,1]\,. \label{eq:CL2-repar}
\end{equation}
\noindent For both models, the term~${1-d_S(\initS)}$ is a measure for the strength of the supportive effect. Furthermore, we see that, the smaller~$\Ddamp$ in comparison to~$\DnormCH$ (i.e. the smaller~$\cdamp$), the stronger the damping. Note that the constraint~${\Ddamp\leq \DnormCH}$\,, which basically assumes a base degree of damping, is not necessary from a mathematical perspective. For larger values of~$\Ddamp$\,, the shape of~$d^\star_S$ progressively approaches the one of the undamped one~$d_S$. However, pre-calibrations gave no indication that~${\Ddamp\leq \DnormCH}$ is too restrictive. Hence, we use the reparametrization~\eqref{eq:CL2-repar} for our calibrations.
\paragraph*{Modeling drug metabolization.} Since, according to~\cite{Ozkan.2021}, the cell line HepG2 actually exhibits a considerable CYP expression, we still want to investigate this effect even though it is not included explicitly in \ref{eq:M-chemo-0-inf}. 
In model~\ref{eq:M-chemo}, the resulting drug metabolization would have been incorporated via exponential decay, i.e. 
\begin{equation*}
	D(t)=\initD\cdot e^{-\metD t}\qquad\text{and}\qquad S(t)=\initS\cdot e^{-\metS t}
\end{equation*}
with the drug-specific metabolization rates~${\metD,\metS\geq 0}$. Without loss of generality, let~${\phi\in\{D,S\}}$ be the dosage of DOX or SOR and~$\met_\phi$ its respective metabolization rate. Then, if we consider a Hill function in~$\phi$ with threshold~${\txtSub{\phi}{thr}>0}$ and Hill coefficient~${m>0}$, drug metabolization would show as
\begin{equation*}
	\frac{\phi(t)^{m}}{\txtSub{\phi}{thr}^{m}+\phi(t)^{m}}=\frac{\init{\phi}^{m}\cdot e^{-m\met_\phi t}}{\txtSub{\phi}{thr}^{m}+\init{\phi}^{m}\cdot e^{-m\met_\phi t}}=\frac{\init{\phi}^{m}}{\left(e^{\met_\phi t}\cdot \txtSub{\phi}{thr}\right)^{m}+\init{\phi}^{m}}\,.
\end{equation*}
This corresponds to a Hill function in~$\phi$ assuming a constant dosage~${\phi(t)=\init{\phi}}$\,, but with an adapted threshold 
\begin{equation*}
	\phi^\text{CYP}_\text{thr}(t)=e^{\met_\phi t}\cdot \txtSub{\phi}{thr}
\end{equation*}
which increases exponentially over time, where~$\txtSub{\phi}{thr}$ is the threshold at the beginning of the treatment~(${t=0}$). Obviously, such a time-dependent dynamic is not considered in ~\ref{eq:M-chemo-0-inf}. Instead, it basically approximates~$\phi^\text{CYP}_\text{thr}$ by a constant. As~$t$ will progress from~$0$ to~${\ttreat\in\{1,2\}}$ during the treatment phase and~\ref{eq:M-chemo-0-inf} does not differentiate~$\phi^\text{CYP}_\text{thr}$ for varying treatment durations, it is reasonable to assume that the calibrations approximate~$t$ with a positive constant $\leq 2$. Hence, we propose
\begin{equation}
	\phi^\text{CYP}_\text{thr}\approx e^{\met_\phi}\cdot \txtSub{\phi}{thr}=\omega_\phi\cdot \txtSub{\phi}{thr}\qquad\Rightarrow\qquad\met_\phi\approx\ln(\omega_\phi)~~\mbox{with }{\omega_\phi\geq 1}\,. \label{eq:CYP-approx}
\end{equation}
As the variables~$D$ and~$S$ do not occur outside from Hill functions in~\ref{eq:M-chemo-0-inf}, the simplified consideration of the drug metabolization with this model can be summarized by
\begin{equation}
\begin{alignedat}{3}
	D^\text{CYP}_{\text{norm},HC}&\approx \MetD\cdot \DnormCH\,,\qquad & S^\text{CYP}_{\text{thr},HC}&\approx \MetS\cdot \SthrCH\,,\\
	D^\text{CYP}_{\text{damp}}&\approx \MetD\cdot \Ddamp\,,\qquad & S^\text{CYP}_{\text{supp}}&\approx \MetS\cdot \Ssupp\,,
\end{alignedat}\label{eq:CYPnoCYPrelation}
\end{equation}
with~${\MetD,\MetS\geq1}$ quantifying the drug metabolization, where a value of one represents no metabolic activity. We expect $\MetD$ and~$\MetS$ to be positively correlated to the CYP expression, i.e. more CYP yields a stronger effect of drug metabolization. In particular, let both
\begin{equation*}
	\MetD=\MetD(\CYP)\qquad\text{and}\qquad\MetS=\MetS(\CYP)
\end{equation*}
be monotonically increasing functions~${[0,\infty)\to[1,\infty)}$ in the CYP concentration~${\CYP\geq 0}$ which we suppose to be constant but potentially dependent on~$H$ and~$C$. We will use the short notations~${\MetDCH=\MetD(\CYP)}$ and~${\MetSCH=\MetS(\CYP)}$. 
Hence, we can obtain estimates for~$\MetDCH$ and~$\MetSCH$ by estimating the remaining parameters of~\eqref{eq:CYPnoCYPrelation}, namely by pairing the estimates of the threshold parameters on the left hand sides (the ones with the superscript \qm{CYP}) using model~\ref{eq:M-chemo-0-inf} and HepG2 data with the respective ones on the right hand sides (without superscript \qm{CYP}) using model~\ref{eq:M-chemo-0-inf} and Hep3B2 data. 

\smallskip
\textsl{Cell line C3Asub28.} There are indications that the supportive effect of SOR for cell line C3Asub28 shows a similar behaviour like HepG2. Hence, it is reasonable to use the enhanced reduced model~\ref{eq:M-chemo-0-inf} for C3Asub28 as well. However, in practice, we observe numerical problems when calibrating any of the above models with the data for C3Asub28. The details towards this issue will be discussed in Section~\ref{sec:DIS-model}.
\paragraph*{Mathematical properties of the models.} For all models, positivity and hence biological meaningfulness can be concluded. Except for the cell density~$V$, all variables are bounded from above. However, unboundedness of~$V$ is acceptable in the time-limited setting (${t\leq\tend}$) of both models, \NEW{which motivated the simplification of the logistic growth to an exponential one (recall~\eqref{eq:ODEVchem})}. The only relevant steady state~${(\bar{V},\,\bar{\stress})}$ of the models is the situation of a dying cell population up to the point of extinction:~${\bar{V}=0}$. Since all drugs are manually removed from the system by the end of the modeled experiment, their steady states are not of interest. As mentioned above, the cell population can only decline if a sufficiently large stress level~$\stress$ is reached during the treatment phase, which gives~${\bar{\stress}=\stress(\ttreat)>\frac{\prol-\nat}{\prol+\indDeath}}$, recall equation~\eqref{eq:maxStressCH}. Due to this condition, the steady state~${(\bar{V},\,\bar{\stress})}$ is stable. If the stress threshold~${\frac{\prol-\nat}{\prol+\indDeath}}$ is not exceeded, there exists no realistic steady state. 

\subsection{Solving the cell line-specific models}
\label{sub:MOD-App2-sol}
Depending on the coupling of the ODEs, it may not be possible to solve each model analytically. Nevertheless, doing some calculations towards solving the initial value problems~\ref{eq:M-chemo}, ~\ref{eq:M-chemo-0} and~\ref{eq:M-chemo-0-inf} are still useful to understand the mathematical structure of the solutions, especially after the treatment phase.
\paragraph*{Towards solving the general model~${\boldsymbol{\mathcal{M}_{DS}^\text{CYP}}}$.} For model calibrations the model~\ref{eq:M-chemo} is solved numerically using the Python function~\texttt{scipy.integrate.odeint}
. Applying separation of variables on each ODE yields, for a time point~${t\geq\ttreat}$ after finishing the treatment:
\begin{align}
	V(t)&=\initV \,\exp\left((\prol-\nat)t-(\prol+\indDeath)\left(\stress(\ttreat)\big(t-\ttreat\big)+\int_{0}^{\ttreat}\stress(\tau)\,\mbox{d}\tau\right)\right), \label{eq:chemo-solV} \\[4pt]
	\stress(t)&=1-(1-\smallunderbrace{\initStCH}_{\leq \frac{\prol-\nat}{\prol+\indDeath}})\exp\left(-\int_{0}^{t}\sensRate{D}^-\,\deact_{D,HC}\big(D(\tau),S(\tau)\big)+\sensRate{S}^-\, \deact_{S,HC}\big(S(\tau)\big)\,\mbox{d}\tau\right) \label{eq:chemo-solStress}\\
	\mbox{with }&D(t)= \initD \,\exp\big(-\metDCH\cdot t\,\big)\cdot\mathbbm{1}_T(t)\quad\mbox{and}\quad S(t)= \initS \,\exp\big(-\metSCH\cdot t\,\big)\cdot\mathbbm{1}_T(t)\,. \nonumber
\end{align}
As mentioned in Subsection~\ref{ssec:expdata}, the model calibrations in this biological context consider the so-called \emph{percentage viability}, i.e. the ratio between the viability of a treated population~$\Vtreat$ and a corresponding
control population~$\Vctrl$ without treatment (i.e. ${\initD ,\initS =0}$ while the remaining environmental conditions and hence all terms not depending on~$D$ and~$S$ are the comparable for both populations). Equation~\eqref{eq:chemo-solV} gives for~${t\geq\ttreat}$:
\begin{equation}
	\frac{\Vtreat}{\Vctrl}(t)=\exp\left(-(\prol+\indDeath)\left(\stress(\ttreat)\big(t-\ttreat\big)-\initStCH\cdot t+\int_{0}^{\ttreat}\stress(\tau)\,\mbox{d}\tau\right)\right)\,. \label{eq:chemo-percV}
\end{equation}
We conclude three observations: First, in~\eqref{eq:chemo-solV}, \eqref{eq:chemo-solStress} and~\eqref{eq:chemo-percV} the growth and death rates show up solely in the form of~${(\prol-\nat)}$ and~${(\prol+\indDeath)}$. For model calibrations, this means that instead of~$\prol$,~$\nat$ and~$\indDeath$ we can only estimate~${(\prol-\nat)}$ and~${(\prol+\indDeath)}$. Second, the term~${(\prol-\nat)}$ does not show up explicitly in~\eqref{eq:chemo-percV} and it is also not implicitly involved, since the stress level~\eqref{eq:chemo-solStress} does not depend on~$V$.
This means~${(\prol-\nat)}$ has no influence on the percentage viability (the quantity of interest, for which we have measurements) and thus cannot be reconstructed by model calibration. For the same reason, the relation~${\initStCH\leq\frac{\prol-\nat}{\prol+\indDeath}}$ cannot be incorporated in the prior information for estimating~$\initStCH$ in Section~\ref{sec:CAL-Prior-App2}. Third, the initial cell density~$\initV$ does not occur in~\eqref{eq:chemo-percV}, neither explicitly nor implicitly, as the ESL~\eqref{eq:chemo-solStress} does not depend on~$V$. Hence, without loss of generality, we can set~$\initV$ to an arbitrary value, which we chose to be~${\initV=1}$ for the simulations. 

\paragraph*{Analytical solution of the reduced models~${\boldsymbol{\mathcal{M}_{DS}^0}}$ and~${\boldsymbol{\mathcal{M}_{DS}^{0,\star}}}$.} Due to the absence of CYP in the system, the present drug concentrations stay constant during the whole treatment phase, i.e.~${D(t)=\initD}$ and~${S(t)=\initS ~\forall t\in T}$. Therefore, using~\eqref{eq:chemo-solStress}, we get
\begin{equation*}
	\stress(t)=1-(1-\initStCH )\exp\bigg(-\Big(\sensRate{D}^-\cdot \deact_{D,HC}(\initD,\initS)+\sensRate{S}^-\cdot \deact_{S,HC}(\initS)\Big)t\bigg)
\end{equation*}
as solution for the ESL. This term is independent from all other variables, especially from~$V$, and easy to integrate. By insertion into~\eqref{eq:chemo-solV}, we achieve the explicit solution of~$V$. Hence, models~\ref{eq:M-chemo-0} and~\ref{eq:M-chemo-0-inf} are analytically solvable (the only difference between them is the definition of~$\deact_{D,HC}$) and the percentage viability~\eqref{eq:chemo-percV} can be calculated explicitly.

\subsection{Bayesian model calibration}\label{ssec:bayesian}

Our next step is to estimate the unknown parameter values on the basis of the experimental data. We follow a probabilistic approach to also quantify the uncertainties and correlations among parameters.  
We start to model a relationship between model parameters and data (Subsection~\ref{sec:PAR-Unc}). This allows to obtain, via Bayes' rule, a probability distribution over the parameters, the so-called posterior distribution, which has been informed by the model and the data (Subsection~\ref{sec:PAR-BAY}) based on prior information not considering the data (Subsection~\ref{sec:CAL-Prior-App2}). Finally, in the subsequent Section~\ref{sec:PAR-SMC}, we describe the Sequential Monte Carlo (SMC) algorithm that we have used to obtain samples from the posterior. 

\subsubsection{Modeling the relationship between parameters and data}
\label{sec:PAR-Unc}
In experiments described in~\cite{Ozkan.2021}, that we have used for calibrating the models, a CellTiter-Blue\textsuperscript{\textregistered} assay was used to monitor the viability of tumor cells. 
The fluorescence intensity produced by viable cells~$I$ at a specific time is assumed to be directly proportional to the corresponding density of viable tumor cells~$V$, with a \emph{proportionality constant} $n_{I/V}$. Incorporating measurement noise, this leads to the relationship
\begin{equation}
	I_i=n_{I/V}\, V_i\cdot\varepsilon_{V,i}\,, \label{eq:intensity} 
\end{equation}
for each intensity measurement $I_i$ (with ${i=1\Ldots M}$), which has been computed as the difference between the measured total and the background fluorescence (more details can be found in our previous work~\cite{Schonfeld.2022}). 

For~${i=1,\ldots,M}$, the quantities $\varepsilon_{V,i}$ denote the measurement noise in each experiment and, in the Bayesian framework, we model them as random variables. We see from \eqref{eq:intensity} that we have used a multiplicative noise model. This can be justified with different reasons. From a mathematical perspective, it allows to preserve positivity of the data. A more practical motivation is the reasonable assumption that the fluorescence noise of the intensity measurements is proportional to the density of viable cells. The validity of using multiplicative noise has already been verified in other related experimental settings~\cite{Sasik.2002} as well as in~\cite{Schonfeld.2022}, where we investigated a very similar experimental setting assuming~$\varepsilon_{V,i}$ (${i=1,\ldots,M}$) as independent, identically distributed random variables $\sim\Gamma(a,b)$ 
with ${a=b>1}$. This showed to be a good noise model for calibration. Since the measurements for the data used here and those used in~\cite{Schonfeld.2022} were carried out in the same way, we still rely here on this assumption.

Differently from~\cite{Schonfeld.2022}, though, our raw data consist of percentage viabilities, which can be used to monitor the efficacy of chemotherapeutic treatment. A percentage viability is given by the ratio between the population sizes of a treated and a corresponding 
untreated (control) cell population, denoted by~$\Vtreat$ and~$\Vctrl$, respectively. Therefore, the \emph{percentage viability data} have the form
\begin{equation}\label{eq:noisemodel}
	I_i^\%=\frac{\txtTop{I}{treat}_i}{\txtTop{I}{ctrl}_i}\overset{\eqref{eq:intensity}}{=}\frac{n_{I/V}\,\Vtreat_i\cdot \varepsilon_1}{n_{I/V}\,\Vctrl_i\cdot \varepsilon_2}=\frac{\Vtreat_i}{\Vctrl_i}\cdot\varepsilon_{\%,i}\,,\quad i=1,\ldots,M.
\end{equation}
We see that, for each $i$, $\varepsilon_{\%,i}$ is the ratio of two independent random variables, which are respectively Gamma-distributed as described above. Assuming that the variance in the measurements is not influenced by the treatment of the cells, i.e. the two Gamma distributions have the same variance, this distribution simplifies to a standard Beta prime distribution:
\begin{equation}
	\varepsilon_{\%,i}\sim\beta'\Big(1/\sigma^2\,,\,1/\sigma^2\Big)\mbox{ with }\sigma^2\in(0,0.5)\,,\quad\text{for }i=1,\ldots,M. \label{eq:noisePerc}
\end{equation}
Since the measurements noises in \eqref{eq:intensity} were independent, we have that also $\varepsilon_{\%,i}$ are independent random variables for~${i=1,\ldots,M}$. Note that in \eqref{eq:noisePerc} we need~${\frac{1}{\sigma^2}>2\Rightarrow\sigma^2<0.5}$ to have a well-defined variance.

The relationship between the model and the data is now given by \eqref{eq:noisemodel} once we consider $\Vtreat$ and $\Vctrl$ to be given by the solutions to the mathematical models from Subsection~\ref{sec:MOD-App2}. More precisely, we collect all parameters that we need to estimate in a vector $\theta$ taking values in a parameter space $\Theta\subseteq \mathbb{R}^d$, where $d\in\mathbb{N}$ is the number of parameters. Similarly to our previous work~\cite{Schonfeld.2022}, we collect all intensity measurements in a vector $\mathcal{I}^{\%}\defas\big({I}^\%_i\big)^M_{i=1}\in\mathbb{R}_+^M$ and introduce the \emph{forward operator}
\begin{equation}
\mathcal{G}_\%^\mathcal{M}:\quad
		\Theta\rightarrow [0,1]^M\,, \quad
		\theta\mapsto\Big(\mathcal{G}_{\%,i}^\mathcal{M}(\theta)\Big)_{i=1}^M~\mbox{with }\mathcal{G}_{\%,i}^\mathcal{M}(\theta)=\dfrac{\txtTop{V}{treat}_i(\theta) }{\txtTop{V}{ctrl}_i (\theta)}\,,\label{eq:forwGperc}
\end{equation}
mapping parameter values to relative intensities, where the scaled cell densities~$\Vtreat_i$ and $\Vctrl_i$ are obtained as the solutions of a model~$\mathcal{M}$ using the parameter values of~$\theta$ (and omitting the chemotherapeutic treatment for $\Vctrl_i$). With the new notation from~\eqref{eq:forwGperc}, we can rewrite \eqref{eq:noisemodel}:
\begin{equation}
	I^\%_i=\mathcal{G}^\mathcal{M}_{\%,i}(\theta)\cdot\varepsilon_{\%,i}\,,\quad\text{for }i=1,\ldots,M,\label{eq:noisePerc2}
\end{equation}
where~$\varepsilon_{\%,i}$ are independent and identically distributed according to~\eqref{eq:noisePerc}.

\paragraph*{Remark (Model inadequacy):} If we use~\eqref{eq:noisemodel} and assume that the viabilities are the solution to our mathematical models, strictly speaking we are assuming that our models are an \textsl{exact} description of the biological process. This is of couse not the case, and in principle we should include and additional term $\chi_i$ in~\eqref{eq:intensity} for the modeling error:
\begin{equation*}
	I_i=(n_{I/V}\, V_i + \chi_i)\cdot\varepsilon_{V,i}\,,\quad\text{for }i=1,\ldots,M.
\end{equation*}
The distribution of~$\chi_i$ could for example be chosen to be a truncated normal, to preserve positivity, or obtained by a multiplicative error like in~\cite{Kennedy.2001}. While the consideration of modeling inadequacy in combination with suitable prior information can be important in some settings~\cite{Brynjarsdottir.2014}, it also increases the dimension of the parameter space as further hyper-parameters like the variance of the modeling error would need to be estimated. This would lead to a higher computational cost in the sampling algorithm as well as require some modeling assumptions on these hyper-parameters. Therefore, here we assume that the measurement noise dominates over the modeling error. Later, we validate this by showing that the calibration results are still reasonable and informative under this assumption. In fact, when neglecting the modeling error in our previous work~\cite{Schonfeld.2022} using a closely related model, we achieved comparable parameter estimates as a similar approach in~\cite{Lima.2018}, where the error was included.

\subsubsection{Informing the parameters with the model and the data}
\label{sec:PAR-BAY}
In the Bayesian setting, we consider the parameter vector~$\theta$ to be a (multi-dimensional) random variables taking values in~$\Theta$. Given some prior knowledge on the parameters before having seen any data, encoded in the so-called prior distribution, we use the information provided by the experimental data, encoded in the likelihood, to update the distribution of the parameters to the so-called posterior distribution. This is expressed by the Bayes' formula
\begin{equation}\label{eq:posterior}
	\pi^{\mathcal{I}}(\theta) = \frac{L(\mathcal{I}\,|\,\theta)\cdot\pi_0(\theta)}{\int_{\Theta}L(\mathcal{I}\,|\,\theta)\cdot\pi_0(\theta)\,\mbox{d}\theta}\propto L(\mathcal{I}\,|\,\theta)\,\pi_0(\theta)\,,
\end{equation}
where~$\pi_0$ and~$\pi^{\mathcal{I}}$ denote the probability densities of the prior and the posterior distribution, respectively. The latter is usually much more concentrated than the prior, meaning that we use the data to reduce the uncertainty. The proportionality constant in \eqref{eq:posterior} is called \emph{model evidence} and it needs to be computed explicitly only when interested in model comparison.

In our application, using the assumption that the noise is Beta prime distributed, we obtain
\begin{equation*}
	\overset{\eqref{eq:noisePerc2}\vphantom{\big|}}{\Longrightarrow}\quad L_i\left(I^\%_i\,\Big|\,\theta\right)\propto \left(\frac{I^\%_i }{\mathcal{G}_{\%,i}(\theta)}\right)^{1/\sigma^2-1}\left(1+\frac{I^\%_i}{\mathcal{G}_{\%,i}(\theta)}\right)^{-2/\sigma^2},\quad \mbox{for }i=1\Ldots M\,,
\end{equation*}
which is nothing but the density of the distribution ${\beta'\left(\frac{1}{\sigma^2},\frac{1}{\sigma^2}\right)}$ evaluated at ${\frac{I^\%_i }{\mathcal{G}_{\%,i}(\theta)}}$\,. The \emph{data likelihood} for a set of intensity measurements is then:
\begin{equation}
	L\Big(\mathcal{I}^\%\,\Big|\,\theta\Big)\propto\prod_{i=1}^M \left(\frac{I^\%_i}{\mathcal{G}_{\%,i}(\theta)}\right)^{1/\sigma^2-1}\left(1+\frac{I^\%_i}{\mathcal{G}_{\%,i}(\theta)}\right)^{-2/\sigma^2}\,. \label{eq:LL-Perc}\tag{$\mathfrak{L}_{\%}$}
\end{equation}
In this and the equation above, we have used the proportionality sign to avoid writing explicitly the normalization constant.

Next, we present our choices for the prior distributions of the parameters. With that and the likelihood \eqref{eq:LL-Perc}, we want to sample from the posterior \eqref{eq:posterior} in order to visualize it and draw biological considerations. To this aim, we use the algorithm described in the subsequent Section~\ref{sec:PAR-SMC}.

\subsubsection{Prior modeling}\label{sec:CAL-Prior-App2} 
Before discussing the prior distributions for the model parameters, we address how to set the hyper-parameter~$\sigma^2$ in the noise distribution \eqref{eq:noisePerc}. Since, from the Beta prime distribution, we can compute~${\Var(\varepsilon_{\%,i})=\frac{\sigma^2(2-\sigma^2)}{(1-2\sigma^2)(1-\sigma^2)^2}}$ (for~$i=1\Ldots M$), we can tune the noise variance with~${\sigma^2\in(0,1/2)}$. Instead of using a fully Bayesian approach with also a prior on~$\sigma^2$, we use a deterministic value for it. To choose the latter, we have oriented ourselves by the magnitude of the estimated noise levels for the calibrations in our previous paper \cite{Schonfeld.2022}, which we remind had data collected in the same setting as those used here. On that basis, we found~$\sigma^2=0.10$ 
resp.~${0.15}$ to be suitable choices for the calibrations to the data of cell lines Hep3B2 resp. HepG2/C3Asub28, achieving a balance between numerical stability of the algorithm and sufficient concentration of the distributions. This results in uncertainty variances of~${\Var(\varepsilon_{\%,i})\approx0.29}$ 
resp.~${0.55}$\,.

We proceed with constructing the prior distributions in the following paragraphs. Beforehand, we note: If we do not give a biological/mathematical criterion to set a distinctive upper bound for a parameter, we choose the corresponding prior support as small as possible without being too restrictive. This allows to initialize the algorithm described in the next section with a computationally manageable sample size. 
In the following, we only present the prior distributions for the reduced models~\ref{eq:M-chemo-0} and~\ref{eq:M-chemo-0-inf}\,, as calibration attempts for the complete model~\ref{eq:M-chemo} showed numerical issues and did not yield reasonable results for mathematical/biological investigation (possible reasons are discussed in Section~\ref{sec:DIS-model}). A summary over the prior distributions to calibrate the models~\ref{eq:M-chemo-0} and~\ref{eq:M-chemo} can be found in Table~\ref{Tab:CalParsChemo}. The corresponding Section~\ref{SUP-sec:CAL-prior} also provides more details on the prior assumptions for the calibration setting of~\ref{eq:M-chemo}.

We calibrate a particular model separately to the measurements for each combination of hypoxia level~${\initH\in\{0,1\}}$ and cirrhosis level~${\initC\in\{0,1\}}$\,, i.e. we perform four individual calibrations per cell line/model. In particular, we start with the case~${\initH,\initC=0}$\,. We will use the following terms and abbreviations to denote the underlying environmental conditions:
\begin{alignat*}{2}
\mbox{\qm{normal conditions} :}&&\quad\hphantom{=0}~\initH,\initC =0~&\overset{\text{abbr.}}{\leadsto}~\mbox{HC0}\,,\\ 
\mbox{\qm{sole hypoxia} :}&&\quad\initH =1, \initC=0~&\hspace*{4pt}\leadsto~\hspace*{4pt} \mbox{H1}\,,\\
\mbox{\qm{sole cirrhosis} :}&&\quad\initH =0,\initC =1~&\hspace*{4pt}\leadsto~\hspace*{4pt} \mbox{C1}\,,\\
\mbox{\qm{cirrhosis and hypoxia} :}&&\quad\hphantom{=0}~\initH,\initC =1~&\hspace*{4pt}\leadsto~\hspace*{4pt} \mbox{HC1}\,.
\end{alignat*}
\paragraph*{Prior distributions for model~${\boldsymbol{\mathcal{M}_{DS}^0}}$\,.}
Given the treatment phase~${T=[0,\ttreat]}$\,, we consider the unknown parameters of system
\leqnomode
\begin{align*}\mbox{\ref{eq:M-chemo-0}}:\,&\!\left\{\begin{aligned}
		\dV &= \Big(\prol-\nat-\big(\prol+\indDeath\big)\stress\Big)\cdot V\,, \\[4pt]
		\dStress &= \left(\vphantom{\frac{\sensRate{{D}}^- \initD^{m_1}}{\big(\txtSub{D}{norm}\cdot\dCH\cdot d_S(\initS) \big)^{m_1}+\initD^{m_1}}}\right.\underbrace{\frac{\sensRate{{D}}^- \initD^{m_1}}{\big(\txtSub{D}{norm}\cdot\dCH\cdot d_S(\initS) \big)^{m_1}+\initD^{m_1}}}_{\deact_{D,HC}(\initD,\initS)}+\underbrace{\frac{\sensRate{S}^- \initS^{m_2}}{\SthrCH^{m_2}+\initS^{m_2}}\vphantom{\frac{\sensRate{{D}}^- \initD^{m_1}}{\big(\txtSub{D}{norm}\cdot\dCH\cdot d_S(\initS) \big)^{m_1}+\initD^{m_1}}}}_{\deact_{S,HC}(\initS)}\left.\vphantom{\frac{\sensRate{{D}}^- \initD^{m_1}}{\big(\txtSub{D}{norm}\cdot\dCH\cdot d_S(\initS) \big)^{m_1}+\initD^{m_1}}}\right) (1-\stress)\,\mathbbm{1}_T(t)\,,\\[4pt]
		&\mbox{with }V(0)=\initV ~~\mbox{and}~~\stress(0)=\initStCH \leq \frac{\prol-\nat}{\prol+\indDeath}\in(0,1)\,.
	\end{aligned} \right.\hspace*{3cm}
\end{align*}
\reqnomode
While the values of~$\ttreat$,~$\initH$,~$\initC$,~$\initD$,~$\initS$ and~$\initV$ are given from the experiment, the remaining parameters need to be calibrated. Recall two observations from Subsection~\ref{sub:MOD-App2-sol}. First, the parameters~$\prol$\,,~$\nat$ and~$\indDeath$ can only be estimated in combination as the terms~${(\prol-\nat)}$ and~${(\prol+\indDeath)}$. Second, the percentage viability (measurements) are modeled by~\eqref{eq:chemo-percV} as
\begin{equation*}
	\frac{\Vtreat}{\Vctrl}(t)=\exp\left(-(\prol+\indDeath)\left(\stress(\ttreat)\big(t-\ttreat\big)-\initStCH\cdot t+\int_{0}^{\ttreat}\stress(\tau)\,\mbox{d}\tau\right)\right)\,.
\end{equation*}
Since this equation is independent from~$(\prol-\nat)$, it is neither necessary nor possible to estimate this term. Consequently, we cannot incorporate the particular upper bound~$\frac{\prol-\nat}{\prol+\indDeath}<1$ in the prior information for~$\initStCH$. Summarizing these observations yields the prior distributions
\begin{equation*}
	\initStCH \sim\Triang(0,0,1)\qquad\mbox{and}\qquad(\prol+\indDeath)\sim\unif(0,3)\,,
\end{equation*}
where $\mathcal{U}(a,b)$ denotes a uniform distribution on the interval $[a,b]$ and ${\Triang(a,h,b)}$ stands for a triangular distribution on~${[a,b]}$ where ${h\in[a,b]}$ is the mode, i.e. the value that we consider to be more likely \textsl{a priori}.
Note that by definition~${\initStCH=0}$ for~${\initH,\initC=0}$~(HC0), i.e.~${\initStCH}$ only needs to be estimated for data with hypoxic and/or cirrhotic conditions~(H1/C1/HC1). Furthermore, we know that the parameter~${(\prol+\indDeath)}$ is uninfluenced by~$\initH$ and~$\initC$\,. Hence, we can use the so-called marginal MAP estimate, which will be introduced in a later Subsection~\ref{sub:RES-ana-est}, for~${(\prol+\indDeath)}$ resulting from the first calibration~(HC0) to set the parameter fixed for the remaining calibrations~(H1/C1/HC1).

Next, we take a closer look at the cells' stress response to DOX and SOR. The sensitivity function~${\deact_{D,HC}}$ is determined by (among others) the unaffected sensitivity threshold~$\Dnorm$ and its shifting terms
\begin{equation*}
	d_S(\initS)=1-\dfrac{\txtSub{a}{max} \initS^{m_3}}{\Ssupp^{m_3}+\initS^{m_3}}\qquad\mbox{and}\qquad \dCH\,.
\end{equation*}
The latter needs to be estimated as the combined term~${\DnormCH=\Dnorm\cdot\dCH}$\,. Based on the experimental design, it is reasonable to assume that significant changes of cytotoxicity of DOX occur in a logarithmic scale in the magnitude of the applied dosages~$\SIrange{0.0001}{1000}{\micro\molar}$~\cite{Ozkan.2021}. We do not expect the threshold~$\DnormCH$ to be very close to the bounds of this range, i.e. we assume 
\begin{equation*}
	\DnormCH\in\big[10^{-4},10^4\,\big]\quad\mbox{and define}\quad\log_{10}(\DnormCH)=\DnormCHlog\sim\Triang(-4,0,4)\,.
\end{equation*}
The Hill coefficient~$m_1$ tunes the slope of~$\deact_{D,HC}$\,. If~$m_1$ is in a high magnitude, large variations are necessary to significantly change the slope. Therefore, we use a logarithmic scale to obtain useful prior samples and set
\begin{equation*}
	m_1\in[1,50]\approx\big[10^0,10^{1.7}\,\big]\quad\Rightarrow~\log_{10}(m_1)=\widehat{m}_1\sim\unif(0,1.7)\,.
\end{equation*}
For the DOX impact rate we assume
\begin{equation*}
	\sensRate{D}^-\sim\unif(0,20)\,,
\end{equation*}
where the upper bound can be interpreted as an immediate reaction to the DOX treatment.

\begin{figure}[H]
	\centering
		\includegraphics[width=\textwidth]{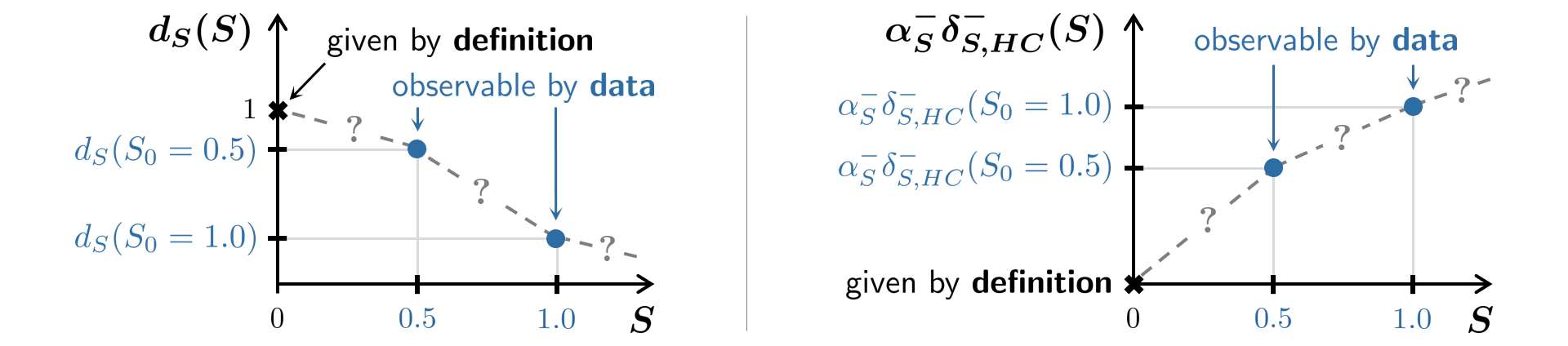}
		\caption[Available information to estimate~${d_S(\initS)}$ resp.~${\adSCH(\initS)}$ for~${\initS \in\{0.5,1\}}$]{The available two data points provide just enough information to estimate~${d_S(\initS)}$ resp.~${\adSCH(\initS)}$ for~${\initS \in\{0.5,1\}}$ (blue markers). Since both functions~${d_S}$ and~${\adSCH}$ have three free parameters each, this data is not enough to reconstruct the curve of the respective function.}
		\label{Fig:infS}
\end{figure}

The supportive influence of SOR is given by~${d_S(\initS)}$ and depends on the parameters~$\txtSub{a}{max}$,~$\Ssupp$ and~$m_3$, i.e. the continuous function~$d_S$ has three degrees of freedom. Analogously, the stress response to SOR~$\adSCH(\initS)$ depends on three parameters ($\sensRate{S}^-$\,,~$\SthrCH$\,,~$m_2$) as well. However, due to the experimental setting, the data only provides additional information for the two non-zero SOR dosages~${\initS\in\{0.5,1.0\}}$, as the values~${d_S(\initS =0)=1}$ and~${\adSCH(\initS=0)=0}$ are given per definition, respectively. This means that for both functions~$d_S$ and~$\adSCH$\,, the two observations of their function value given by~${\initS\in\{0.5,1.0\}}$ are not enough to distinctively reconstruct their continuous form and hence estimate the involved parameters. This issue is illustrated in Figure~\ref{Fig:infS}.
\noindent To counteract this issue, we make use of the fact that in the reduced model~\ref{eq:M-chemo-0} the SOR concentration remains constant during treatment. Then, instead of estimating the continuous shape of~$d_S$ or~${\adSCH}$ with its respective parameters, we only consider the discrete values~$d_S(\initS )$ resp.~${\adSCH(\initS )}$ for~${\initS \in\{0,0.5,1\}}$ (blue markers in Figure~\ref{Fig:infS}). Since~${d_S:\mathbb{R}_+\to[0,1]}$ is strictly monotonically declining with the relationship~${0< d_S(1)<d_S(0.5)<1}$, we define
\begin{equation*}
	d_S(0.5)\sim\unif(0,1)\qquad\mbox{and}\qquad d_S(1)=c_d\cdot d_S(0.5)\mbox{ with }c_d\sim\unif(0,1)\,.
\end{equation*}
We can use a similar approach to reparametrize~${\adSCH}\,$, since~${\deact_{S,HC}\in[0,1]\Rightarrow \adSCH\leq \sensRate{S}^-}$. As~$\deact_S$ is strictly monotonically increasing, we define
\begin{equation*}
	\adSCH(1)\sim\unif(0,2)\quad\mbox{and}\quad\adSCH(0.5)=c_\delta\cdot \adSCH(1)\quad\mbox{ with }c_\delta\sim\Triang(0,0,1)\,.
\end{equation*}
The choice of the upper bound of~${\adSCH(1)}$ follows the assumption that the tumor cells are more sensitive to treatment with DOX than with SOR, as the latter mainly serves as a supportive drug. According to~\cite{Ozkan.2022}, even high dosages of SOR show relatively low cytotoxicity compared to DOX. Furthermore, the cells show no obvious reaction to the treatment with a standard SOR dosage~(${\initS=0.5}$) for low~$\initD$, motivating the triangular prior for~$c_\delta$. 

\paragraph*{Prior distributions for the enhanced reduced model~${\boldsymbol{\mathcal{M}_{DS}^{0,\star}}}$\,.} For the model~\ref{eq:M-chemo-0-inf}, which used to calibrate the data of cell line HepG2, we again use the same prior assumptions as for the reduced model except for the following adaptions:
\begin{alignat*}{6}
	\log_{10}(m_1)=\widehat{m}_1&\sim\unif(0,1.7)\quad&&\overset{~\text{instead}~}{\longsquiggly}\quad &m_1~&&\sim\unif(0,6)\,,\\
	d_S(0.5)&\sim\unif(0,1)\quad&&\overset{~\text{instead}~}{\longsquiggly}\quad &1-d_S(1.0)~&&\sim\unif(0,1)\,,\\
	\frac{d_S(1.0)}{d_S(0.5)}=c_d&\sim\unif(0,1)\quad&&\overset{~\text{instead}~}{\longsquiggly}\quad& \frac{1-d_S(0.5)}{1-d_S(1.0)}=\tilde{c}_d~&&\sim\unif(0,1)\,.
\end{alignat*}
The adjusted prior for the Hill coefficient~$m_1$ is motivated by the estimated marginal posteriors from the calibration of cell line Hep3B2 (see Figure~\ref{A-Fig:CL1-marg}). The remaining adaptions result from the fact that the damping factor is applied to the supportive effect term~${1-d_S(S)}$ and that~${1-d_S(0.5)\leq1-d_S(1.0)}$. With the damping factor comes along another parameter: the damping threshold~${\Ddamp\leq\DnormCH}$\,. By using the reparametrization~\eqref{eq:CL2-repar} in log scale, i.e.
\begin{equation*}
	(0,1]\ni\cdamp=10^{\txtSub{\widehat{c}}{damp}}\qquad\Rightarrow~\log_{10}(\cdamp)=\txtSub{\widehat{c}}{damp}\leq 0\,,
\end{equation*}
we can assume the prior distribution
\begin{equation*}
	\txtSub{\widehat{c}}{damp}\sim\unif(-3,0)\,.
\end{equation*}
The support of this uniform distribution translates to~${\cdamp\in(0.001,1]}$ and is motivated by the observation that even smaller values for~$\cdamp$ would only marginally change the damping effect (see right subplot of Figure~\ref{Fig:CL2-damp}, where~${\cdamp=1/50}$ already almost completely damped the supportive effect of SOR).
\subsection{Sequential Monte Carlo (SMC) algorithm}
\label{sec:PAR-SMC} 
To sample from the posterior measure~$\mu^\mathcal{I}$, we use the \emph{Sequential Monte Carlo} (SMC) method,  
which we explain in this section based on~\mbox{\cite{Kantas.2014, Chopin.2002, Beskos.2015}}. Another possibility would be to use Markov Chain Monte Carlo (MCMC)~\mbox{\cite[Ch. 6-7]{Robert.2004}} methods, but we opt for SMC because, as it will become clearer from the explanation of the algorithm, SMC is more efficient when one can split the data. In that case, one needs to compute only the incremental likelihoods rather than the full likelihood, requiring less evaluations of the forward model.

In SMC, one considers a sequence of~${K\in\mathbb{N}}$ intermediate distributions~${(\mu_k)_{k=0}^K}$\,, such that~$\mu_0$ is the prior and~${\mu_K=\mu^{\mathcal{I}}}$ coincides with the posterior distribution. In particular, over the course of~$K$ so-called~\emph{SMC steps} we approximate the respective intermediate distribution~$\mu_k$ at the~$k$-th step (${k=1\Ldots K}$). In the following, we describe the algorithm and then discuss enhancements with an adaptive choice of some algorithmic parameters that made it more practical and efficient for our application.

\subsubsection{Description of the algorithm}

The SMC method sequentially draws from the intermediate measures~$\mu_k$ using a swarm of samples, so-called \emph{particles}~${\{\theta_p\}_{p=1}^{P}}$ with associated \emph{weights}~${\{W^k_p\}_{p=1}^{P}}$\,, where~${P\in\mathbb{N}}$ is the \emph{sample size} (i.e. the number of particles). This particle approximation gives the probability density of~$\mu_k$~(${k=1\Ldots K}$) by
\begin{equation}
	\pi_k(\theta)\approx \sum_{p=1}^{P} W^k_p\delta_{\theta_p}(\theta)\propto \mathcal{L}_k\cdot \pi_0(\theta)\,, \label{eq:partAprx}
\end{equation}
where~$\mathcal{L}_k=\mathcal{L}_k(\,\cdot\,|\,\theta)$ is an appropriate \emph{intermediate data likelihood} \NEW{(its explicit argument depends on the method to construct~$\pi_k$, as we explain later)}. Each SMC iteration consists of three steps: the \emph{reweighting step}, to update the particle distribution, the \emph{resampling step}, to improve the particle representation, and the \emph{mutation step}, to better explore the parameter space. We know describe each step in more details.

\paragraph*{Reweighing step.} We describe two methods to reweigh the particles, which we later combine (see Subsection \ref{ssec:algsummary}): \emph{data splitting}~\cite{Chopin.2002} and \emph{likelihood tempering}~\cite{Beskos.2015}. Note that, as there is no commonly used distinctive name, the term \qm{data splitting} is used here for the purpose of being able to terminologically distinguish between the reweighting methods. One can think of the data splitting as a filtering approach. It divides up the available data set~$\mathcal{I}$, such that the intermediate likelihoods~$\mathcal{L}_k$ in~\eqref{eq:partAprx} consider the respective chunks over the course of the SMC steps until the full data set~$\mathcal{I}$ is taken into account. In contrast, likelihood tempering makes use of~${L(\mathcal{I}\,|\,\theta)^{\nu=0}\,\pi_0(\theta)=\pi_0(\theta)}$ \NEW{and~${L(\mathcal{I}\,|\,\theta)^{\nu=1}\,\pi_0(\theta)\propto\pi^\mathcal{I}(\theta)}$ from~\eqref{eq:posterior}} to tune the data likelihood~${L(\mathcal{I}\,|\,\theta)}$ by step-wisely increasing the exponent~${\nu\in[0,1]}$ over the course of the SMC steps. In general, each reweighing method solely changes the weights and not the position of the particles.

\smallskip
\textsl{Data splitting.} The idea of data splitting is based on dividing the set of all available measurements~$\mathcal{I}$ into progressively increasing data sets:
\begin{equation}\label{eq:datasplitting}
	\emptyset=\mathcal{I}_{:0}\subset\calIk{1}\subset\calIk{2}\subset\ldots\subset\hspace*{-1.1cm}\underbrace{\calIk{k}}_{\substack{\text{\emph{calibration data}}\\\text{at the~$k$-th SMC step}}}\hspace*{-1.1cm}\subset\ldots\subset\calIk{K}=\mathcal{I}\,.
\end{equation}
Then, the intermediate distribution~$\mu_k$~(${k=1\Ldots K}$) is approximated at the~$k$-th SMC step using the data likelihood of~$\calIk{k}$\,. In particular, analogously to~\eqref{eq:posterior}, Bayesian rule gives 
\begin{equation*}
	\pi_k(\theta)\propto L\big(\calIk{k}\,|\,\theta\big)\cdot \pi_0(\theta)\,,
\end{equation*}
which relates the prior and the~$k$-th intermediate distribution. Analogously, with the \emph{increment data}~${\calIIk{k-1}{k}=\calIk{k}\hspace*{-3pt}\setminus\hspace*{-1pt}\calIk{k-1}}$ at the~$k$-th SMC step and the associated disjoint data sets~${\{\calIIk{k-1}{k}\}_{k=1}^K}$, we can write
\begin{equation}\label{eq:pit}
	\pi_k(\theta) \propto \prod_{s=1}^k L\big(\mathcal{I}_{s-1:s}\,|\,\theta\big)\cdot \pi_0(\theta)~~\mbox{or, equivalently,}~~\pi_k(\theta) \propto L\big(\calIIk{k-1}{k}\,|\,\theta\big)\,\cdot \pi_{k-1}(\theta)\,,
\end{equation} 
where~$L(\calIIk{k-1}{k}\,|\,\theta)$ is the \emph{incremental data likelihood} as defined in~\eqref{eq:LL-Perc}. We recall \eqref{eq:partAprx}, i.e the representation of the intermediate distribution at the~$k$-th SMC step by a collection of weighted particles:~${\pi_k(\theta)\approx \sum_{p=1}^{P} W^k_p\delta_{\theta_p}(\theta)}$. With~\eqref{eq:pit}, we now have a relation between two consecutive intermediate densities~$\pi_k$ and~$\pi_{k-1}$\,, which is used to construct the weight~${W^{k}_p}$ of each particle~${\theta_p}$ by updating its previous weight~${W^{k-1}_p}$ with importance sampling:
\begin{equation}
	W_p^k = \frac{w_p^k}{\sum_{p=1}^{P} w_p^k}\,,\quad\mbox{with}\quad
	w_p^k \overset{\eqref{eq:pit}}{=} L(\calIIk{k-1}{k}\,|\,\theta_p)\,W^{k-1}_p\quad \mbox{for }p=1\Ldots P\,. \label{eq:RW-I} \tag{$\text{RW}_\mathcal{I}$}
\end{equation} 
\NEW{We see that the self-normalization~${\sum_{p=1}^{P}W_p^k=1}$ avoids the need to compute normalization constants explicitly.}

\smallskip
\textsl{Likelihood tempering.} In likelihood tempering~\cite{Beskos.2015}, instead, one sequentially scales the effect of the likelihood in order to ensure a smooth transition from the prior to the posterior. Namely, the intermediate distributions are given by
\begin{equation}
	\pi_{k}(\theta)\propto L\big(\mathcal{I}\,|\,\theta\big)^{\nu_k}\cdot \pi_0(\theta)\propto L\big(\mathcal{I}\,|\,\theta\big)^{\nu_k-\nu_{k-1}}\cdot \pi_{k-1}(\theta)\,, \label{eq:temp}
\end{equation}
for a sequence of \emph{inverse temperatures}~${0=\nu_0<\nu_1<\ldots<\nu_K=1}$. Analogously to~\eqref{eq:RW-I}, the reweighting of each particles~$\theta_p$ is now achieved by
\begin{equation}
	W_p^k = \frac{w_p^k}{\sum_{p=1}^{P} w_p^k}\,,\quad\mbox{with}\quad
	w_p^k \overset{\eqref{eq:temp}}{=} L(\mathcal{I}\,|\,\theta_p)^{\nu_k-\nu_{k-1}}\cdot W^{k-1}_p\quad \mbox{for }p=1\Ldots P\,. \label{eq:RW-L} \tag{$\text{RW}_\nu$}
\end{equation} 
In Subsection \ref{ssec:adaptive}, we explain how the sequence~$(\nu_k)_{k=1}^{K-1}$ can be selected adaptively on the fly to ensure efficiency of the algorithm.

\paragraph*{Resampling step.} In the reweighing step, independentily of the reweighing method, some particles might have very small weight after the update. It is therefore reasonable to discard those to focus the computational effort on the high portability region~\cite{Doucet.2009}. This can be done by \emph{resampling}, which replaces particles according to their weights. Whether resampling is necessary, can be decided with the \emph{effective sample size} (ESS), denoted here by $\ESS$\,,
which, at a given SMC iteration $k$, can be computed as~\cite{Kantas.2014}
\begin{equation*}
    \ESS = \left(\sum_{p=1}^P \left(W^{k}_p\right)^2\right)^{-1}.
\end{equation*}
The effective sample size quantifies how many independent, identically distributed samples (the best one could hope for) would be needed to represent the current probability distribution. If it is small, it means that the particle representation is poor. To improve the particle representation, resampling can be done by comparing~$\ESS$ with a user-set threshold~${P^*=\tau\cdot P,~\tau\in(0,1)}$:
\begin{equation*}
	\ESS\,\begin{cases}
		\,< P^* & \Rightarrow \mbox{resample } \{\theta_p\}_{p=1}^{P} \mbox{ according to }\{W^{k}_p\}_{p=1}^{P} \mbox{ and uniformly weigh new particles,}\\
		\,\geq P^* & \Rightarrow \mbox{do not resample.} \\
	\end{cases}
\end{equation*}
We make use of two different resampling schemes (other choices are possible, see e.g.~\mbox{\cite{Bulte.2020, R.Douc.2005, Doucet.2009, Gerber.2019}}). The first approach, which we refer to as \emph{random resampling}~(RR), is given in~\cite{Bulte.2020} and basically draws~$P$ i.i.d. samples with replacement from the set of indices~$\{p\}_{p=1}^P$, where~$p$ has the probability~$W_p^k$. Alternatively, we apply \emph{systematic resampling}~(SR) as described in~\cite{Doucet.2009}. This method has the advantage that it returns the same collection of input samples, if they all have the same weight, i.e. no samples get lost~\cite{Thrun.2006}. Recalling that~${\ESS\approx P}$ indicates almost uniformly weighted particles, we observed this feature to be especially useful for a relatively large ESS~$\ESS$ at the moment of resampling.

\paragraph*{Mutation step.} Performing only importance sampling \NEW{(known as Sequential Importance Sampling, or SIS for short, see e.g.~\cite{Doucet.2009})} will eventually lead to degeneracy in the diversity of the particle population, since resampling will discard more and more low-weighted particles in the course of the algorithm. To prevent this, we scatter the particles by applying a \emph{Markov kernel}~${\kappa_k(\,\cdot\,,\,\cdot\,)}$. Such a kernel is~$\mu_k$-invariant, i.e. it preserves the particle distribution. A large variety of possible choices for~$\kappa_k$ can be found in the literature on so-called Markov Chain Monte Carlo (MCMC) methods (see e.g.~\mbox{\cite{Robert.2004,Liu.2001}}).

For our application, we adopt the adaptive strategy developed in~\cite{Beskos.2015}, i.e. we use a random walk Metropolis-Hastings (RWMH) proposal on each univariate component, conditionally independently. Remembering from Section~\ref{sec:PAR-Unc} that each particle is a vector~${\theta_p\in\mathbb{R}^d}$, RWMH proposes another set of samples~${\{q_p\}_{p=1}^{P}}$ with~${q_p\in\mathbb{R}^d}$ by computing its~$j$-th component via
\begin{equation}
	(q_p)_j = (\theta_p)_j + \xi_j\quad\mbox{with random walk step size }\xi_j\sim\mathcal{N}(0,\epsilon_j^2)\,. \label{eq:randomWalk}
\end{equation} 
Eventually, the positions of the current particles are adapted by randomly accepting the proposed samples, i.e.
\begin{equation*}
	\mbox{set }\theta_p = q_p\quad\text{ with probability }\quad \min\left\{\frac{\pi_k(q_p)}{\pi_k(\theta_p)}\,,\,1\right\}.
\end{equation*}
To further improve the particle diversity, this process can be repeated more than once by applying~$\kappa_k$ again on the moved particles. We refer to each application of~$\kappa_k$ as an \emph{MCMC update}. For~$H\in\mathbb{N}$ such updates, we can define the \emph{average acceptance ratio}
\begin{equation}
	a_k = \frac{1}{H}\sum^{H}_{h=1} \frac{A_h}{P}\,, \label{eq:accR}
\end{equation}
with~${A_h\in\{1\Ldots P\}}$ being the number of accepted samples in the~$h$-th MCMC update. The step size~$\xi_j$ of~\eqref{eq:randomWalk} as well as the number of MCMC updates~$H$ affect to which extent the proposed particles are able to explore the parameter space. In the same spirit as in~\cite{Beskos.2015}, we explain in the following Subsection~\ref{ssec:adaptive} how to choose these adaptively on the fly.
\subsubsection{Adaptive choice of algorithmic parameters}\label{ssec:adaptive}

In the previous paragraphs, we did not specify how to choose some algorithmic parameters. These are chosen adaptively as the algorithm proceeds. The reason for this is algorithmic efficiency, namely, having an algorithm that gives robust results for a computationally manageable number of particles~$P$. In the following, we describe the adaptive choice of the inverse temperatures~$\nu_k$ when using likelihood tempering in the reweighing step, of the step size~$\xi_j$ in the random walks, and of the number of Markov chain updates~$H$ in the mutation step. The strategy that we have used is the one in~\cite{Beskos.2015}, with some further tuning.

\paragraph*{Selection of the inverse temperatures~$\boldsymbol{\nu_k}$.} These are chosen so as to make the largest possible step while keeping a good effective sample size, which we remind being a measure for the representativity of the particles. Therefore, at each SMC iteration, given the previous exponent~$\nu_{k-1}$, we select~$\nu_k$ such that, after having updated the weights, we still have an effective sample size $\ESS=P^*$ with a user-set threshold~${P^*=\tau\cdot P,~\tau\in(0,1)}$, where usually~${\tau\geq 0.5}$. In practice, this can be done e.g. by the bisection method.

\paragraph*{Selection of the step size~$\boldsymbol{\xi_j}$ in the random walks.} The goal of adjusting this algorithmic parameter is to achieve a balanced acceptance ratio in the Metropolis-Hastings updates. A balanced acceptance ratio finds a compromise between sufficiently exploring the parameter space while moving enough particles. According to~\cite{Beskos.2009}, it is reasonable to aim for~${a_k\approx 0.23}$ (for more than two scalar parameters). To achieve that, following \cite{Beskos.2015}, we tune~$\epsilon_j^2$ from~\eqref{eq:randomWalk} based on the empirical marginal variance~${\widehat{\text{Var}}_k}$ of the~$j$-th component of all current particles~${\underline{\theta}_j=\{(\theta_1)_j\Ldots (\theta_P)_j\}}$ via
\begin{equation}
	\epsilon_j^2 = \rho^2\cdot \widehat{\text{Var}}_k\big(\underline{\theta}_j\big)\,\label{eq:scaleVar},
\end{equation}
\NEW{with a \emph{scaling factor}~${\rho>0}$.} 
To appropriately select~$\rho$, we use two different adaptive approaches according to necessity for obtaining an efficient algorithm. The first adaptive scheme (which we denote by~$\text{MH}_{k-1}$) is based on~\cite{Beskos.2015} and described in full detail in our previous paper~\cite{Schonfeld.2022}. Extending this scheme yields another approach, in which we choose the scaling factor~$\rho$ adaptively by allowing it to change also over different Metropolis-Hastings updates within the same SMC step. In the following, we explain the details to this adaptive scheme.

Let~${H^*\in\{1\Ldots H\}}$ be the number of MCMC updates, after which we want to check the acceptance ratio and potentially adapt the step size. Then, we can split the~$H$ MCMC steps into~$\lceil H/H^*\rceil$ compartments, which we denote by~$\mathfrak{H}_l$\,,~${l=1\Ldots \lceil H/H^*\rceil}$\,. Analogously to~\eqref{eq:accR}, we define the \emph{partial acceptance ratio} for the~$l$-th compartment as
\begin{equation*}
	a_{k,l}^* = \frac{1}{|\mathfrak{H}_l|}\sum_{h\in\mathfrak{H}_l} \frac{A^*_h}{P}\,,
\end{equation*}
where~$A^*_h\in\{1\Ldots P\}$ is the number of accepted proposals in the~$h$-th MCMC update. We set the variance of the random walk for the~$h$-th MCMC update at the~$k$-th SMC step as
\begin{equation}
	\epsilon_j^2 = \rho_{k,h}^2\cdot \widehat{\text{Var}}_k\big(\underline{\theta}_j\big)\,, \label{eq:varMH-h}
\end{equation}
where the scaling factor~$\rho_{k,h}$ is defined recursively as follows. After each compartment~$\mathfrak{H}_l$ of MCMC updates is done, we adapt the scaling~$\rho_{k,h}$ according to the partial acceptance ratio of the recently completed compartment of steps. For the first MCMC update of a SMC step, we adapt the scaling factor according to the partial acceptance ratio of the last compartment of the previous SMC step. \NEW{The mathematical description of this recursive scheme is given by:}
\begin{equation}
	\begin{aligned}
		\rho_{k,s+1}&=\begin{cases}
			\,\rho_{k,s}\cdot 2 &\text{if }s=lH^*\text{ and }a_{k,l}^*>0.30,\\
			\,\rho_{k,s}\,/\,2&\text{if }s=lH^*\text{ and }a_{k,l}^*<0.15,\\
			\,\rho_{k,s}& \text{otherwise},
		\end{cases}\\
		\rho_{k+1,1}&=\begin{cases}
			\,\rho_{k,H}\cdot 2 &\text{if }a_{k,\lceil H/H^*\rceil}^*>0.30,\\
			\,\rho_{k,H}\,/\,2&\text{if }a_{k,\lceil H/H^*\rceil}^*<0.15,\\
			\,\rho_{k,H}& \text{otherwise},
		\end{cases}
	\end{aligned} \qquad\begin{aligned}
		\mbox{for }k&=1\Ldots K-1\,, \\
		s&=1\Ldots H-1 \\
		\mbox{and }l&=1\Ldots \lceil H/H^*\rceil-1\,.
	\end{aligned}\label{eq:MH-h}\tag{$\text{MH}^*$}
\end{equation}
Section~\ref{SUP-sec:CAL-MH} provides an exemplary demonstration of~\eqref{eq:MH-h}, see Table~\ref{Tab:MH-h}, and explains by this example why further enhancement of the above scheme is reasonable. In particular, there are situations, where the scaling factor proposed by~\eqref{eq:MH-h} will jump back and forth between two values due to alternating division and multiplication with the factor~$2$ from one MCMC update to another. This indicates that by picking a scaling factor between the two alternating values, we could achieve an acceptance ratio in the desired range~$[0.15,0.30]$\,. We make use of this observation by enhancing~\eqref{eq:MH-h}, resulting in the following adaptive scheme to set the variance~\eqref{eq:varMH-h} of the random walk:
\begin{equation*}
	\begin{aligned}
		\rho_{k,s+1}&=\begin{cases}
			\,\rho_{k,s}\cdot 2 &\text{if }s=lH^*\text{ and }a_{k,l}^*>0.30,\\
			\,\rho_{k,s}\,/\,2&\text{if }s=lH^*\text{ and }a_{k,l}^*<0.15,\\
			\,\rho_{k,s}& \text{otherwise},
		\end{cases}\\
		\tilde{\rho}_{k+1,1}&=\begin{cases}
			\,\rho_{k,H}\cdot 2 &\text{if }a_{k,\lceil H/H^*\rceil}^*>0.30,\\
			\,\rho_{k,H}\,/\,2&\text{if }a_{k,\lceil H/H^*\rceil}^*<0.15,\\
			\,\rho_{k,H}& \text{otherwise},
		\end{cases} \\
		\rho_{k+1,1}&=\begin{cases}
		\,\frac{1}{2}\big(\rho_{k,H}+\tilde{\rho}_{k+1,H}\big) &\text{if }\bar{a}_{k}^*\in[0.15,0.30],\\
		\,\tilde{\rho}_{k+1,1}& \text{otherwise},
	\end{cases}
	\end{aligned} \qquad\begin{aligned}
		\mbox{for }k&=1\Ldots K-1\,, \\
		s&=1\Ldots H-1 \\
		\mbox{and }l&=1\Ldots \lceil H/H^*\rceil-1\,,
	\end{aligned}\label{eq:MH-h-m}\tag{$\text{MH}^*_{k,h}$}
\end{equation*}
where~$\bar{a}_{k}^*=\frac{1}{2}\big(a_{k,\lceil H/H^*\rceil-1}^*+a_{k,\lceil H/H^*\rceil}^*\big)$ is the mean of the last two partial acceptance ratios of the~$k$-th SMC step and~$\rho_{1,1}=1$\,. 
\paragraph*{Selection of the number of MCMC updates~$\boldsymbol H$.} So far, we have considered the number of MCMC updates~$H$ to be constant for the whole SMC algorithm. However, the more often the Markov kernel is applied, the further we can reach to explore the parameter space. This becomes especially relevant if the selected variance~$\epsilon_j^2$ and hence the step size~$\xi_j$ of the random walk are small. Therefore, it can be reasonable to adaptively choose the number of MCMC updates in dependence of the scaling factor~$\rho_{k,h}$ for~${h=1\Ldots H}$. Based on~\cite{Beskos.2015} we can use for example
\begin{equation}
	H_{k,h}=\min\left\{\,\max\left\{\,\left\lfloor \frac{\zeta}{\rho_{k,h}^2} \right\rfloor\,,\, \txtSub{H}{min}\right\}\,,\,\txtSub{H}{max}\,\right\}\,, \label{eq:adjH}\tag{$\text{MH}_H$}
\end{equation}
where~${\zeta>0}$ is a global parameter and~$\txtSub{H}{min}$ resp.~$\txtSub{H}{max}$ are lower/upper bounds. The latter should enable a minimum extent of particle movement (lower bound) as well as avoid an exploding number of MCMC updates (upper bound). Note that the price to pay for a better exploration of the parameter space given by this adaptive choice is a potential increase in the computational cost of the algorithm for a fixed number of particles, since the forward model has to be solved for each MCMC update. 

\subsubsection{Summary of the algorithm}\label{ssec:algsummary}

For our calibrations of the chemoresistance models, sole data splitting was not sufficient to achieve reasonable posteriors with a practical sample size. Hence, we use a nested approach of performing likelihood tempering within an external data splitting scheme to calibrate these models with a sample size of~${P=10\,000}$. The approach is summarized in Algorithm~\ref{alg:SMC-chemo2}.
\begin{algorithm}[H]
	\small
	\caption{-- SMC with nested \NEW{reweighting}\\[4pt] {\small Sampling settings: sample size~$P=10\,000$ (particle index~$p=1\Ldots P$), threshold~$P^*=0.75\cdot P$\,;\\ MCMC settings:~$\txtSub{H}{min}=8$\,,~$\txtSub{H}{max}=64$\,,~$\zeta=1$}}\label{alg:SMC-chemo2}
	\begin{algorithmic}[1]
		\State $k=0:~$sample~$\theta_p\sim\mu_0$ and set~$W^0_p=1/P$,~$H_{0,H}=\txtSub{H}{min}$\label{algline:start2}
		\vspace*{4pt}\For{$k=1\Ldots K$}\Comment{\textbf{external: data splitting}}\label{algline:datasplitting}
		\vspace*{2pt}\State $s=0:~$set~$\tilde{\mathcal{I}}=\calIIk{k-1}{k}$\,,~$\tilde{W}^0_p=W^{k-1}_p$\,,~$\nu_s=0$ \Comment{\textbf{internal: likelihood tempering}}\label{algline:RW2}
		\vspace*{2pt}\While{$\nu_s<1$}\Comment{reweighting with~\eqref{eq:RW-L} and~\eqref{eq:LL-Perc}}
		\State determine~$\nu_{s+1}\in(\nu_s,1]$ such that~$\ESS\big(\{\tilde{W}^{s+1}_p\}_{p=1}^P\big)=P^*$
		\Statex \hspace*{1.1cm}for~$\tilde{W}_p^{s+1} = \tilde{w}_p^{s+1}/\big(\sum_{p=1}^{P} \tilde{w}_p^{s+1}\big)$ with~$\tilde{w}_p^{s+1} = L(\tilde{\mathcal{I}}\,|\,\theta_p)^{\nu_{s+1}-\nu_{s}}\, \tilde{W}^{s}_p$
		\vspace*{4pt} \State resample particles acc. to weights~$\{\tilde{W}^{s+1}_p\}_{p=1}^P$\Comment{resampling}\label{algline:RS2}
		\vspace*{2pt}\State $h=1~$: set~$H=H_{k-1,H}$ acc. to~\ref{eq:adjH} \Comment{MCMC updates}
		\vspace*{2pt}\While{$h\leq H$} \label{algline:MCMC2}
		\State move~$\theta_p\sim\kappa_k(\theta_p\,,\,\cdot\,)$ with adaptive MCMC scheme while updating~$H=H_{k,h}$\label{algline:MCMC3}
		\State $h=h+1$
		\EndWhile
		\vspace*{2pt}
		\State $s=s+1$
		\vspace*{4pt}
		\EndWhile
		\vspace*{2pt}\State set~$W^k_p=\tilde{W}^s_p$
		\vspace*{2pt}\EndFor
	\end{algorithmic}
\end{algorithm}
\noindent The choice of the resampling method in line~\ref{algline:RS2} and of the adaptive MCMC scheme in line~\ref{algline:MCMC3} depends on the investigated data/cell line according to necessity for robust calibrations. For Hep3B2, we used random resampling and the MCMC scheme~$(\text{MH}_{k-1})$ (for details on this scheme, see our paper~\cite{Schonfeld.2022}), while for HepG2/C3Asub28, we used selective resampling and the MCMC scheme~\eqref{eq:MH-h-m}. The algorithm has been implemented in Python~3 adapting the code structure from~\cite{Bulte.2020}.

Note that Algorithm~\ref{alg:SMC-chemo2} assumes that we have ordered our data according to~\eqref{eq:datasplitting}, see line \ref{algline:datasplitting} in the algorithm. We now describe how this has been done. We remind from Subsection \ref{ssec:expdata} that the calibration data describes the dose-response over~$\initD$ for a given combination of~$\ttreat$,~$\initS$,~$\initH$ and~$\initC$\,. We have to consider that tumor cells of distinct cell lines can react differently to the treatment, especially depending on the present oxygen supply and ECM stiffness. Hence, we calibrate the corresponding model separately to the measurements for each combination of~${\initH\in\{0,1\}}$ and~${\initC\in\{0,1\}}$, i.e. four individual calibrations are performed per cell line.

\begin{figure}[H]
	\centering
		\includegraphics[width=\textwidth]{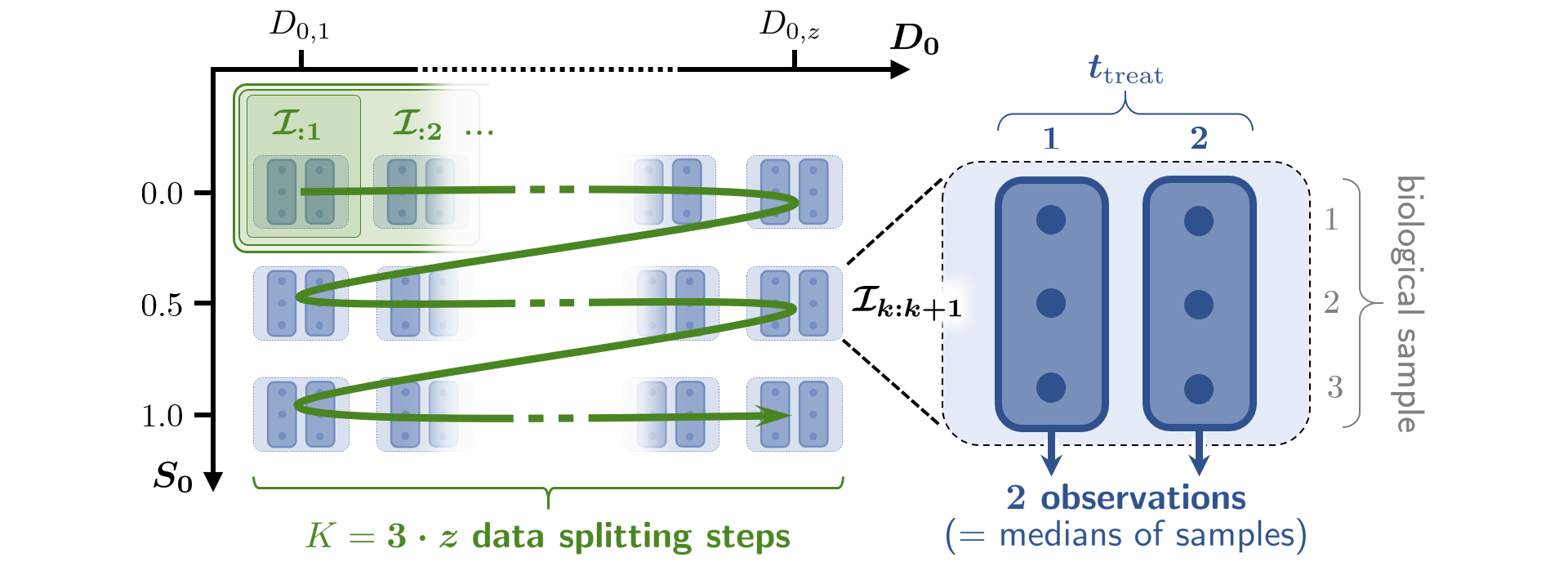}
		\caption[Data utilization to calibrate the models~\ref{eq:M-chemo-0} and~\ref{eq:M-chemo}]{Schematic description of the data utilization to calibrate the models~\ref{eq:M-chemo-0} and~\ref{eq:M-chemo}. The data splitting steps~${k=1\Ldots K}$ iterate over the observations for a selection of DOX dosages~${D_{0,1}\Ldots D_{0,z}}$\,,~${z\in\{1\Ldots 11\}}$ (inner loop) and all SOR dosages~${\initS\in\{0.0,\,0.5,\,1.0\}}$ (outer loop).}
		\label{Fig:MCsteps2}
\end{figure}

Figure~\ref{Fig:MCsteps2} illustrates how the measurements are used for each calibration to construct the consecutive data sets~${\{\calIk{k}\}_{k=1}^K}$\,. Over the course of the~$K$ data splitting steps, we iterate over the drug dosages~$\initS$ and~$\initD$\,, where a data increment~$\calIIk{k}{k+1}$ consists of the two respective measurements for~${\ttreat\in\{1,2\}}$ (right side of Figure~\ref{Fig:MCsteps2}). The algorithm starts with data for~${\initS=0}$ and iterates over the relevant values of~$\initD$ in ascending order, which is repeated for~${\initS=0.5}$ and eventually~${\initS=1}$ (left side of Figure~\ref{Fig:MCsteps2}). 
\noindent This order has the advantage that the data of the first third of SMC steps~(${\initS=0}$) contains no information about any SOR-related parameter. Hence, before considering the first measurements with~${\initS>0}$ we can reset the values of these parameters to their prior samples without changing the current intermediate distribution. This improves the sample diversity of the particle approximation.

\begin{figure}[h]
	\centering
		\includegraphics[width=\textwidth]{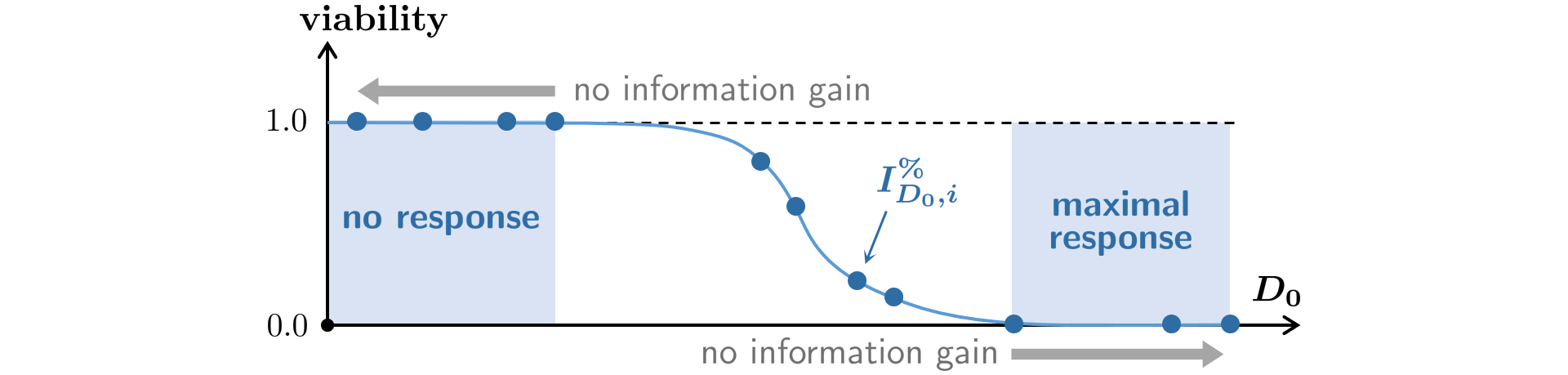}
		\caption[Informative content of exemplary (undisturbed) dose-response data~${\big\{I^\%_{\initD,i}\big\}_i}$]{Illustration of the informative content of exemplary (undisturbed) dose-response data~${\big\{I^\%_{\initD,i}\big\}_i}$\,. Once the measurements suggest no resp. maximal response to the applied dosage~$\initD$ (i.e.~${I^\%_{\initD}=1}$ resp.~${I^\%_{\initD}=0}$), there is no significant information gain from considering data for smaller resp. larger dosages in the calibration.}
		\label{Fig:doseresponse}
\end{figure}

When iterating along the~$\initD$-axis for calibration, there are situations where particular data points do not contribute to significantly improve the estimation of the posterior distribution. In this context, we can make the following observation (illustrated in Figure~\ref{Fig:doseresponse}). If the calibration has already incorporated the information \qm{no response for dosages~$D_{0,1}$ and~$D_{0,2}$} with~${D_{0,1}<D_{0,2}}$\,, then a data point~${\initD\in(D_{0,1},D_{0,2})}$ will not provide additional information to that statement, since the cells will show no response for this dosage either. An analogous observation can be made for dosages, which trigger a maximal response. Hence, such data points are feasible candidates to be omitted for the calibration process. The particular omitted measurements for each data set are mentioned while presenting the calibration results in Section~\ref{sec:RES-pred}.

\section{Results}\label{sec:results}
Recall the investigated treatments with combined drug dosages of
\begin{align*}
	\mbox{DOX (main drug):}&\quad\initD\in\{0.0001,0.001,0.01,0.1,0.5,1,5,10,50,100,1000\}~(\mbox{in }\SI{}{\micro\molar}),\\
	\mbox{SOR (supportive drug):}&\quad\,\initS\in\{0.0,0.5,1.0\}~(\mbox{in }\SI{22}{\micro\molar}),
\end{align*}
as well as the terms and abbreviations for the underlying environmental settings:
\begin{alignat*}{2}
\mbox{\qm{normal conditions} :}&&\quad\hphantom{=0}~\initH\,,\initC =0~&\overset{\text{abbr.}}{\leadsto}~\mbox{HC0}\,,\\ 
\mbox{\qm{sole hypoxia} :}&&\quad\initH =1\,, \initC=0~&\hspace*{4pt}\leadsto~\hspace*{4pt} \mbox{H1}\,,\\
\mbox{\qm{sole cirrhosis} :}&&\quad\initH =0\,,\initC =1~&\hspace*{4pt}\leadsto~\hspace*{4pt} \mbox{C1}\,,\\
\mbox{\qm{cirrhosis and hypoxia} :}&&\quad\hphantom{=0}~\initH\,,\initC =1~&\hspace*{4pt}\leadsto~\hspace*{4pt} \mbox{HC1}\,.
\end{alignat*}
The following sections display the results of the investigative model calibrations: We compare the model predictions with the corresponding data (Section~\ref{sec:RES-pred}) and show the the obtained estimates regarding drug-independent cell dynamics (Section~\ref{sec:RES-nodrug}), the treatment response in different environments (Section~\ref{sec:RES-treat}) and the drug metabolization (Section~\ref{sec:RES-met}). First, we present some choices that were taken to run the investigative model calibrations on the basis of preliminary calibration results for cell lines Hep3B2, HepG2 and C3Asub28.  
\paragraph*{Applicability of the models.} 
Recall that only cell line Hep3B2 exhibits no significant CYP expression~\cite{Ozkan.2021}, i.e. for these data we use the reduced model~\ref{eq:M-chemo-0}\,. However, attempts to calibrate the remaining data sets for HepG2 and C3Asub28 with model~\ref{eq:M-chemo} did not give reasonable posterior distributions to calculate an appropriate model solution reproducing the measurements. In particular, the algorithm shows difficulties to estimate the drug impact rates~$\sensRate{D}^-$ and~$\sensRate{S}^-$ as well as the metabolization rates~$\metD$ and~$\metS$\,. Possible reasons for this are discussed further in Section~\ref{sec:DIS-model}. On that basis, we calibrate the data for HepG2 and C3Asub28 with the enhanced reduced model~\ref{eq:M-chemo-0-inf} instead and focus our investigations on potential correlations between parameters and the available measurements of CYP expression from~\cite{Ozkan.2021}. 
\paragraph*{Applicability of the data.} 
The percentage viability data was clipped to the interval~${[0.001,1]}$ to accommodate the models' limited co-domain~${(0,1]}$. 
This should omit characteristics of enhanced growth observed in some data sets, where the percentage viability was strongly overshooting the mark of~$100\%$\,. In particular, such characteristics can be seen for the cell lines HepG2 (C1, HC1) and C3Asub28 (all). However, attempting model calibrations with these data sets resulted in degenerating variances of the intermediate distributions and unreasonable posteriors. These issues are caused by the inclusion of specific data points. A closer analysis of the respective measurements suggests that obvious growth enhancement only occurs in combined chemotherapy, i.e. in the presence of SOR. This leads to situations where, for a particular DOX dosage, the viability of cells which underwent combined treatment is significantly higher than for cell exposed to sole DOX treatment. Since the models do not consider this synergistic effect and assume that SOR can only decrease the cell viability, such measurements appeared to be impossible to reproduce without disproportionately increasing the noise variance. Therefore, it is not possible to reasonably investigate the concerned data sets (HepG2:~C1,~HC1 and C3Asub28:~all) with our models. The following Sections~{\ref{sec:RES-pred}--\ref{sec:RES-met}} present the calibration results of the remaining data sets Hep3B2~(all) and HepG2~(HC0,~H1). Complementary material is provided in Section~\ref{APP-Res2.1}. 

\subsection{Model predictions}\label{sec:RES-pred}
The model solution is calculated based on the parameter estimates, for which we use the so-called marginal \emph{maximum a posteriori probability estimates (MAP)}, i.e. the respective modes of the marginal distributions (for details see Subsection~\ref{sub:RES-ana-est}). As the marginal posteriors partially showed strong skewness (see Section~\ref{sec:othercal}), it is more reasonable to use the marginal MAPs instead of the marginal means to obtain parameter estimates. Note that using the MAP of the joint posterior to calculate the corresponding model solution would be more appropriate. However, its derivation is highly non-trivial due to the high dimensionality. As we need it for visualization purposes only, we use the marginal MAPs which are much easier to compute.

\begin{figure}[h]
	\centering
	\includegraphics[width=\linewidth]{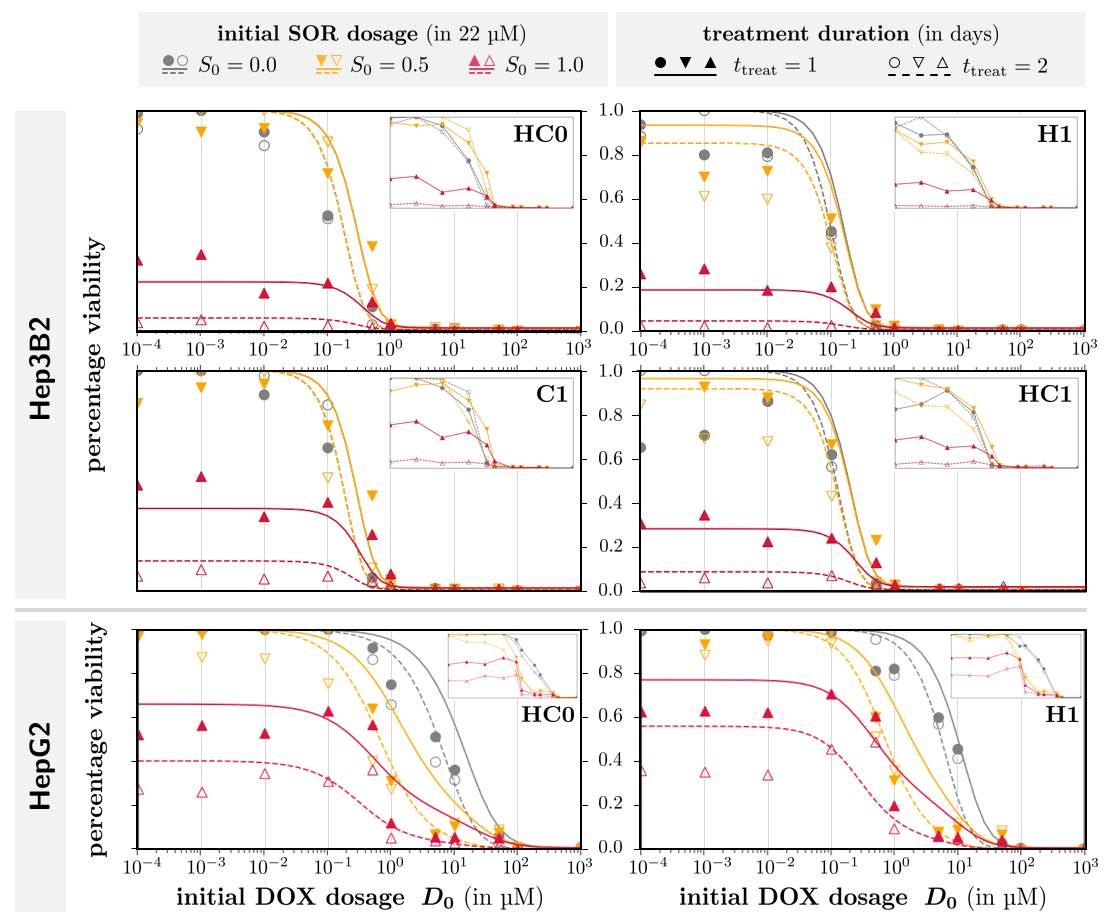}
	\caption[Comparison of measurements and estimated model solutions]{Comparison of the measured percentage viability for cell lines Hep3B2 (row one and two) resp. HepG2 (bottom row) with the corresponding estimated model solutions using model~\ref{eq:M-chemo-0} resp.~\ref{eq:M-chemo-0-inf} for different environmental settings. Note that for Hep3B2, some of the gray line plots might not be visible, as they overlap with the yellow ones.}
	\label{Fig:percV-all}
\end{figure}

For both cell lines Hep3B2 and HepG2, Figure~\ref{Fig:percV-all} shows the data (markers) with the corresponding model solution (line plots) based on the estimated marginal MAPs of the parameters. For an easier comparison, the small plots in the upper right corners illustrate the trend of the data. Following the approach mentioned in Subsection~\ref{ssec:algsummary} (recall also Figure~\ref{Fig:doseresponse}), we omitted the measurements for~${\initD\geq 10}$ resp.~${\initD\in\{10^{-3},10^{-2}\}}$ for the calibration data set of cell line Hep3B2 resp. HepG2.
\noindent In general, we get reasonable fits of the model solutions to the measurements. For both cell lines, we see a tendency of the model to overestimate the solution. 
For HepG2, we observe more asymmetry in the viability-dose relationship of the measurements in contrast to the solution, which is only clearly visible for the gray markers/lines, i.e. without the treatment support of SOR. We return to these aspects in Section \ref{sec:DIS-res}.

The following Sections~{\ref{sec:RES-nodrug}--\ref{sec:RES-met}} respectively present the estimated model parameters and discuss them in the underlying biological context. Marginal posteriors are mostly depicted as violin plots and obtained by a an adapted version of a Gaussian kernel density estimation~(KDE), which we refer to as \emph{truncated KDE} or \emph{trKDE} for short (for details see Section~\ref{SUP-sec:CAL-KDE}). An overview over all marginal distributions, MAPs and corresponding numerical deviations from cross-validation can be found in Figures~\ref{A-Fig:CL1-marg} (Hep3B2) and~\ref{A-Fig:CL2-marg} (HepG2). \NEW{We present numerical variations with the cross-validated 95\% confidence interval, visualized by error bars.} In general, the cross-validation showed high robustness and only small numerical variations of the calibration results. Therefore, it is reasonable to base our investigations on the estimates of the posterior obtained by collecting the particles of all four runs and do the significance check as proposed in Subsection~\ref{sub:RES-ana-var} and Figure~\ref{Fig:sigCheck}. 

\subsection{Drug-independent cell dynamics}\label{sec:RES-nodrug}
\begin{figure}[H]
	\centering
	\includegraphics[width=0.96\linewidth]{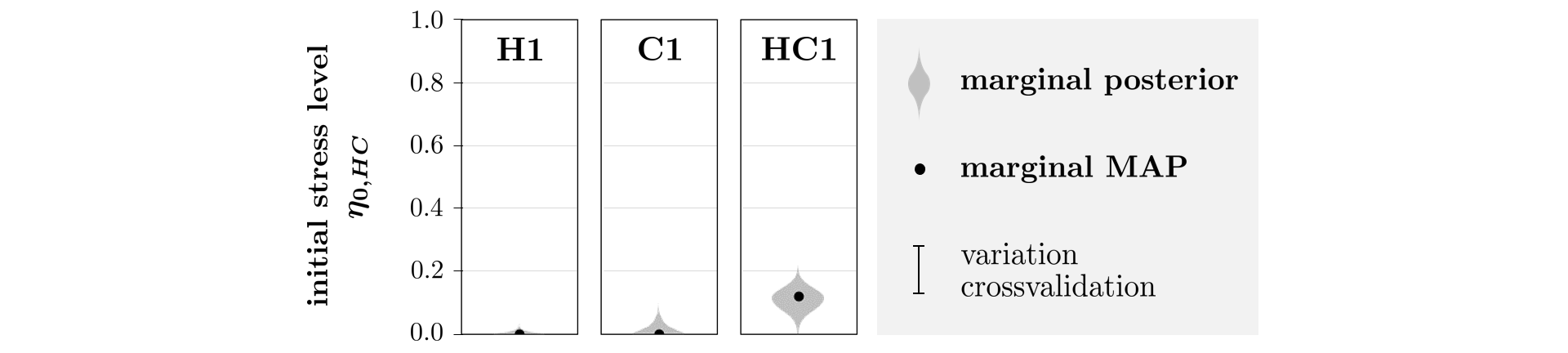}
	\caption[Hep3B2: Marginal posteriors and MAPs of~$\initStCH$]{Comparison of marginal posteriors (violin plots) and corresponding MAPs of~$\initStCH$ for Hep3B2. Due to minor numerical variations, the cross-validated 95\% confidence intervals (error bars) are not visible.}
	\label{Fig:CL1-stress0}
\end{figure}

In all chemoresistance models, the direct influence of hypoxia and ECM stiffness on the tumor growth as well as the growth and death rates are cell dynamics which do not depend on the applied treatment. The corresponding estimates are presented in this section.
\paragraph*{Stress induction by hypoxia and/or cirrhosis.} The direct influence of hypoxia and ECM stiffness on the tumor growth is modeled by the initial ESL~${\initStCH=\initSt(H,C)\in[0,1]}$\,. Its obtained marginal MAP estimates are given by
\begin{align}
\text{Hep3B2 (for H1, C1, HC1):}\qquad&0.2\cdot 10^{-6}\,,\quad4.5\cdot 10^{-6}\,,\quad0.119\,,\label{eq:RES-stress0-CL1}\\
	\text{HepG2 (for H1):}\qquad&1.2\cdot 10^{-6}\label{eq:RES-stress0-CL2}\,,
\end{align}
Recall that by definition~${\initStCH=0}$ for HC0. For HepG2, the MAPs show no significant stress induction by sole hypoxia (H1). For Hep3B2, Figure~\ref{Fig:CL1-stress0} compares the influence of the environment on the initial ESL by depicting the corresponding marginal posteriors. The distributions indicate that sole hypoxia or cirrhosis~(H1/C1) does not induce a significant level of stress on the tumor cells, while a combination of both~(HC1) does~($\AAAst$\,, see Figure~\ref{A-Fig:CL1-P-stress0}). 
\paragraph*{Cell growth and death.} Due to the context of percentage viability, we were not able to directly estimate the growth/death rates~$\prol$,~$\indDeath$ and~$\nat$ (recall Subsection~\ref{sub:MOD-App2-sol}). In particular, we could only estimate the term~${\prol+\indDeath}$\,, which was done just for the setting HC0 and fixed to its marginal MAP for the respective remaining calibrations~(Hep3B2: H1/C1/HC1, HepG2: H1).

\begin{figure}[H]
	\centering
	\includegraphics[width=0.96\linewidth]{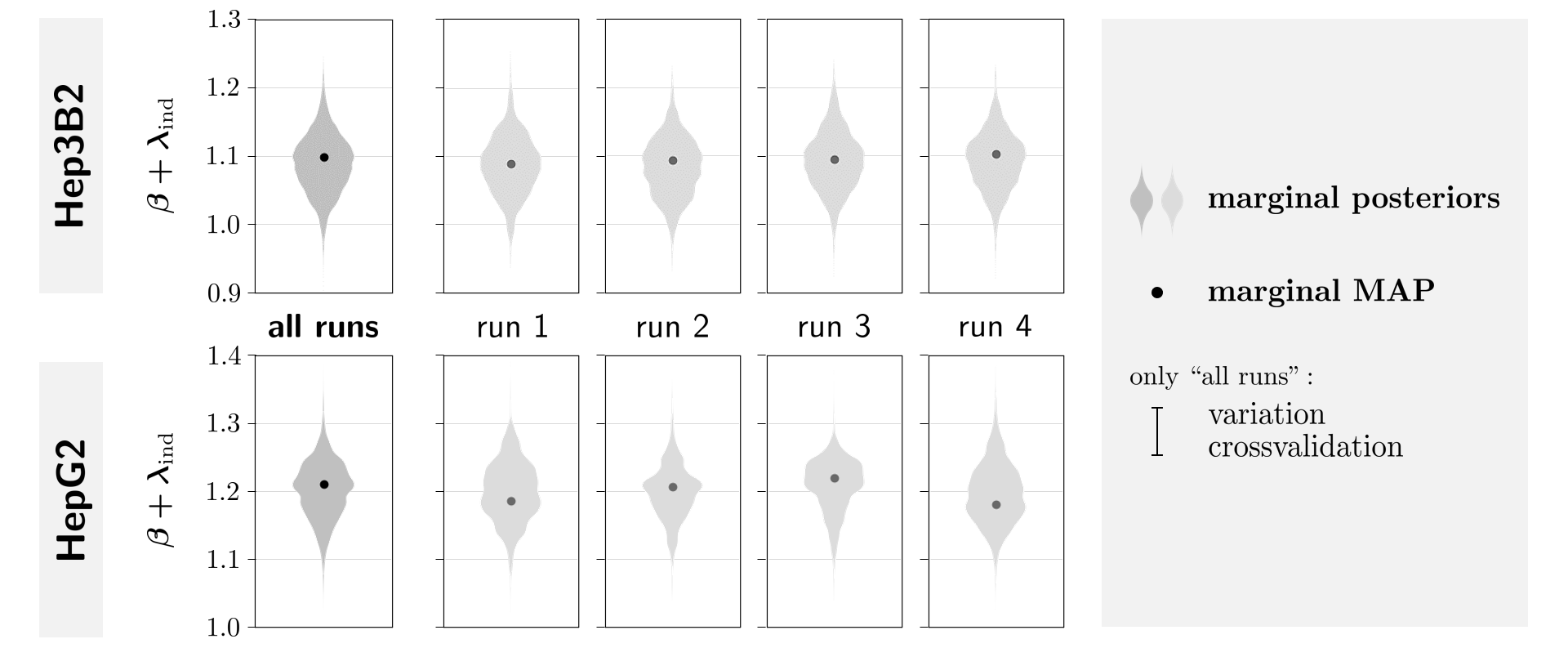}
	\caption[Estimations of~${\prol+\indDeath}$ in the setting HC0]{Estimations of~${\prol+\indDeath}$ for Hep3B2 (top) and HepG2 (bottom) in the setting HC0: Marginal posteriors as violin plot and corresponding MAPs (used deterministically in each run for calibrating the remaining settings). Due to minor numerical variations, the cross-validated 95\% confidence interval (error bar) is not visible. Note the different scaling of the vertical axis for Hep3B2/HepG2.}
	\label{Fig:RES-prolInd}
\end{figure}

For both cell lines, Figure~\ref{Fig:RES-prolInd} shows the estimates of this term for and each SMC run in comparison with the respective estimation resulting from the particle approximation collecting all runs. While the run-specific MAPs were used to perform further calibrations, the MAP considering all runs was used to calculate the new weights of the posterior approximation collecting the particles of all runs.
\noindent We see that the estimation of~${\prol+\indDeath}$ is very consistent for both cell lines, i.e. all remaining calibrations suppose virtually the same value for each run. Using the MAPs~${\prol+\indDeath\approx1.099}$ (Hep3B2) resp.~${\approx1.210}$ (HepG2) from the particle approximation combining all SMC runs, we can conclude\vspace*{-2pt}
\begin{equation*}
	\prol+\indDeath>\prol-\nat \quad\Rightarrow\quad\prol\,,\indDeath\,,\nat<1.099\text{ (Hep3B2)~~resp.~~}\prol\,,\indDeath\,,\nat<1.210\text{ (HepG2)}\,.\vspace*{-2pt}
\end{equation*}
Furthermore, we know that~${\initStCH\leq \frac{\prol-\nat}{\prol+\indDeath}}$ holds for all environmental conditions \NEW{(recall bounds in~\eqref{eq:initStress})}. For Hep3B2, by using the largest observed estimate of~${\initStCH}$ from~\eqref{eq:RES-stress0-CL1}, we obtain a lower bound for~${\prol-\nat}$\,:\vspace*{-2pt}
\begin{equation*}
	0.119\overset{\text{HC1}}{\approx}\initStCH\leq \frac{\prol-\nat}{~\underbrace{\,\prol+\indDeath\,}_{\approx~1.099}~} \quad\Rightarrow~\prol-\nat\gtrsim\frac{0.119}{1.099}\approx 0.108\,.\vspace*{-2pt}
\end{equation*}
For HepG2, we cannot 
derive a \NEW{numeric} lower bound for~${\prol-\nat}$\,, since we do not have a sufficiently non-zero estimate of~${\initStCH}$ for this cell line, see~\eqref{eq:RES-stress0-CL2}.
\subsection{Influence of oxygen supply and ECM stiffness on the treatment response}\label{sec:RES-treat}
\begin{figure}[h]
	\centering
	\includegraphics[width=0.96\linewidth]{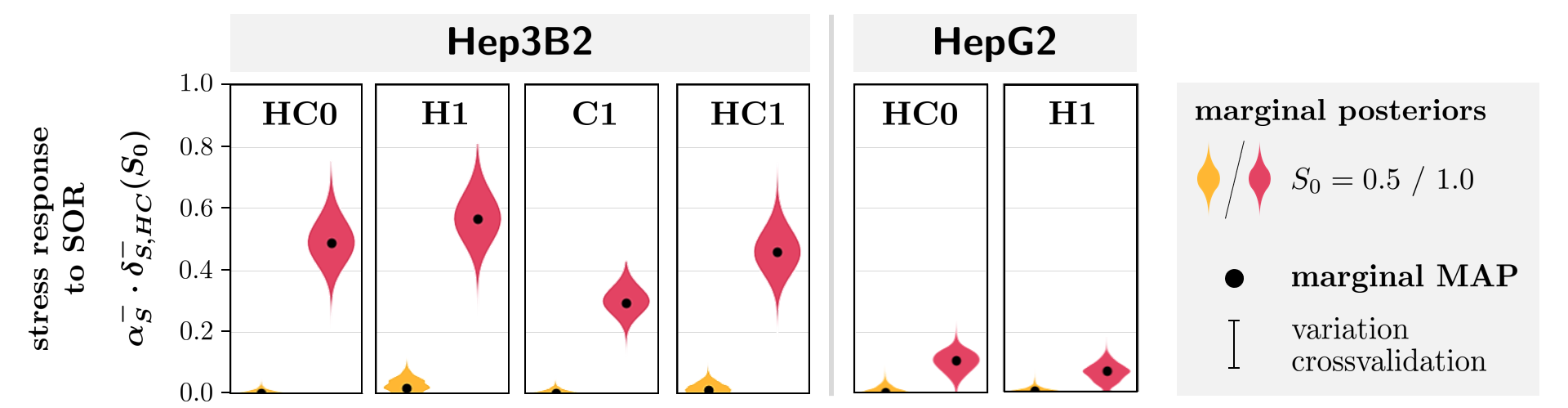}
	\caption[Hep3B2: Marginal posteriors and MAPs of~${\adSCH(\initS)}$]{Comparison of marginal posteriors (violin plots) and corresponding MAP estimates of~${\adSCH(\initS)}$ for Hep3B2 (left) resp. HepG2 (right). Note that for~${\initS=1.0}$ it was directly calibrated, while for~${\initS=0.5}$ the estimates are obtained by calibration of~${c_\delta=\adSCH(0.5)\,/\,\adSCH(1.0)}$. Due to minor numerical variations, the cross-validated 95\% confidence intervals (error bars) are not visible.}
	\label{Fig:CL1-aS}
\end{figure}

In this section we present the estimates regarding the cells' stress response to the chemotherapeutic treatment and how it is influenced by the environment.
\paragraph*{Cytotoxic efficacy of SOR for Hep3B2 and HepG2.}
Besides its supportive effect, SOR can be directly cytotoxic for the tumor cells. 
This would show in significantly non-zero values of the stress response function \vspace*{-2pt}
\begin{equation*}
	\adSCH(S)=\sensRate{S}^- \deact_{S}(S,H,C)=\frac{\sensRate{S}^-\cdot S^{m_2}}{\Sthr(H,C)^{m_2}+S^{m_2}} \vspace*{-2pt}
\end{equation*}
for~${S=\initS >0}$\,. Recall from Figure~\ref{Fig:infS} 
that we could only estimate the function values of~${\adSCH}$ pointwise for~${\initS\in\{0.5,1.0\}}$ due to the limited number of experimental SOR dosages. Thus we lack quantitative information on how fast the SOR treatment~($\sensRate{S}^-$) or the dose-response~($\deact_{S,HC}$) for continuous~$S$ impacts the cells. Figure~\ref{Fig:CL1-aS} shows the marginal posteriors of the stress response as violin plots as well as the estimated marginal MAPs for both cell lines.  \enlargethispage{\baselineskip}
\noindent We proceed with a closer investigation of the individual estimates of Figure~\ref{Fig:CL1-aS}, where all mentioned significance results are derived from Figure~\ref{A-Fig:CL1-P-aS} (Hep3B2) resp. Figure,~\ref{A-Fig:CL2-P-SOR} (HepG2). 

\mypar{High SOR dosage} In general, the cytotoxic efficacy of a high SOR dosage appears to be weaker for cell line HepG2 compared to Hep3B2.

For Hep3B2, we see a clear stress response for~${\initS =1.0}$ under all environmental conditions where 
cirrhosis and hypoxia have a significant influence on the drug efficacy. In particular, oxygen deprivation increases the response to SOR~($\AAAst$), while stiffening the ECM decreases it~($\AAAst$):
\begin{align*}
	\text{sole hypoxia (HC0${\,\leadsto\,}$H1):}\quad0.489\nearrow0.566~~(+15.7\%)\,,\\
\text{hypoxia in cirrhosis (C1${\,\leadsto\,}$HC1):}\quad0.294\nearrow0.458~~(+55.8\%)\,,\\[4pt]
	\text{sole cirrhosis (HC0${\,\leadsto\,}$C1):}\quad0.489\searrow0.294~~(-39.9\%)\,,\\
	\text{cirrhosis in hypoxia (H1${\,\leadsto\,}$HC1):}\quad0.566\searrow0.458~~(-19.1\%)\,.
\end{align*}
We see that the degree of the increase/decrease of the response to SOR is different for sole cirrhosis/hypoxia compared to a combination of both. Furthermore, the combination of high ECM stiffness and hypoxia appears to slightly diminish the response compared to normal conditions:
\begin{equation*}
	\text{HC0}{\,\leadsto\,}\text{HC1}:\quad0.489 \searrow 0.458~~(-6.3\%)\,.
\end{equation*}
However, we see no statistical significance of latter observation, i.e. its relevance is unclear.

In contrast, for HepG2, we see a significant~($\AAAst$) reduction of the stress response to SOR for~${\initS=1.0}$ due to
\begin{equation*}
    \text{sole hypoxia (HC0${\,\leadsto\,}$H1):}\quad0.103\searrow0.063~~(-38.8\%)\,.
\end{equation*}
This does not necessarily indicate a contradictory behaviour compared to Hep3B2, as the discrete representation of the stress response to SOR limits the comparability of the corresponding results for different cell lines. More details to that manner will be provided in Section~\ref{sec:DIS-res}. 

\mypar{Standard SOR dosage} Overall, we get relatively small marginal MAPs for~${\adSCH(\initS =0.5)}$:
\begin{align*}
\text{Hep3B2 (for HC0, H1, C1, HC1):}\qquad&4.4\cdot 10^{-9}\,,\quad\hspace*{19pt}0.019\,,\quad5.0\cdot 10^{-7}\,,\quad0.011\,,\\
	\text{HepG2 (for HC0, H1):}\qquad&1.6\cdot 10^{-6}\,,\quad0.9\cdot 10^{-6}\,.
\end{align*}
A non-zero stress response is only visible for Hep3B2 and if hypoxia involved~(H1/HC1).

For Hep3B2, we again see an increasing~($\AAAst$) resp. decreasing (unclear significance) influence of hypoxia resp. cirrhosis on the treatment response. The fact that we only see a reaction under hypoxic conditions, even in combination with a stiff ECM (HC1), indicates a dominating increase of SOR efficacy due to hypoxia compared to the decrease by tissue stiffening~($\AAAst$).

For HepG2, the estimates indicate a similar tendency like for a high SOR dosage, namely that hypoxia appears to reduce the stress response to SOR. However, the significance of this observation is unclear.
\paragraph*{Cytotoxic efficacy of DOX for Hep3B2.} Next, we analyze the estimated stress response to DOX given by
\begin{equation*}
	\sensRate{D}^-\cdot \deact_{D,HC}(\initD,\initS)=\sensRate{D}^-\cdot\frac{\initD ^{m_1}}{\big(\DnormCH\cdot\, d_S(\initS )\big)^{m_1}+\initD ^{m_1}}
\end{equation*}
for varying DOX and SOR dosages~$\initD$ resp.~$\initS$ and different environmental conditions~$\initH$\,,~$\initC$\,. 
As expected, the particle approximation indicates relevant correlations of the DOX susceptibility threshold~$\DnormCH$ and the remaining parameters of~${\sensRate{D}^-\cdot\deact_D}$ (see Figure~\ref{A-Fig:CL1-corr2}). 
\NEW{To consider these,} we calculate~${\adDCH(\initD,\initS)}$ for each particle and each combination of~$\initD$ and~$\initS$\,. 

\begin{figure}[h]
	\centering
	\includegraphics[width=\linewidth]{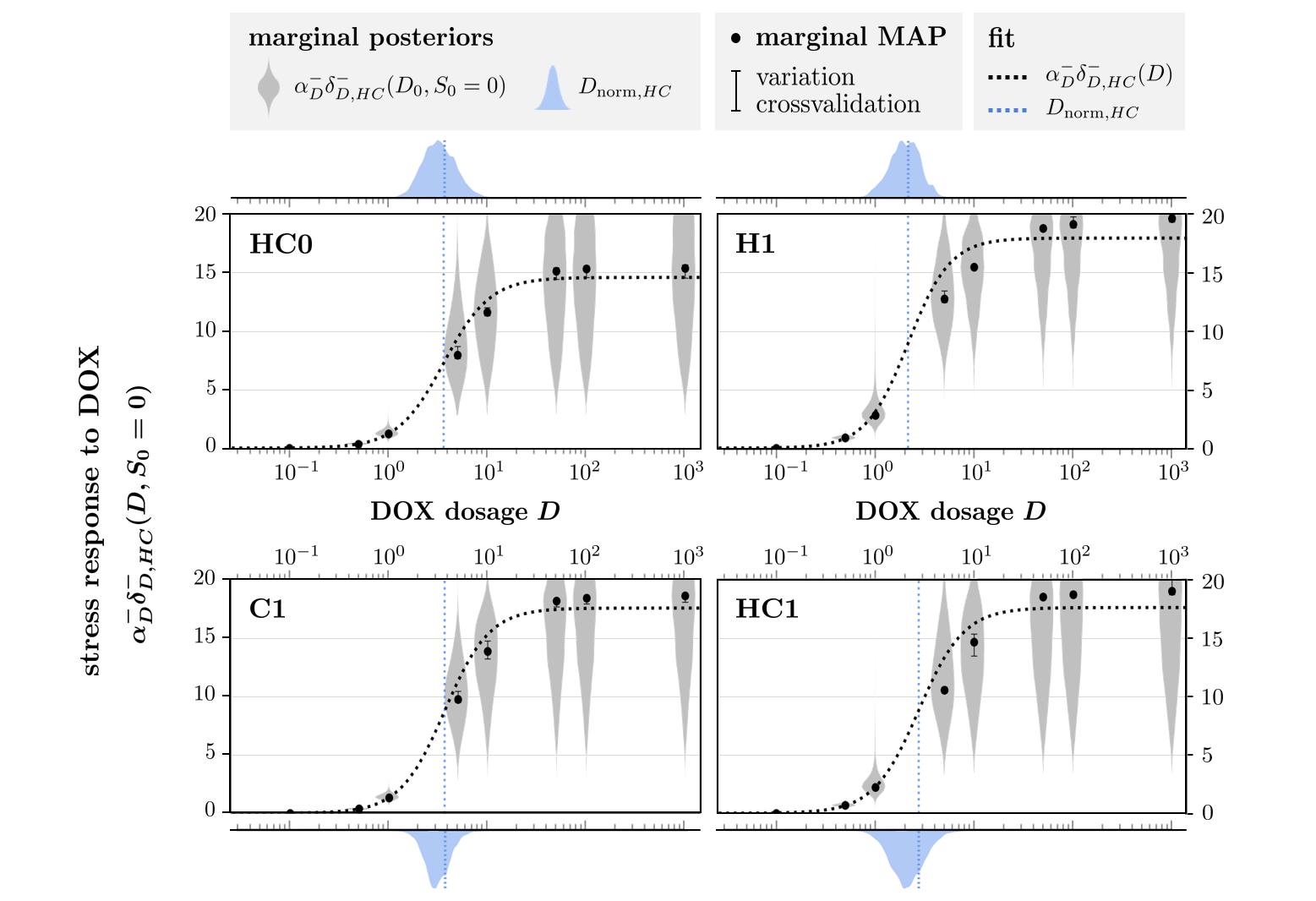}
	\caption[Hep3B2: Marginal posteriors and MAPs of~${\adDCH(\initD,0)}$]{Comparison of marginal posteriors (violin plots) and corresponding MAP estimates of the unsupported DOX efficacy~${\adDCH(\initD,0)}$ for Hep3B2. The sigmoidal line is the corresponding least square fit. Note the only measurements for DOX dosages up to~${\initD=5}$ were actually used for the model calibration\NEW{, hence the marginals for larger~${\initD}$ are extrapolated to reach the horizontal asymptote for reasonable fitting}. Due to minor numerical variations, some cross-validated 95\% confidence intervals (error bars) are not visible.}
	\label{Fig:CL1-aD1}
\end{figure}

Figure~\ref{Fig:CL1-aD1} depicts the resulting distributions (gray violin plots) for particular DOX dosages~$\initD$ and without the support of SOR (${\initS =0}$). Furthermore, it shows a weighted least square fit of~${\adDCH}$ (dotted black line) to the MAPs of these distributions under consideration of the respective variances. We use the Python function \texttt{scipy.optimize.curve\_fit(..., sigma=stdev, absolute\_sigma=True)}, where~\texttt{stdev} are the standard deviations obtained from the respective marginal variances. The resulting fitted values of~$\DnormCH$ are illustrated (dotted blue line) in comparison with the marginalized posterior distribution of~$\DnormCH$ (blue shaded truncated KDE plots on the top/bottom).
\noindent In general, we achieve good 
and reasonable fits where the resulting values for the susceptibility threshold~$\DnormCH$ are in the high probability region of the respective marginal posteriors, i.e. close to the global modes of the truncated KDEs. Furthermore, the estimated sensitivity parameter~$m_1$ (steepness of the stress response) shows no obvious differences for all environmental conditions:
\begin{align*}
	\text{HC0: }~& 1.858~(\text{fit}),~ 1.772~(\text{MAP})\,,\qquad\text{\hphantom{C}H1: }~ 2.059~(\text{fit}),~ 1.807~(\text{MAP})\,,\\
	\text{C1: }~& 1.992~(\text{fit}),~ 1.898~(\text{MAP})\,,\qquad\text{HC1: }~ 1.907~(\text{fit}),~ 1.857~(\text{MAP})\,.
\end{align*}
\NEW{For such small variations, a closer biological interpretation of the differences is of minor interest}. Hence, we focus the following investigations on the susceptibility threshold~$\DnormCH$ and the impact rate~$\sensRate{D}^-$\,.

\mypar{DOX susceptibility threshold} We analyze the fitted estimates of~$\DnormCH$ as they are obtained in consideration of the parameter correlations. However, they do not provide a sample representation of the estimates and hence we cannot judge the statistical significance of our observations in contrast to the approach of investigating the marginal distributions (without considering correlations) of~$\DnormCH$ instead. The latter is done for comparison in Subsection~\ref{APP-Res2.1-MARG} and will be referred to as \emph{marginalized investigation} for the rest of this paragraph. \NEW{The significance checks for both fitted and marginal estimates can be found in Subsection~\ref{APP-Res2.1-MARG} as well.}

The fitted values of~$\DnormCH$ are:
\begin{equation}
\begin{alignedat}{3}
	\text{sole hypoxia (HC0${\,\leadsto\,}$H1):}\quad3.646\searrow2.161~~(-40.7\%)\,,\\
	\text{hypoxia in cirrhosis (C1${\,\leadsto\,}$HC1):}\quad3.689\searrow2.762~~(-25.1\%)\,,\\[4pt]
	\text{sole cirrhosis (HC0${\,\leadsto\,}$C1):}\quad3.646\nearrow3.689~~(+\hphantom{0}1.2\%)\,,\\
	\text{cirrhosis in hypoxia (H1${\,\leadsto\,}$HC1):}\quad2.161\nearrow2.762~~(+27.8\%)\,.
\end{alignedat}\label{eq:CL1-DnormHC}
\end{equation}
The smaller~$\DnormCH$\,, the more susceptible the cells' are to DOX. Hence, the above estimates show an increase/decrease of the cells' susceptibility to DOX due to hypoxia/cirrhosis. Qualitatively, the same result is obtained by the marginalized investigation. The degree of increase/decreases of~$\DnormCH$ appears to be different for sole and combined environmental factors, indicating a synergistic effect between hypoxia and cirrhosis. Furthermore, comparing the fitted values
\begin{equation*}
	\text{HC0}{\,\leadsto\,}\text{HC1}:\quad3.646 \searrow 2.762~~(-24.2\%)
\end{equation*}
suggests that the increase of susceptibility due to hypoxia prevails the decreasing effect of high ECM stiffness, which is consistent with the marginalized investigations.

\mypar{DOX impact} Figure~\ref{Fig:CL1-aD1} shows that the MAPs as well as the fit of the stress response~${\adDCH}$ tend to a horizontal asymptote for large DOX dosages. Mathematically, this asymptote is given by the DOX impact rate~$\sensRate{D}^-$ scaling the sigmoid function~${\deact_{D,HC}\in[0,1)}$. Due to the observable asymptotic behavior, it is reasonable to assume that~${\deact_{D,HC}\approx 1}$ for the largest applied dosage~${\initD=10^3}$. Hence, the estimates of~${\adDCH(10^3,0)}$ can be seen as estimates for~$\sensRate{D}^-$ (i.e. the rightmost gray marginals in each subplot in Figure~\ref{Fig:CL1-aD1}).

In general, we see relatively large estimates for~$\sensRate{D}^-$\,, i.e. a fast reaction to the DOX treatment. The corresponding MAPs as well as the fitted value of~$\sensRate{D}^-$ appear to be smaller for HC0 than for the other environmental conditions:
\begin{align*}
	\text{HC0: }~&14.564~(\text{fit}),~15.365~(\text{MAP})\,,\qquad\text{\hphantom{C}H1: }~17.919~(\text{fit}),~19.619~(\text{MAP})\,,\\
	\text{C1: }~&17.507~(\text{fit}),~18.553~(\text{MAP})\,,\qquad\text{HC1: }~17.551~(\text{fit}),~18.978~(\text{MAP})\,,
\end{align*}
which suggests a faster reaction to the DOX treatment if hypoxia and/or cirrhosis is involved.

\mypar{Supportive effect of SOR} Additionally to direct cytotoxicity, SOR can have a supportive effect on the DOX treatment. This is modeled by the quantity~${d_S(S)\in[0,1]}$\,, which potentially shifts the DOX susceptibility threshold of the dose-response function~$\deact_{D,HC}$ via
\begin{equation*}
	\Dthr(S,\initH,\initC)\overset{\eqref{eq:Dthr}}{=}\DnormCH\cdot d_S(S)=\DnormCH\cdot \left(1-\frac{\txtSub{a}{max}\cdot S^{m_3}}{\Ssupp^{m_3}+S^{m_3}}\right)\,.
\end{equation*}
Again, due to the limited number of experimentally investigated SOR dosages we could only estimate the values of~${d_S(S)}$ for~${S\in\{0.5,1.0\}}$ instead of the separate parameters~$\txtSub{a}{max}$\,,~$m_3$ and~$\Ssupp$ (recall Figure~\ref{Fig:infS} in Subsection~\ref{sec:CAL-Prior-App2}). The smaller the value of~${d_S(S)}$\,, the stronger the supportive influence of SOR. Therefore, we can take the term~${1-d_S(S)\in[0,1]}$ as a measure for the supportive effect.

\begin{figure}[H]
	\centering
	\includegraphics[width=\linewidth]{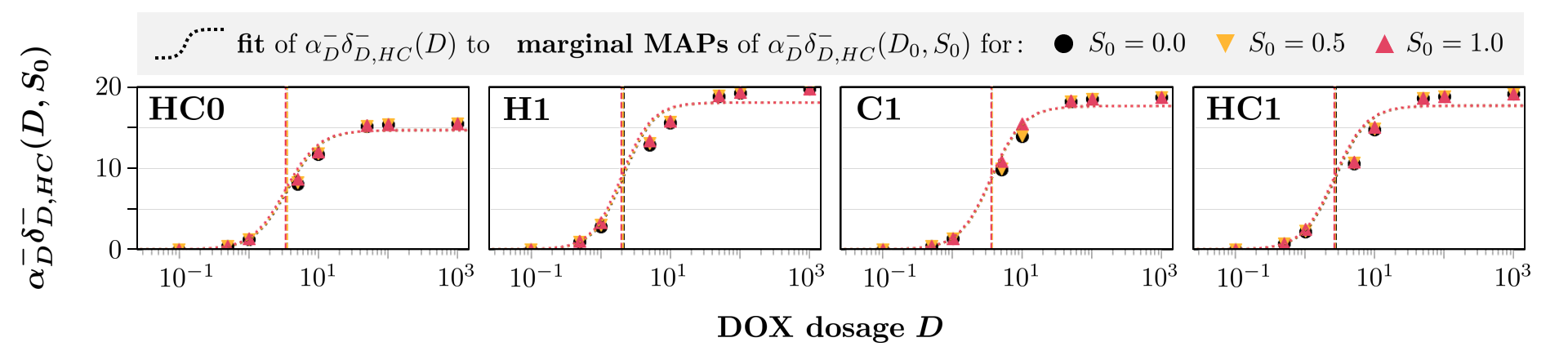}
	\caption[Hep3B2: Fitted DOX efficacy~${\adDCH(D,\initS)}$ for varying SOR dosages]{Comparison of the fitted DOX efficacy~${\adDCH(D,\initS)}$ (sigmoidal lines) for Hep3B2 and the resulting shifted DOX susceptibility threshold~${\DnormCH\cdot d_S(\initS)}$ (vertical lines) for varying SOR dosages~${\initS\in\{0.0,0.5,1.0\}}$ to support the DOX treatment.}
	\label{Fig:CL1-aD-SOR}
\end{figure}

In order to judge if the obtained estimates for~${1-d_S(\initS)}$ actually result in considerable support of the DOX treatment, we set them into relation with the estimates of the unsupported stress response~$\adDCH(\initD ,\initS=0)$. More precisely, we calculate~$\adDCH(\initD ,\initS)$ for each particle and each combination of~$\initH$\,,~$\initC$\,,~$\initD$ and~${\initS\in\{0.5,1.0\}}$ for comparison. This way we also consider the observed correlations between~$d_S$ and~$\Dnorm$ (see Figure~\ref{A-Fig:CL1-corr2}). Analogously to the fits of the unsupported stress response in Figure~\ref{Fig:CL1-aD1}, we can now fit~$\adDCH(D,\initS)$ for~${\initS\in\{0.5,1.0\}}$ to the respective MAP estimates at~${D=\initD}$\,. For better comparability, we fix the parameters~$\sensRate{D}^-$ and~$m_1$ to the calculated values from the fit of~${\adDCH(D,0)}$\,. Eventually, the resulting fits as well as the MAPs do not show an obvious supportive effect of SOR on the DOX treatment, which is illustrated in Figure~\ref{Fig:CL1-aD-SOR}. Although the estimates for~${1-d_S(\initS)}$ do not yield a noticeable treatment support, it might be interesting to take a closer look at its marginals, which is done based on Figure~\ref{Fig:CL1-dS} in Section~\ref{APP-Res2.1-MARG}. \enlargethispage{\baselineskip}

\begin{figure}[h]
	\centering
	\includegraphics[width=\linewidth]{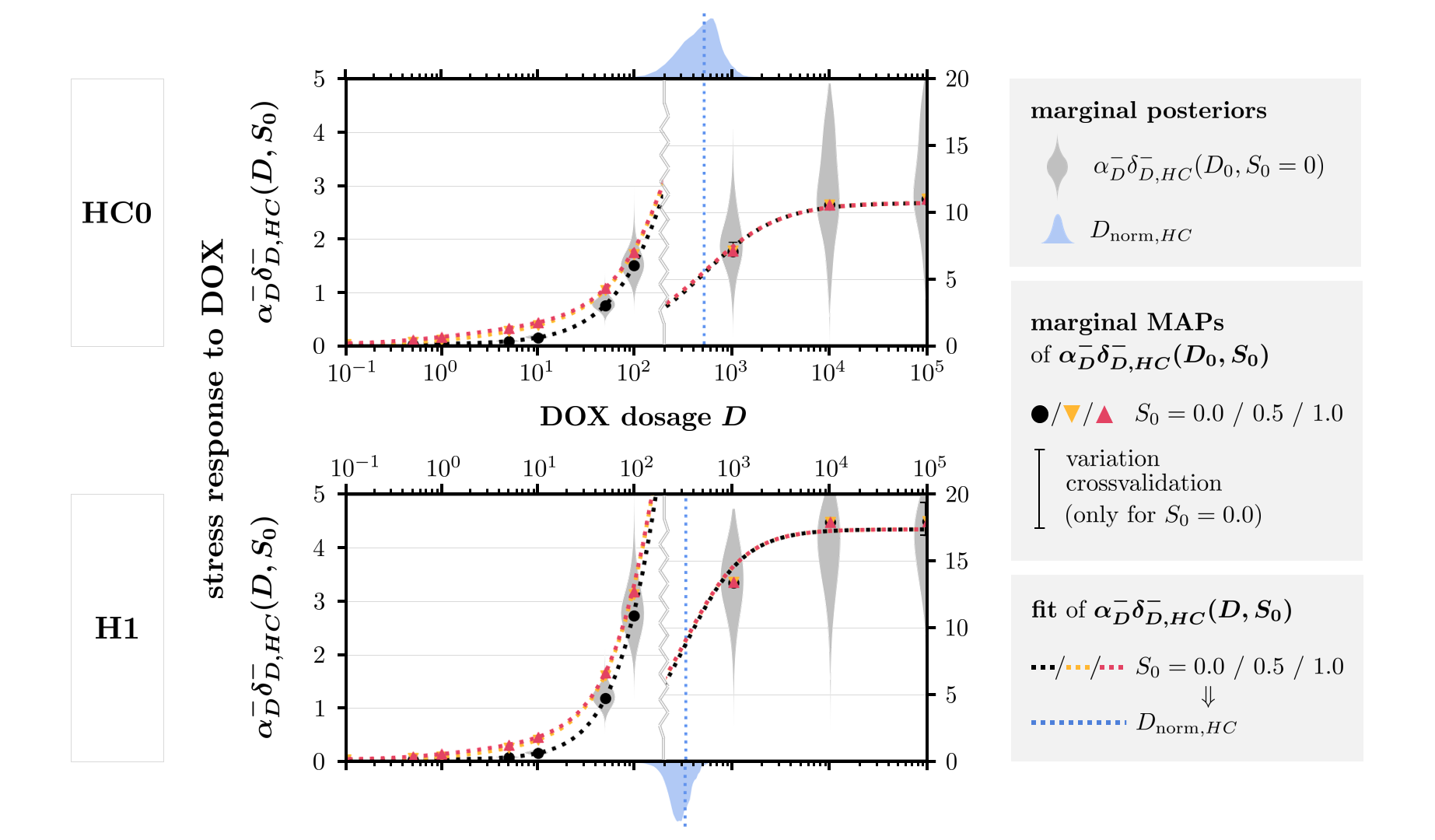}
	\caption[HepG2: Marginal posteriors and MAPs of~${\adDCH(\initD,0)}$]{Comparison of distributions (violin plots) and corresponding MAP estimates (black markers) of the unsupported DOX efficacy~${\adDCH(\initD,0)}$ for HepG2 for particular dosages~$\initD$. Note the respectively different scaling of the vertical axis for the left and right part of the plot (separated by the vertical light gray wavy line). The non-black markers are the MAPs for combination therapy (corresponding violin plots are not shown; they do not bring more clarity as they have a similar shape to the gray ones). The dotted curves are the least square fits to the respective MAPs and the fitted values of~$\DnormCH$ are illustrated as light blue dotted lines in comparison with the marginalized posterior distribution of~$\DnormCH$ (light blue shaded truncated KDE plots on the top/bottom). Due to minor numerical variations, some cross-validated 95\% confidence intervals (error bars) are barely visible.}
	\label{Fig:CL2-aD}
\end{figure}

\paragraph*{Cytotoxic efficacy of DOX for HepG2.}
Like for Hep3B2, we calculate stress response to DOX
\begin{equation*}
	\adDCH(\initD,\initS)=\frac{\sensRate{D}^-\cdot\initD ^{m_1}}{\left(\DnormCH\cdot\, \left(1-\big(1-d_S(\initS)\big)\cdot\left(1-\frac{D}{\Ddamp+D}\right)\right)\right)^{m_1}+\initD ^{m_1}}\vspace*{-2pt}
\end{equation*}
for each particle and each combination of~$\initD$ and~$\initS$ to consider the correlations (see Figure~\ref{A-Fig:CL2-corr1}) of the involved parameters. Figure~\ref{Fig:CL2-aD} shows the weighted least square fits (dotted curves) of~${\adDCH(D,\initS)}$ to the MAPs (markers) of the resulting distributions. Note that the marginals of~${\initD> 10^3}$ are extrapolated to reach the horizontal asymptote for reasonable fitting.
\noindent We achieve good and reasonable fits 
with a susceptibility threshold~$\DnormCH$ (blue dotted line) close to the global modes of the corresponding marginal truncated KDEs (blue distributions). Table~\ref{Tab:CL2-parsDOX} summarizes the fitted values of the parameters involved in the stress response in comparison with their marginal MAPs and shows a high consistency.

\begin{table}[h]
	\caption[HepG2: Parameter estimates regarding the stress response to DOX]{Comparison of the parameter estimates regarding the stress response to DOX for HepG2 in normal (HC0) and sole hypoxic (H1) environment. The 95\%~confidence intervals of the fitted values can be found in Table~\ref{A-Tab:CL2-fitsSTD}.}\label{Tab:CL2-parsDOX}
	\begin{subtable}{\textwidth}
		\centering	
		\caption{Parameters of the unsupported~(${\initS=0}$) stress response. The statistical significance of differences between HC0 an H1 is checked with Figure~\ref{A-Fig:CL2-P-DOX}.} \label{Tab:CL2-parsDOX-1}
		\begin{tabular}{rrrcrl} \toprule
			~&~&\textbf{HC0}&$\boldsymbol{\leadsto}$&\textbf{H1}& significant?\\\midrule
			\multirow{2}{*}{$\boldsymbol{\sensRate{D}^-}$}&\textbf{fitted}&10.694&\multirow{2}{*}{$\nearrow$}&17.413&$\AAAst$\\
			&\textbf{MAP}&11.043&&17.997&$\AAAst$\\[4pt]
			\multirow{2}{*}{$\boldsymbol{\DnormCH}$}&\textbf{fitted}&501.122&\multirow{2}{*}{$\searrow$}&318.290&$\AAAst$\\
			&\textbf{MAP}&574.812&&272.357&$\AAAst$\\[4pt]
			\multirow{2}{*}{$\boldsymbol{m_1}$}&\textbf{fitted}&1.128&\multirow{2}{*}{$\nearrow$}&1.433&$\AAAst$\\
			&\textbf{MAP}&1.074&&1.405& $\AAAst$\\
			\bottomrule
		\end{tabular}
	\end{subtable}
	\newline
	\vspace{10pt}
	\newline
	\begin{subtable}{\textwidth}
		\centering	
		\caption{Parameters of the supportive influence of SOR on the stress response for~${\initS\in\{0.5,1.0\}}$, where~${1-d_S(\initS)\in[0,1)}$ shows the undamped supportive effect and~${1/\cdamp}$ the strength of damping. For~${1/\cdamp}$ we have two different fits for~${\initS=0.5}$~(top value) and~${\initS=1.0}$~(bottom value). The statistical significance of differences between HC0 an H1 is checked with Figure~\ref{A-Fig:CL2-P-SOR}.} \label{Tab:CL2-parsDOX-2}
		\begin{tabular}{rrrcrl} \toprule
			~&~&\textbf{HC0}&$\boldsymbol{\leadsto}$&\textbf{H1}&significant?\\\midrule
			\multirow{2}{*}{$\boldsymbol{1-d_S(0.5)}$}&\textbf{fitted}&0.879&\multirow{2}{*}{$\approx$}&0.873&potentially insignificant\\
			&\textbf{MAP}&0.903&&0.910&potentially insignificant\\[4pt]
			\multirow{2}{*}{$\boldsymbol{1-d_S(1.0)}$}&\textbf{fitted}&0.948&\multirow{2}{*}{$\approx$}&0.942&potentially insignificant\\
			&\textbf{MAP}&0.960&&0.963&potentially insignificant\\[4pt]
			\multirow{3}{*}{$\boldsymbol{\frac{1}{\cdamp}}$}&\multirow{2}{*}{\textbf{fitted}}&22.542&\multirow{3}{*}{$\searrow$}&18.864&potentially significant\\[-2pt]
			&&22.467&&19.128&$\Ast$\\
			&\textbf{MAP}&20.532
			&&16.742
			&potentially significant\\ \bottomrule
		\end{tabular}
\end{subtable}
\end{table}
\noindent When comparing the parameter values for HC0 and H1, Table~\ref{Tab:CL2-parsDOX-1} shows an increase of~$\sensRate{D}^-$ and a decrease of~$\DnormCH$, which both indicate an enhanced cytotoxic efficacy of DOX due to hypoxia. This is similar to the observations for cell line Hep3B2. We also see a slight increase of~$m_1$\,, i.e. a steeper dose-response relationship.

The estimated values of~${1-d_S(\initS)}$ in Table~\ref{Tab:CL2-parsDOX-2} as well as the left side of Figure~\ref{Fig:CL2-aD} exhibit a considerable supportive effect for both a standard and high SOR dosage (in contrast to cell line Hep3B2). The effect is respectively similar under normal and hypoxic conditions. Furthermore, we see~${1/\cdamp\gg 1}$ for both HC0 and H1, which yields a considerable DOX dose-dependency of the supportive effect. This is also visible on the right side of Figure~\ref{Fig:CL2-aD}, where the supportive effect vanishes for large DOX dosages.

\subsubsection{Summary and comparison of the treatment response for Hep3B2 and HepG2}
\label{sub:RES-Chemo-SUM}
We conclude with a brief overview over the obtained calibration results for cell lines Hep3B2 and HepG2 regarding the treatment response. In general, both cell lines show a qualitatively similar behavior, except for the supportive influence of SOR. The latter appears to be cell line-specific as we only observe a considerable effect for HepG2. Table~\ref{Tab:CL12} summarizes the observed influences of hypoxia (for both cell lines) and high ECM stiffness (for Hep3B2).
\begin{table}[h]
\caption[Hep3B2/HepG2: Influence of hypoxia and/or tissue stiffening on the treatment response]{Influence of introducing hypoxia (H), tissue stiffening (C) or a combination of both (HC) on the treatment response of the investigated cell lines. Recall that for HepG2 we could not apply model calibrations on any data considering high stiffness.}\label{Tab:CL12}
\centering
\begin{tabular}{cccll} \toprule
	~&~&&\textbf{Hep3B2}&\textbf{HepG2}\\\toprule
	\textbf{DOX impact rate}& \textbf{H} && faster reaction & faster reaction \\
	\multirow[t]{2}{*}{$\sensRate{D}^-$} & \textbf{C} && faster reaction & \multirow{2}{*}{$\left[\vphantom{\begin{array}{c}
				~ \\[-3pt] ~ \\[-3pt] ~
		\end{array}}\mbox{N/A}\right]$} \\
	& \textbf{HC} && \makecell[tl]{faster reaction, \\[-2pt] synergism} & \\[2pt]
	\midrule
	\vspace*{2pt}
	\textbf{DOX susceptibility}& \textbf{H} && higher susceptibility & higher susceptibility \\
	\multirow[t]{2}{*}{$\DnormCH$} & \textbf{C} && lower susceptibility & \multirow{2}{*}{$\left[\vphantom{\begin{array}{c}
				~ \\[-3pt] ~ \\[-3pt] ~
		\end{array}}\mbox{N/A}\right]$} \\
	& \textbf{HC} && \makecell[tl]{hypoxia prevails, \\[-2pt] synergism} & \\[2pt]
	\midrule
	\vspace*{2pt}
	\textbf{SOR support}& \textbf{H} && \multirow{3}{*}{$\left[\begin{array}{c}
			 ~ \\[-7pt] \substack{\mbox{unclear due to} \\[2pt] \mbox{inconsiderable} \\[2pt] \mbox{support}} \\[-7pt] ~
		\end{array}\right]$} & similar support \\
	$d_S(\initS)$ resp. $d^\star_S(\initS,D)$ & \textbf{C} &&& \multirow{2}{*}{$\left[\vphantom{\begin{array}{c}
				~ \\[-5pt] ~ 
		\end{array}}\mbox{N/A}\right]$} \\
	& \textbf{HC} &&& \\[2pt]
	\midrule
	\vspace*{2pt}
	\textbf{SOR stress response}& \textbf{H} && increased response & decreased response \\
	\multirow[t]{2}{*}{$\adSCH(\initS)$} & \textbf{C} && decreased response & \multirow{2}{*}{$\left[\vphantom{\begin{array}{c}
				~ \\[-3pt] ~ \\[-3pt] ~
		\end{array}}\mbox{N/A}\right]$} \\
	& \textbf{HC} && \makecell[tl]{hypoxia prevails for~${\initS=0.5}$, \\[-2pt] synergism} & \\
	\bottomrule
\end{tabular}	
\end{table}
\noindent Overall, the calibration results consistently show an enhancing effect of hypoxia on the treatment efficacy of both drugs, whereas tissue stiffening increases the chemoresistance (for Hep3B2). The only apparent discrepancy appears for the stress response to SOR under hypoxia. As mentioned in Section~\ref{sec:RES-treat}, this does not necessarily imply a generally opposite reaction to the SOR treatment, which we discuss in more detail in Section~\ref{sec:DIS-res}.

\subsection{Drug metabolization and CYP expression}\label{sec:RES-met}
Recall that one of the main differences between the cell lines Hep3B2 and HepG2 is their observed expression of the drug-metabolizing enzyme CYP3A4~\cite{Ozkan.2021}: Hep3B2 shows none, while HepG2 does. Although the model~\ref{eq:M-chemo-0-inf} used to calibrate the data for HepG2 does not include explicitly CYP expression, we can estimate this via the relationships~\eqref{eq:CYPnoCYPrelation} derived at the end of Subsection~\ref{sec:MOD-App2}.
Moreover, we make considerations based on the measured CYP data.

\mypar{Estimated drug metabolization} For all SOR-related thresholds in~\eqref{eq:CYPnoCYPrelation}, we cannot reconstruct the corresponding drug metabolization~$\MetSCH$ as we do not have estimates for~$\SthrCH$ and~$\Ssupp$. Hence, it is not possible to investigate the role of SOR metabolization with the calibration results. For the DOX-related thresholds, it is reasonable to assume that, for cell line HepG2, we actually estimated~$D^\text{CYP}_{\text{norm},HC}$ and~$D^\text{CYP}_{\text{damp}}$ instead of~$\DnormCH$ resp.~$\Ddamp$\,. Therefore, the estimates of these parameters from Table~\ref{Tab:CL2-parsDOX} could be used to reconstruct~$\MetDCH$\, via \eqref{eq:CYPnoCYPrelation}. However, we have only sufficient information to do so if we take two modeling assumptions as a starting point. Both suppose that cell lines Hep3B2 and HepG2 respond comparably to the DOX treatment. According to~\cite{Qiu.2015}, it is reasonable to assume some degree of similarity between those two cell lines although we have to keep in mind that there are also several differences. Hence, the following investigations only make sense under the premise that these differences are not relevant in our modeling context, which we cannot validate yet with the available data.

First assumption: The environmental factors have the same effect on the stress response to DOX for cell lines Hep3B2 and HepG2. In sole hypoxic conditions~(H1:~${H=1}$,~${C=0}$), this means that we let the value of~${d_H(H=1)}$ from the relation~${\DnormCH=\Dnorm\cdot d_H(H=1)\cdot d_C(C=0)}$ be the same for both cell lines. Second assumption: Except for the influences of drug metabolization and the environment (in particular hypoxia), Hep3B2 and HepG2 have the same DOX susceptibility threshold~$\Dnorm$. Then, combining the calibration results from both cell lines and using \eqref{eq:CYPnoCYPrelation} yields
\begin{alignat}{3}
	\text{HC0:}&\qquad &\overbracket[0.187ex]{\,D^\text{CYP}_{\text{norm},HC}\,\vphantom{\overbrace{\,\Dnorm\cdot d_H(H=0)\,}^{\Dnorm}}}\,&=\,\overbracket[0.187ex]{\,\MetDCH\,\vphantom{\overbrace{\,\Dnorm\cdot d_H(H=0)\,}^{\Dnorm}}}\,\cdot\, \overbracket[0.187ex]{\overbrace{\,\Dnorm\cdot d_H(H=0)\,}^{\DnormCH}}\,\cdot\, \overbracket[0.187ex]{\overbrace{\,d_C(C=0)\,}^{=\,1\vphantom{\DnormCH}}}\,\,,\nonumber\\[-7pt]
	&&&\label{eq:CYP-assump}\\[-7pt]
	\text{H1:}&\qquad& \underbracket[0.187ex]{\,D^\text{CYP}_{\text{norm},HC\,}\vphantom{\underbrace{\,\Dnorm\cdot d_H(H=1)}_{\DnormCH}}\,}_{\substack{\text{estimated}\\\text{with HepG2}}}\,&=\,\underbracket[0.187ex]{\,\MetDCH \vphantom{\underbrace{\,\Dnorm\cdot d_H(H=1)\,}_{\DnormCH}}\,}_{\substack{\text{unknown}\\\text{(HepG2)}}}\,\cdot\,\underbracket[0.187ex]{ \underbrace{\,\Dnorm\cdot d_H(H=1)\,}_{\DnormCH}}_{\substack{\text{estimated}\\\text{with Hep3B2}}}\,\cdot\, \underbracket[0.187ex]{\underbrace{\,d_C(C=0)\,}_{=\,1\vphantom{\DnormCH}}}_{\text{by definition}}\,.\nonumber
\end{alignat}
Solving each of the above equations for~${\ln(\MetDCH)}$ eventually provides environment-specific estimates for the involved DOX metabolization~$\metDCH$ based on the approximating assumption~\eqref{eq:CYP-approx}. Using the fitted values of~$D^\text{CYP}_{\text{norm},HC}$ (Table~\ref{Tab:CL2-parsDOX-1}) and of~$\DnormCH$ (see~\eqref{eq:CL1-DnormHC}), we get
\begin{equation}
\begin{alignedat}{3}
	\text{HC0:}\qquad \metDCH&=\metD(\CYP)\approx 4.904\pm0.930\,,\\
	\text{H1:}\qquad \metDCH&=\metD(\CYP)\approx 5.022\pm0.705\,.
\end{alignedat}\label{eq:MetEstimates}
\end{equation}
In particular, we draw~$10^7$ samples of~$D^\text{CYP}_{\text{norm},HC}$ and~$\DnormCH$ respectively according to a normal distribution (truncated to $\mathbb{R}_+$) which is centered at their fit and using the corresponding standard deviation. Then, the presented 95\%~confidence interval of~$\metDCH$ is obtained by calculating its value for each sample. The estimates~\eqref{eq:MetEstimates} show a considerable DOX metabolization~(${\metDCH\gg0}$) for HepG2 and the metabolic activity appears to be stronger under sole hypoxic conditions (potentially statistically insignificant, see Figure~\ref{A-Fig:CL2-P-CYP}).

We could now proceed with using the estimates of~$\MetDCH$ to reconstruct the threshold~$\Ddamp$ from ${D^\text{CYP}_\text{damp}=\MetDCH\cdot\Ddamp}$\,. However, there are indications that in this context the cell lines Hep3B2 and HepG2 are not sufficiently comparable as we observe a clear supportive effect for HepG2 in contrast to Hep3B2. Intuitively, we would expect the opposite due to potentially lower SOR concentrations in the presence of CYP, which suggests that the support of SOR might occur differently for the respective cell lines.

\begin{figure}[H]
	\centering
	\includegraphics[width=\linewidth]{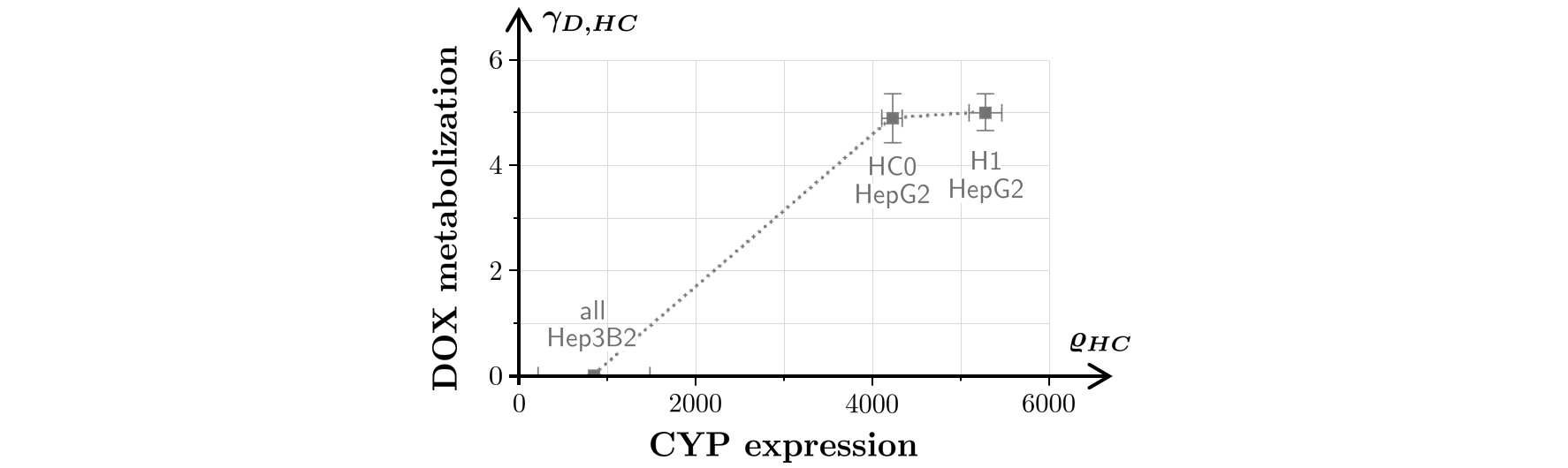}
	\caption[Relation between estimated DOX metabolization~${\metDCH}$ and measured CYP expression]{Relation between the estimated DOX metabolization~${\metDCH}$ and corresponding measured CYP expression~$\CYP$ (dotted line: trendline). The error bars show the 95\% confidence interval of the CYP data (horizontal) and of the estimated values of~$\metDCH$ (vertical).}
	\label{Fig:CL2-CYP}
\end{figure}

\mypar{Measured CYP expression} Since~${\metDCH=\metD(\CYP)}$ is defined as a function of the CYP expression~$\CYP$\,, we can compare those quantities using the corresponding CYP data from~\cite{Ozkan.2021}, given as quantities which are assumed to be directly proportional to the CYP expression~\cite{Promega.2023}. Since the proportionality constant has no qualitative influence on the results, we treat the measurements as direct quantities for~$\CYP$\,. Figure~\ref{Fig:CL2-CYP} visualizes the relation between the derived estimates of~$\metDCH$ (for Hep3B2 we assume~${\metDCH=0}$, i.e. no DOX metabolization) with the measured CYP expression.
\noindent The observed relation suggests a positive and potentially non-linear correlation between~$\metDCH$ and~$\CYP$\,, which fits to~$\metD$ actually being a monotonically increasing function in~$\CYP$\,. This implies that the two assumptions for approach~\eqref{eq:CYP-assump} can make sense from an experimental perspective. Obviously, the validity is quite limited by being based on only three observations. The relationship between $\metDCH$ and~$\CYP$ is further discussed in Section \ref{sec:DIS-res}. 

\section{Discussion}\label{sec:discussion}
We have developed a mathematical model to study the effect of environmental factors, in particular tissue stiffness and oxygen level, on the response of liver tumor cells to chemotherapeutic treatment, the latter consisting of a main drug (DOX) and a supportive drug (SOR). Using Bayesian statistics, we calibrated the model with experimental data on different cell lines. With the mathematical model, we have been able to systematically unveil biological mechanisms expressed by the data, motivating possible causes and synergistic effects. The Bayesian approach has allowed us to quantify the uncertainty and statistical significance of the observations drawn from the calibrations. In the following, we discuss our methodology and findings regarding the mathematical modeling (Section~\ref{sec:DIS-model}) and the calibration results (Section~\ref{sec:DIS-res}).

\subsection{Considerations on the mathematical model}\label{sec:DIS-model}

We calibrated successfully the reduced model~\ref{eq:M-chemo-0} for the cell line Hep3B2 with negligible CYP expression. For the other two cell lines, HepG2 and C3Asub28, expressing the CYP enzyme, two considerations are in order. First, the experimental data show enhanced cell growth  (cirrhotic environment for HepG2 and all environments in C3Asub28), with a percentage viability exceeding the mark of $100\%$. As explained at the beginning of Section \ref{sec:results}, this occurs when combined treatment is applied, suggesting a synergistic effect of the two drugs which is not accounted for in any of our models. These are thus not suited for being calibrated in these scenarios, where indeed preliminary calibrations were not successful. Second, preliminary calibrations of the complete model~\ref{eq:M-chemo} with data for the cell line HepG2 showed a difficulty in the calibration of the drug impact rates and the metabolization rates. A possible reason is the sparse time resolution of the available data: Viability was measured only once, three days after a one- or two-day treatment. Hence, the only time-related information is given by the variation in viability for different treatment durations. It appears that this information is not sufficient to properly estimate the rates. This conjecture is supported by trying to calibrate the Hep3B2 data with the complete model~\ref{eq:M-chemo}, which did not result in vanishing metabolization rates as we would expect knowing that the cells express almost no CYP. Calibrations of model~\ref{eq:M-chemo-0-inf}, an enhanced version of the reduced model~\ref{eq:M-chemo-0}, to the data of HepG2 (non-cirrhotic environments) were instead successful. The features added to the reduced model were necessary to model the asymetric supportive effect of this cell line, meaning a weaker supportive effect for larger dosages of the main drug.

The unsatisfactory fitting of the \textsl{mathematical model} in the scenarios mentioned above could have been compensated by modifying the \textsl{statistical model}, including an additive misfit term due to model discrepancy \cite{Brynjarsdottir.2014}, see Remark at the end of Subsection \ref{sec:PAR-Unc}. Although we then might be able to fit the data, this would not contribute to our goal to gain biological insight via the mathematical model, since it would be hard to interpret the calibration results if the model discrepancy is too large. Also, including the model discrepancy would increase the number of parameters to be estimated from our data, making it harder for the calibrations to obtain informative results. We see the inclusion of model discrepancy in the statistical model as an intermediate tool to possibly improve our mathematical model in the case that more experimental data become available. For instance, correlations between the model discrepancy and some model parameters might hint on which biological phenomena need to be modeled and how.

For the successful calibrations presented in this paper, the marginal posterior distributions of the model parameters are unimodal and rather concentrated with respect to the priors, see Figures \ref{A-Fig:CL1-marg} and \ref{A-Fig:CL2-marg}. This suggests that the model parameters used are \textsl{practically identifiable} \mbox{\cite{chis2011structural,Falco.2023,siekmann2012mcmc}} in the settings considered.

\subsection{Considerations on the calibration results}
\label{sec:DIS-res}
The following paragraphs further discuss the key observations of Sections~{\ref{sec:RES-pred}--\ref{sec:RES-met}}. 

\paragraph*{Model predictions.} For both cell lines Hep3B2 and HepG2 (Figure~\ref{Fig:percV-all}) we observe reasonable fits by the calibrated models. The fact that, especially for small DOX dosages~${\initD\leq 10^{-2}}$, the model tends to overestimate the data in some cases could be a result of the uncertainty modeling. On the one hand, the assumed Beta prime distribution of the noise from~\eqref{eq:noisePerc} is
\begin{equation*}	
	\varepsilon_\%=\frac{~\overbracket[0.187ex]{\,\txtTop{I}{treat}/\txtTop{I}{ctrl}\,}^{\text{measured}}~}{~\underbracket[0.187ex]{\,\Vtreat/\Vctrl\,}_{\text{modeled}}~}\sim\beta'\left(\frac{1}{\sigma^2},\frac{1}{\sigma^2}\right)\mbox{ with }\sigma^2=0.1\,,
\end{equation*}
which is positively skewed with a mode (maximum of the density function) of approximately~$0.82$\,. Hence, a modeled percentage viability of  \enlargethispage{\baselineskip}
\begin{equation*}
	\frac{\Vtreat}{\Vctrl}=\frac{1}{0.82}\cdot\frac{\txtTop{I}{treat}}{\txtTop{I}{ctrl}}\approx1.22\cdot\frac{\txtTop{I}{treat}}{\txtTop{I}{ctrl}}\,,
\end{equation*}
i.e. larger than the corresponding measurement, maximizes the data likelihood function. On the other hand, the multiplicativity of the noise allows larger deviations between solution and data for larger measurements. Therefore, the calibration algorithm might to some extent \qm{prioritize} a closer fit to small data. These observations suggest the potential to improve the statistical model via the noise term, in case the available data is sufficiently informative to calibrate this.

Another reason for the observed overestimation could be the modeling of the DOX stress response~${\sensRate{D}^-\cdot \deact_{D}}$ by a symmetric Hill function~$\deact_{D}$\,. The line plots of the solutions in Figure~\ref{Fig:percV-all} show that this leads to a symmetric viability-dose relationship as well. However, the overestimated measurements (e.g. H1/HC1 for Hep3B2) indicate a less steep slope for smaller DOX dosages, which the model design cannot reproduce. The potential of improving the model by reconsidering the symmetry assumption for the DOX dose-response function~$\deact_{D}$ is supported by fact that, for cell line HepG2, enhancing the model~\ref{eq:M-chemo-0} to~\ref{eq:M-chemo-0-inf} by allowing some asymmetry in the dose-response function enabled the calibrations to be successful.

\paragraph*{Drug-independent cell dynamics.} For both cell lines, we can observe consistent estimates for the stress-independent term~${\prol+\indDeath}$ (Figure~\ref{Fig:RES-prolInd}), suggesting robustness of the mathematical model and validating to fix this term to its estimate for parts of the calibrations. Although the model design did not allow for estimating the individual growth/death rates~$\prol$,~$\indDeath$ and~$\nat$\,, we were able to give upper bounds with the estimated values of~${\prol+\indDeath}$\,. Furthermore, for Hep3B2, the availability of a significantly non-zero estimate for the initial stress level~$\initStCH$ allowed us to obtain a lower bound for the term~${\prol-\nat}$\,.

For both cell lines, we cannot observe a significant increase of the stress level of the cells when only hypoxia or only cirrhosis are present. We could observe instead a significant increase of the stress when both hypoxia and cirrhosis are present, at least for the cell line Hep3B2 which could be calibrated in this scenario. This suggests a synergistic effect of those two environmental factors, since, for Hep3B2, the estimates of~$\initStCH$ for H1 and C1 clearly do not add up to the one of HC1. Mathematically, this means our modeling assumption in~\eqref{eq:initStress}, i.e.
\begin{equation*}
	 \initSt(\initH,\initC)=\initSt(\initH\,,\,0)+\initSt(0\,,\,\initC)=\frac{\sensRate{H}\cdot\deactFct{H}{\initH }}{\alpha_H+\alpha_C}+\frac{\sensRate{C}\cdot\deactFct{C}{\initC }}{\alpha_H+\alpha_C}\,,
\end{equation*}
has to be reconsidered, and we should allow both influence functions~$\deact_H$ and~$\deact_C$ to respectively depend on both variables~$H$ and~$C$. Since we do not have any suitable data to estimate these functions or the impact rates~$\sensRate{H}$ and~$\sensRate{C}$\,, we cannot make any further conclusions about the influence of hypoxia and high ECM stiffness on the tumor growth.

\paragraph*{Cytotoxic efficacy of the supportive drug SOR.} We remind that, due to the limited number of experimentally investigated SOR dosages, we could only estimate point values of~${\adSCH}$ (at~${\initS\in\{0.5,1.0\}}$) rather than the parameters for the sigmoid stress response function. Nevertheless, for both cell lines, we can still get a feeling for the rough shape of the sigmoid from the calibration results. Let us turn our attention to Figure~\ref{Fig:CL1-aS}, starting with the estimates for Hep3B2. First, the fact that a standard dosage of SOR~(${\initS =0.5}$) barely shows any stress response compared to a high dosage~(${\initS =1.0}$) in all environments indicates a SOR susceptibility threshold with
\begin{equation*}
\Sthr(H,C)>0.5\qquad\forall H,C\in\{0,1\}\,. 
\end{equation*}
Second, due to~${0\leq\deact_{S,HC}(S)<1~\forall S\geq0}$\,, the response for a high dosage~${\adSCH(\initS=1.0)}$ yields a lower bound for the respective SOR impact rate
\begin{equation*}
	\sensRate{S}^-\geq\adSCH(\initS=1.0)\,,
\end{equation*}
for given~$H$ and~$C$ and up to variations in the magnitude of the corresponding marginal variance. Similarly to cell line Hep3B2, we can follow the parameter bounds \enlargethispage{\baselineskip} \vspace*{-2pt}
\begin{equation*} 
	\Sthr(H,0)>0.5~\mbox{for }H\in\{0,1\}\qquad\mbox{and}\qquad
	\sensRate{S}^-\geq\adSCH(\initS=1.0) \vspace*{-2pt}
\end{equation*}
for given~$H$ and~${C=0}$ and up to variations in the magnitude of the corresponding marginal variance.

Looking now at the individual estimates for Hep3B2 and a high SOR dosage, the fact that the relative increase/decrease of the response to SOR is different for sole cirrhosis/hypoxia compared to a combination of both hints on a synergistic effect of those environmental factors. For HepG2, this effect could not be investigated as we could only calibrate the settings HC0 and H1. In general, the estimates for a high SOR dosage indicate a weaker cytotoxic efficacy for cell line HepG2 compared to Hep3B2.
Interestingly, for cell line HepG2, we observe a decreasing effect of hypoxia on the stress response, which is just the opposite influence as we observe for cell line Hep3B2 (see Table~\ref{Tab:CL12}). However, one should keep in mind that there is a difference between the stress response~${\adSCH(S)}$ and the cells' susceptibility threshold~$\SthrCH$\,. While the first describes the treatment response for a particular dosage~${S=\initS}$, the latter gives a more general measure independent from~$S$. Therefore, the apparently contrary influence of hypoxia observed for~${\adSCH(\initS)}$ does not necessarily imply that the cells react oppositely to the SOR treatment in general. For comparison, we check the unsupported stress response~${\adDCH(\initD,0)}$ to DOX in Table~\ref{Tab:CL12-stressSOR}.
\begin{table}[H]
	\caption[Hep3B2/HepG2: Comparison ({HC0\,$\leadsto$\,H1}) of the MAP estimates of~${\adDCH(\initD,0)}$]{Comparison ({HC0\,$\leadsto$\,H1}) of the MAP estimates of~${\adDCH(\initD,0)}$ given particular~$\initD$. }\label{Tab:CL12-stressSOR}
	\centering
	\begin{tabular}{rcccc} \toprule
 $\boldsymbol{\initD}$ && \textbf{Hep3B2}  && \textbf{HepG2}  \\\midrule
$\boldsymbol{ 0.5 }$ && $ \hphantom{0}0.368 ~\nearrow~ \hphantom{0}0.889 $ && $ 0.005 ~\searrow~ 0.001 $ \\
$\boldsymbol{ 1.0 }$ && $ \hphantom{0}1.223 ~\nearrow~ \hphantom{0}2.804 $ && $ 0.010 ~\searrow~ 0.004 $ \\
$\boldsymbol{ 5.0 }$ && $ \hphantom{0}7.971 ~\nearrow~ 12.761 $ && $ 0.060 ~\searrow~ 0.047 $ \\
$\boldsymbol{ 10.0 }$ && $ 11.607 ~\nearrow~ 15.463 $ && $ 0.130 ~\searrow~ 0.127 $ \\
\bottomrule
\end{tabular}	
\end{table}
\noindent In fact, analogously to the stress response to SOR, we also observe an increasing resp. decreasing effect of hypoxia on the response to DOX for Hep3B2 resp. HepG2. Nevertheless, the impact rate~$\sensRate{D}^-$ and susceptibility threshold~$\DnormCH$ in Table~\ref{Tab:CL12} both indicate consistency regarding the DOX treatment response for both cell lines. Hence, the alleged discrepancy does not imply a different influence of hypoxia on the SOR dose-response.

\paragraph*{Cytotoxic efficacy of the main drug DOX.} For the cell line Hep3B2, we compared the fitted estimates for the DOX susceptibility threshold~$\DnormCH$ with their corresponding MAPs (marginalized investigation in Subsection~\ref{APP-Res2.1-MARG}). We obtained consistent qualitative results with both estimates. The only exception: The relative increase/decrease of~$\DnormCH$ differs when both hypoxia and cirrhosis are present in comparison to sole hypoxia resp. cirrhosis, indicating a synergistic effect, is only observable for the fitted vales and not by the marginalized investigation. However, this does not necessarily imply contradicting results as the marginals of~$\DnormCH$ neglect the parameter correlations. Since incorporation of the correlations gives a more holistic approach, it is reasonable to assume a higher validity for the observations obtained with fitting.

Concerning the DOX impact, for both cell lines the calibration results suggest a faster reaction to the DOX treatment if hypoxia and/or cirrhosis is involved. For Hep3B2, comparing the marginal distributions does not imply statistical significance of this observation, see Figure~\ref{A-Fig:CL1-P-aD1000}. However, recall that this does not necessarily indicate that the impact rate is actually the same in all environments, i.e. that there is no environmental effect. The distributions for~$\sensRate{D}^-$ show relatively high variances, which might play a role for the unclear statistical significance. A possible reason for the large variances is the sparse time resolution of the data. Recalling that viability was measured once at~$\tend$ (three days after a one- or two-day treatment), the only time resolution of the data is given by the variation in viability for different treatment durations. For large DOX dosages, the percentage viability is generally close to zero, i.e. the correspondingly small variation might be hardly distinguishable from measurement inaccuracy. Furthermore, if the cells' reaction to the DOX treatment is considerably fast (and for a sufficiently large induced death rate~$\indDeath$), the cell population \NEW{might go extinct} at some time point before viability is measured. Then, the single measurement at~$\tend$ has no information about when in particular the cells have ceased. 

Regarding the supportive effect of SOR, the calibration results suggest a significant effect only for HepG2, where significant support is observed for both for standard and high SOR dosages.

\paragraph*{Drug metabolization and CYP expression} For the CYP-expressing cell line HepG2, we have compared, in Figure~\ref{Fig:CL2-CYP}, the relationship between the DOX metabolization and CYP expression as given by the calibrations and the experimental values. Although the small number of points does not allow to meaningfully fit and compare different possibilities for the monotonic function~${\metD(\CYP)}$, we can conjecture its rough shape of the function (dotted trendline in Figure~\ref{Fig:CL2-CYP}). Due to the monotonicity and the considerable distance of the left-most marker (Hep3B2) from the origin, it appears that~$\metD$ stays close to zero for sufficiently small values, i.e. approx.~${\CYP\in[0,1000]}$. Furthermore, the markers for HepG2 indicate the function's slope to be relatively flat in the range of~${\CYP\in[4000,5000]}$\,, especially in comparison to the increase between the Hep3B2 and HepG2 markers. Combining these observations to a smooth, monotonically increasing function suggests e.g. a Hill-type function
\begin{equation*}
	\metD(\CYP)=\frac{\tilde{\met}_D\cdot \CYP^m}{\txtSub{\varrho}{thr}^m+\CYP^m}\,,
\end{equation*}
with a threshold~${\txtSub{\varrho}{thr}\in[1000,4000]}$, a horizontal asymptote~${\tilde{\met}_D\gtrsim 5}$ and a Hill coefficient~${m\geq 2}$ for sigmoidal shape (supposing the non-zero CYP data for Hep3B2 is not an effect of sole measurement inaccuracy). On the one hand, this would mean that there needs to be a sufficient amount of CYP present to achieve considerable DOX metabolization. On the other hand, the metabolic activity appears to not increase unlimitedly the higher the CYP expression.

In general, this shape of the function~$\metD(\CYP)$ seems reasonable considering the experimentally observed treatment effect for Hep3B2 and HepG2~(see Figure 4 in \cite{Ozkan.2021}). In particular, the experiments show a considerable shift of the DOX effect if we compare Hep3B2 and HepG2, which fits to~$\metD$ being significantly larger for HepG2 (HC0, H1) compared to Hep3B2. However, the treatment effect for HepG2 in different environments (i.e. comparing HC0/H1/C1/HC1 with each other) appears to be less drastic, although the cells exhibit nearly double the amount under cirrhotic conditions (${\CYP\approx 9000}$ resp.~$10600$ for C1/HC1). This supports the indication of an upper bound for~$\metD(\CYP)$ and that it is almost reached by the CYP expression for HC0 and H1.

It might be interesting to note that calibration attempts with the complete model~\ref{eq:M-chemo} resulted in fairly concentrated marginal distributions for~$\metDCH$ (in contrast to other parameters) and the corresponding estimates are actually in a magnitude of~${\metDCH\approx 5}$ (see Figure~\ref{A-Fig:CL2-metRate}). Hence, there could indeed be potential in calibrating~\ref{eq:M-chemo} to gain deeper understanding of the drug metabolization.

\subsection{Outlook}

The calibration results show good capabilities of our modeling and calibration approaches. As already mentioned in the previous section, the future availability of more experimental data would allow to calibrate the full model~\ref{eq:M-chemo} for the CYP-expressing cell lines, leading to more insights into the biological mechanisms. It would also be interesting to incorporate growth-enhancing influences of the environmental factors, as observed experimentally for the measurements of the cell line C3Asub28 and half of the ones for HepG2, and calibrate the resulting model if the experimental data are informative enough. Finally, we note that the modeling approach based on the environmental stress level can easily accommodate spatially-heterogeneous settings, by replacing ordinary with partial differential equations.

\section*{Acknowledgements}
This paper was supported by the Deutsche Forschungsgemeinschaft (DFG) through the TUM International Graduate School of Science and Engineering (IGSSE), GSC 81.

\printbibliography[heading=bibintoc] 

%% file: only_supplementary.tex
\section{Model variables and parameters}
The following tables summarize the variables (Table~\ref{Tab:model2Vars}) and parameters (Table~\ref{Tab:model2Pars}) of model~\ref{eq:M-chemo}.
\begin{table}[H]
\centering
		\caption[Variables of model~\ref{eq:M-chemo}]{Variables of model~\ref{eq:M-chemo}.}
		\resizebox{\textwidth}{!}{%
		\begin{tabular}{rclc} \toprule
			& &\textbf{Meaning}& \textbf{Scaling} \\\midrule
			\textbf{General} &$V=V(t)$ & Density of viable tumor cells& -- \\
			&$\stress=\stress(t)$ & Environmental stress level & --\\[8pt]
			\textbf{Environmental} &$D=D(t)$ & Doxorubicin (DOX) concentration & \SI{1}{\micro\molar} \\
			&$S=S(t)$ & Sorafenib (SOR) concentration & \SI{22}{\micro\molar}\\[4pt]
			&$H=\initH ~~\,$ & {Hypoxia level} (constant), representing & -- \\
			& & oxygen supply:~$21\%\,\chem{O_2}$ ($H\equiv0$) or~$1\%\,\chem{O_2}$ ($H\equiv 1$)& \\[2pt]
			&$C=\initC ~~\,$ & {Cirrhosis level} (constant), representing & -- \\
			& & ECM stiffness: normal ($C\equiv0$) or cirrhotic ($C\equiv1$)& \\
			\bottomrule
		\end{tabular}}
		\label{Tab:model2Vars}
\end{table}
\begin{table}[H]
\centering
		\caption[Parameters of model~\ref{eq:M-chemo}]{Parameters of model~\ref{eq:M-chemo}.}
		\begin{tabular}{clc} \toprule
			&\textbf{Meaning}& \textbf{Scaling} \\\midrule
			& Time point of & \\
			$\initt=0$ & ~~the end of adaption phase / beginning of treatment phase & \SI{}{\day}\\
			$\ttreat$ & ~~the end of treatment phase / beginning of growth phase & \SI{}{\day}\\
			$\tend$ & ~~the end of growth phase & \SI{}{\day}\\[8pt]
			$\initV$ & Number of viable tumor cells at~$\initt$& -- \\
			$\initStCH$ & Environmental stress level at~$\initt$& --\\
			$\initD$ & Doxorubicin (DOX) concentration at~$\initt$& \SI{1}{\micro\molar} \\
			$\initS$ & Sorafenib (SOR) concentration at~$\initt$ & \SI{22}{\micro\molar}\\[8pt]
			$\prol$ & Maximal possible growth rate (under optimal growth conditions) & \SI{}{1/\day}\\
			$\indDeath$ & Maximal possible induced death rate (under stressful conditions) & \SI{}{1/\day} \\
			$\nat$ & Natural death rate & \SI{}{1/\day}\\[4pt]
			$\metDCH$ & Drug metabolization rate of DOX (can be influenced by~$H$ and~$C$)& \SI{}{1/\day}\\
			$\metSCH$ & Drug metabolization rate of SOR (can be influenced by~$H$ and~$C$)& \SI{}{1/\day}
			\\[4pt]
			& Impact rate of cells reacting to & \\
			$\sensRate{H}$, 	$\sensRate{C}$ & ~~changes in {hypoxia level} resp. {cirrhosis level} & \SI{}{1/\day}\\
			$\sensRate{D}^-$,~$\sensRate{S}^-$ & ~~stressful changes in DOX resp. SOR concentration & \SI{}{1/\day}\\
			[4pt]
			& Sensitivity threshold of cells reacting to stressful changes & \\
			$\txtSub{D}{norm}$& ~~in DOX concentration, if~$S,H,C=0$ & \SI{1}{\micro\molar} \\
			$\DthrCH(S)$& ~~in DOX concentration (can be influenced by~$S,H$ and~$C$) & \SI{1}{\micro\molar} \\
			$\SthrCH$& ~~in SOR concentration (can be influenced by~$H$ and~$C$) & \SI{22}{\micro\molar} \\[4pt]
			$\Ssupp$ & SOR concentration threshold to support DOX treatment & \SI{22}{\micro\molar} \\[4pt]
			& Hill-coefficient regarding the & \\
			$m_1$,~$m_2$ & ~~influence function~$\deact_{D,HC}$ resp.~$\deact_{S,HC}$ & -- \\
			$m_3$ & ~~supportive effect of SOR, i.e. function~$d_S(S)$ & -- \\[4pt]
			$\txtSub{a}{max}$ & Scaling parameter for bounding the supportive effect & -- \\
			\bottomrule
		\end{tabular}
		\label{Tab:model2Pars}
\end{table}

\section{Model calibration and post processing}

We provide additional information regarding the model calibration procedure and the post processing of the results. In particular, we summarize the prior distributions for the model calibrations and we explain how we obtain a kernel density estimation (KDE) for a marginal posterior distribution on the bounded support of its respective prior.
\subsection{Prior distributions for model calibrations}
\label{SUP-sec:CAL-prior}
The following Table~\ref{Tab:CalParsChemo} summarizes the prior distributions to perform the calibrations for models~\ref{eq:M-chemo-0} and~\ref{eq:M-chemo}. Afterwards, we provide  details regarding the construction of the priors for~\ref{eq:M-chemo} used for pre-calibrations.
\begin{sidewaystable}[hp!]
		\begin{table}[H]
		\caption[\textit{A priori} information to calibrate model~\ref{eq:M-chemo-0} resp.~\ref{eq:M-chemo}]{Overview of the \textit{a priori} information used to calibrate model~\ref{eq:M-chemo-0} resp.~\ref{eq:M-chemo} for different combinations of~${\initH\in\{0,1\}}$ and~${\initC\in\{0,1\}}$\,. The following notations are used:\\ \qm{N/A} (not calibrated), \qm{$a,b$} (uniform prior on~$[a,b]$), \qm{$a,h,b$} (triangular prior on~$[a,b]$ with mode~$h$),\\ \qm{$= a$} (not calibrated, fixed to~$a$), \qm{$=\mbox{MAP}$} (not calibrated, fixed to MAP estimate from calibration for~$\initH ,\initC =0$, i.e. HC0).} \label{Tab:CalParsChemo}
		\begin{subtable}{\textwidth}
			\captionsetup{singlelinecheck = false, justification=justified}
			\caption{Priors for calibration of model~\ref{eq:M-chemo-0} using Hep3B2 data.} \label{Tab:CalParsChemo0}
			\begin{tabular}{c!{\vrule width 1pt}c|c}
				\toprule
				& \textbf{cell dynamics} & \textbf{ESL dynamics} \\
				\begin{tabular}{c}
					\textbf{environment} \\
					\midrule
					\textbf{HC0}\\	
					\textbf{H1/C1/HC1}
				\end{tabular} &
				\begin{tabular}{cc}
					$(\prol-\nat)$ &$(\prol+\indDeath)$ \\
					\midrule
					\multirow{2}{*}{N/A} &$0,3$ \\
					&$=\mbox{MAP}$ 
				\end{tabular} & \begin{tabular}{cccccccc}
					$\initStCH$ &$\sensRate{D}^-$ &$\widehat{D}_{\text{norm},HC}$ &$\widehat{m}_1$&$\sensRate{S}^-\deactFct{S}{1}$ &$c_\delta$ &$d_S(0.5)$ &$c_d$ \\
					\midrule
					$=0$ & \multirow{2}{*}{$0,20$} &\multirow{2}{*}{$-4,0,4$} & \multirow{2}{*}{$0,1.7$} &\multirow{2}{*}{$0,2$} &\multirow{2}{*}{$0,0,1$}&\multirow{2}{*}{$0,1$} &\multirow{2}{*}{$0,1$} \\
					$0,0,1$ && & & & & &
				\end{tabular} \\
				\bottomrule
			\end{tabular}
		\end{subtable}
		\newline
		\vspace{15pt}
		\newline
		\begin{subtable}{\textwidth}
			\captionsetup{singlelinecheck = false, justification=justified}
			\caption{Priors for calibration of model~\ref{eq:M-chemo} using HepG2 or C3Asub28 data.} \label{Tab:CalParsHepG2}
			\begin{tabular}{c!{\vrule width 1pt}c|c|c}
				\toprule
				& \textbf{cell dynamics} & \textbf{ESL dynamics} & \textbf{drug dynamics} \\
				\begin{tabular}{c}
					\textbf{environment} \\
					\midrule
					\textbf{HC0}\\	
					\textbf{H1/C1/HC1}
				\end{tabular} &
				\begin{tabular}{cc}
					$(\prol-\nat)$ &$(\prol+\indDeath)$ \\
					\midrule
					\multirow{2}{*}{N/A} &$0,3$ \\
					&$=\mbox{MAP}$ 
				\end{tabular} & \begin{tabular}{ccccccccc}
					$\initStCH$ &$\sensRate{D}^-$ &$\widehat{D}_{\text{norm},HC}$ &${m}_1$ &$\sensRate{S}^-$ &$\SthrCH$ &$\widehat{m}_2$ & $\Ssupp$ &$\widehat{m}_3$ \\
					\midrule
					$=0$ & \multirow{2}{*}{$0,20$} &\multirow{2}{*}{$-4,0,4$} & \multirow{2}{*}{$0,6$} &\multirow{2}{*}{$0,10$} &\multirow{2}{*}{$0,3$} & \multirow{2}{*}{$0,1.7$} &\multirow{2}{*}{$0,1$} & \multirow{2}{*}{$0,1.7$}\\
					$0,0,1$ && & & & & & &
				\end{tabular} & \begin{tabular}{cc}
					~~\,$\metD$~\,~ &~\,~$\metS$\,~~ \\
					\midrule
					\multirow{2}{*}{$0,20$} & \multirow{2}{*}{$0,20$} \\
					& \\
				\end{tabular}\\ 
				\bottomrule
			\end{tabular}
		\end{subtable}
	\end{table}
\end{sidewaystable}
\clearpage
\noindent For the complete model
\leqnomode
\begin{align*}\mbox{\ref{eq:M-chemo}}:&\left\{\begin{aligned}
		\dV &= \Big(\prol-\nat-\big(\prol+\indDeath\big)\stress\Big)\cdot V\,, \\[4pt]
		\dStress &= 
		\left(\frac{\sensRate{D}^- D^{m_1}}{\left(\DnormCH \cdot\left(1-\dfrac{\txtSub{a}{max}S^{m_3}}{\Ssupp^{m_3}+S^{m_3}}\right)\right)^{m_1}+D^{m_1}}+\frac{\sensRate{S}^- S^{m_2}}{\SthrCH^{m_2}+S^{m_2}}\right)(1-\stress)\,,\\[4pt]
		D(t)&= \initD \,\exp\big(-\metDCH\cdot t\,\big)\cdot\mathbbm{1}_T(t)\,,\\[2pt]
		S(t)&= \initS \,\,\exp\big(-\metSCH\cdot t\,\big)\cdot\mathbbm{1}_T(t)\,,\\
		&\mbox{with }V(0)=\initV ~~\mbox{and}~~\stress(0)=\initStCH \leq \frac{\prol-\nat}{\prol+\indDeath}\in(0,1)\,,
	\end{aligned} \right. \hspace*{-.5cm}&
\end{align*}
\reqnomode
we take the prior distributions of the previous model~\ref{eq:M-chemo-0} as an orientation (see Section~\ref{sec:CAL-Prior-App2}). In particular, while it is still not possible to estimate~${(\prol-\nat)}$\,, we adopt the priors
\begin{equation*}
	\initStCH \sim\Triang(0,0,1)\qquad\text{and}\qquad(\prol+\indDeath)\sim\unif(0,3)\,.
\end{equation*}
Again,~${\initStCH}$ resp.~${(\prol+\indDeath)}$ are only calibrated under hypoxia and/or cirrhosis~(H1,C1,HC1) resp. under normal environmental conditions~(HC0).

Furthermore, as~${S=S(t)}$ is now explicitly time-dependent, we cannot employ a discrete approach for estimating the SOR-related dynamics anymore. Hence, for the stress reaction parameters we set
\begin{alignat*}{5}
	\DnormCHlog&\sim\Triang(-4,0,4)\,,&\qquad	{m}_1&\sim\unif(0,6)\,,&\qquad\sensRate{D}^-&\sim\unif(0,20)\,,\\
	\SthrCH&\sim\unif(0,3)\,,&\qquad	\widehat{m}_2&\sim\unif(0,1.7)\,,&\qquad\sensRate{S}^-&\sim\unif(0,10)\,,\\
	\Ssupp&\sim\unif(0,1)\,,&\qquad	\widehat{m}_3&\sim\unif(0,1.7)\,,&&
\end{alignat*}
where~${\DnormCHlog=\log_{10}(\DnormCH)}$ and~${\widehat{m}_i=\log_{10}(m_i)}$\,,~${i\in\{2,3\}}$. For the DOX-related parameters~$\DnormCHlog$, $\sensRate{D}^-$ and~$m_1$, we adopt the priors from~\ref{eq:M-chemo-0}, except that we choose a smaller prior support for the Hill coefficient~$m_1$. This was found to be sufficient, especially after seeing the calibration results from model~\ref{eq:M-chemo-0} using the Hep3B2 data. For the SOR impact rate~$\sensRate{S}^-$, we enlarged the prior support of~${\adSCH(1)}$ from model~\ref{eq:M-chemo-0}, since by definition it holds~${\adSCH(1)\leq\sensRate{S}^-}$. The supports of the thresholds~$\SthrCH$ and~$\Ssupp$ are based on experimental observations from~\cite{Ozkan.2021}. Additionally to these parameters, model~\ref{eq:M-chemo} considers the drug metabolization rates, for which we set
\begin{equation*}
	\metD\sim\unif(0,20)\qquad\mbox{and}\qquad\metS\sim\unif(0,20)\,.
\end{equation*}
%
\subsection{Adaptive selection of the random walk step size}\label{SUP-sec:CAL-MH} 
Table~\ref{Tab:MH-h} demonstrates the adaptive MCMC scheme~\eqref{eq:MH-h} from Subsection~\ref{ssec:adaptive} for exemplary MCMC updates with~${H^*=4}$ and~${H=10}$\,.
\begin{table}[H]
	\centering
	\caption[Exemplary application of the adaptive MCMC scheme~\eqref{eq:MH-h}]{Application of the adaptive scheme~\eqref{eq:MH-h} for the~$k$-th SMC step and~$h$-th MCMC update, given exemplary acceptance ratios~$A^*_h/P$. It splits the~$H=10$ MCMC steps into compartments~$\mathfrak{H}_l$ of maximal size~$H^*=4$ and checks the respective partial acceptance ratios~$a^*_{k,l}$ to readjust the scaling factor~$\rho_{k,h}$ in between MCMC steps.} \label{Tab:MH-h}
	\begin{tabular}{c|cccccccccccc|ccc} 
		\toprule
		&\multicolumn{2}{l}{$\boldsymbol{k=1}$}&&&&&&&&&&&&\multicolumn{2}{l}{$\boldsymbol{k=2}$} \\
		\midrule
		$\boldsymbol{h}$ &$\boldsymbol{1}$ &$\boldsymbol{2}$&$\boldsymbol{3}$&$\boldsymbol{4}$&$\boldsymbol{5}$&$\boldsymbol{6}$&$\boldsymbol{7}$&$\boldsymbol{8}$&\multicolumn{3}{c}{~~$\boldsymbol{9}~~~~~~~\boldsymbol{10}$}&&&$\boldsymbol{1}$&$\cdots$\\[-8pt]
		&\multicolumn{4}{c}{$\underbrace{\hspace*{3.7cm}}_{\boldsymbol{\mathfrak{H}_{l=1}}}$}&\multicolumn{4}{c}{$\underbrace{\hspace*{3.7cm}}_{\boldsymbol{\mathfrak{H}_{l=2}}}$}&\multicolumn{3}{c}{\hspace*{6pt}$\underbrace{\hspace*{2cm}}_{\boldsymbol{\mathfrak{H}_{l=3}}}$} &&&& \\
		\midrule
		$\boldsymbol{A^*_h/P}$ & 0.10&0.12&0.09&0.10&0.35&0.36&0.33&0.30&\multicolumn{3}{c}{~\,0.22~~~~0.21}&&&\multicolumn{2}{c}{$\cdots\cdots$}\\[-6pt]
		&\multicolumn{4}{c}{$\underbrace{\hspace*{3.7cm}}_{a^*_{k,l=1}~<~0.15\hspace*{9pt}}$}&\multicolumn{4}{c}{$\underbrace{\hspace*{3.7cm}}_{a^*_{k,l=2}~>~0.30\hspace*{9pt}}$}&\multicolumn{3}{c}{$\underbrace{\hspace*{2cm}}_{0.15\,<\,a^*_{k,l=3}\,<\,0.30}$\hspace*{-0.2cm}} &&&& \\
		\midrule
		$\boldsymbol{\rho_{k,h}}$ & 1.0&1.0&1.0&1.0&0.5&0.5&0.5&0.5&\multicolumn{3}{c}{~\,1.0~~~~~1.0}&&&1.0&$\cdots$\\
		
		&&&&\multicolumn{2}{c}{$\underarrow{1.2cm}{\cdot 1/2}$}&&&\multicolumn{2}{c}{\hspace*{8pt}$\underarrow{1.5cm}{\cdot \,2}$}&\multicolumn{5}{c}{\hspace*{10pt}$\underarrow{2.1cm}{\text{keep}}$}& \\
		\bottomrule
	\end{tabular}
\end{table}
\noindent Note that Table~\ref{Tab:MH-h} is just a demonstration example. In practice, the acceptance ratios in the last compartment~$\mathfrak{H}_3$ will not lie in the range~${[0.15,0.30]}$ after rescaling~${\rho_{1,8}\leadsto\rho_{1,9}}$\,. Instead, we expect them to be similar to the ones of~$\mathfrak{H}_1$\,, since the scaling factor has the same value. In such situations, where consecutive partial acceptance ratios~${a^*_{k,l}\leadsto a^*_{k,l+1}}$ switch from being too small (${<0.15}$) to being too large (${>0.30}$), or vice versa, the scaling factor will jump back and forth between two values due to alternating division and multiplication with the factor~$2$ (e.g.~Table~\ref{Tab:MH-h}:~${1.0{\leadsto}0.5{\leadsto}1.0}$ for~${\mathfrak{H}_1\leadsto\mathfrak{H}_2\leadsto\mathfrak{H}_3}$). This indicates that by picking a scaling factor between the two alternating values, we could achieve an acceptance ratio in the desired range~$[0.15,0.30]$\,. We make use of this observation by enhancing~\eqref{eq:MH-h} to scheme~\eqref{eq:MH-h-m} in Subsection~\ref{ssec:adaptive}. 
\subsection{Post-processing of calibration results}\label{ssec:postproc}

Once a model calibration has finished, we can mathematically post-process the resulting approximation of the (relatively high-dimensional) posterior distribution to get insight into the modeling and the biological interpretation. Here, we describe the post-processing techniques that we have used to obtain the results in Section~\ref{sec:results}. For the following, the SMC algorithm was performed with a sample size~$P$ to calibrate~$d$ parameters over the course of~$K$~steps~(${P,d,K\in\mathbb{N}}$). 
\subsubsection{Estimated parameters values}
\label{sub:RES-ana-est}
Several quantities characterize the estimates of the parameter values resulting from the posterior approximation. Recalling the definition of the particle approximation of the posterior~$\pi_K=\pi^\mathcal{I}$\,, the marginal distribution~${\pi^\mathcal{I}_j}$ of the $j$-th parameter~(${j=1\Ldots d}$) is obtained by 
\begin{equation*}
	\pi_K(\theta)\approx \sum_{p=1}^{P} W_p\cdot\delta_{\theta_p}(\theta)\quad\Rightarrow\quad\pi^\mathcal{I}_j\big((\theta)_j\big)\approx \sum_{p=1}^{P} W_p\cdot\delta_{(\theta_p)_j}\big((\theta)_j\big)\,,
\end{equation*}
where~${W_p=W^K_p}$ is the posterior weight of particle~$\theta_p$ and~$(\theta_p)_j$ resp.~$(\theta)_j$ is the $j$-th component of a particle~$\theta_p$ resp. of an element~$\theta$ of the parameter space~$\Theta$\,. Due to the discrete nature of the particle approximation, the statistics of the marginal distributions tend to be more robust than the ones of the global distribution. Hence, it is reasonable to access the parameter estimates by the marginals.

We take a look at the most interesting moments of the distributions. The marginal mean resp. variance of the~$j$-th parameter~(${j=1\Ldots d}$) can be calculated directly from the particle approximation:
\begin{equation*}
	\mean\big((\theta)_j\big)=\sum_{p=1}^P (\theta_p)_j\cdot W_p\qquad\mbox{and}\qquad\Var\big((\theta)_j\big)=\sum_{p=1}^P \Big((\theta_p)_j-\mean\big((\theta)_j\big)\Big)^2\cdot W_p\,. 
\end{equation*}
These quantities can give first insight into the parameter estimates and their uncertainty. If the corresponding marginal distribution is unimodal and symmetric (not skewed), the marginal mean is a good approximate for the mode of the distribution, i.e. the maximum of the probability density function. Otherwise, the mode of the posterior has to be accessed differently. It was found that some kind of smoothing of the particle approximation is necessary to get a robust estimate of the mode. We obtain the corresponding probability density function by a \emph{kernel density estimation}~(KDE) on~${\{(\theta_p)_j\}_{p=1}^P}$ with Gaussian kernels and Scott's bandwidth selection method~\cite{Scott.2015}, executed with the Python function \texttt{scipy.stats.gaussian\_kde} with~$\{W_p\}_{p=1}^P$ as the \texttt{weights} argument. Evaluating this function on the marginalized particle population and selecting the~$\arg\max$ gives an estimate of the marginal mode. In the context of Bayesian statistics, this quantity is also called marginal \emph{maximum a posteriori probability (MAP) estimate}. \NEW{Note that we adapted the KDE, if needed, to ensure that the support of the obtained density function lies within the bounds of the prior support -- we will refer to this adjusted routine as \emph{truncated KDE} or \emph{trKDE} for short (details can be found in the the supplemented Section~\ref{SUP-sec:CAL-KDE}).} By combining the marginal means resp. MAPs of all calibrated parameters, we can calculate a corresponding model solution and investigate the underlying biological processes.
\subsubsection{Variation of the estimates and statistical significance}
\label{sub:RES-ana-var}
The SMC algorithm is performed several times for each application to investigate dispersion of the results. For biological and numerical investigation we want to compare different estimates of the same quantity (e.g. a parameter) and judge whether an observed difference is systematic or a result of sampling noise (biological, numerical or experimental variation). We can assess the statistical significance of such observations by the so-called~$p$-value~${\pval }$\,. We consider a value smaller than 0.05 as significant and classify the significance level with the widely used \qm{star notation}:
~$\Ast$/$\AAst$/$\AAAst$ denoting~{${\pval <0.05~/~0.01~/~0.001}$}, respectively. It is important to note that~${\pval \geq 0.05}$ does not lead to the conclusion that there is actually no difference between two estimates. It merely states that if there is a difference, it cannot be distinctively distinguished from the sample noise.

For our application, we did~${q=4}$ repetitions of the SMC algorithm for each cell line and environmental setting. To have more statistically stable results, we collect all the weighted particles from the individual runs in one set of samples. As their weights are normalized on the run-specific particle sets, we have to make the collected particles comparable by updating their weights appropriately. We do this by removing the duplicates and recalculating the posterior densities of the remaining particles with equation~\eqref{eq:posterior}:~${\pi^{\mathcal{I}}(\theta) = L(\mathcal{I}\,|\,\theta)\,\pi_0(\theta)}$. After normalizing these calculated densities and associating them as weights to the respective particles, we get a new particle approximation of the posterior with a larger sample size. The model parameters are then investigated based on the estimates of this posterior. 

We can do a cross-validation to check the robustness of the results obtained by collecting the particles of all SMC runs. Numerical variation is assessed by applying the same reweighting approach as described above on a set of particles obtained by collecting all except one run. This yields four different \qm{extended} particle representations of the posterior. Suppose we want to investigate the numerical variation of a quantity~$x$ (e.g. the marginal mean of a parameter), each of these posteriors gives an estimate~$x_i$~(${i=1\Ldots q}$ and~${q=4}$) and we can state the \emph{cross-validated 95\% confidence interval}:~${\bar x\pm1.96 \sigma_xq^{-1/2}}$ with~${\bar x}$ being the average over~${x_1\Ldots x_4}$ and~$\sigma_x$ the corresponding standard deviation. This notation will be used in Section~\ref{sec:results} to depict the dispersion of the calibration results in the context of marginal MAPs.

\begin{figure}[H]
	\centering
		\includegraphics[width=\textwidth]{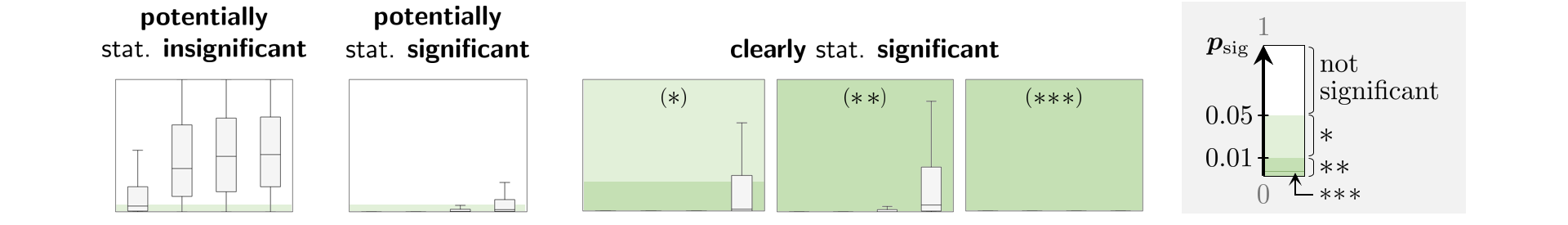}
		\caption[Different scenarios for boxplots of the \mbox{$p$-values} obtained by significance checks]{\NEW{Different scenarios for} the boxplots (box: first to third quartile; horizontal line: median) of the 1000 \mbox{$p$-values} obtained by significance checks with sample sizes 500, 100, 50 and 30 (from left to right per subplot, i.e. horizontal axis). The shaded background gives the significance level and determines the upper limit of the vertical axis (the lower limit is always zero): mainly white (upper limit:~1), two-shaded green \NEW{marked with~{\tiny (}$\Ast${\tiny )}} (upper limit:~0.05), dark green marked with~{\tiny (}$\AAst${\tiny )} or~{\tiny (}$\AAAst${\tiny )} (upper limit:~0.01 resp.~0.001). \NEW{Note that some boxplots might not be visible due to vanishing $p$-values.}}
		\label{Fig:sigCheck}
\end{figure}

We want to compare the estimates of the same quantity (e.g. a particular parameter) obtained in different environmental settings. Such an estimate is given by the particle representation of the respective marginal posterior distribution. There are several possibilities (e.g. Student's~\cite{Student.1908} or Welch's~\cite{WELCH.1947} $t$-test, Mann–Whitney $U$-test~\mbox{\cite{Wilcoxon.1945,Mann.1947}}, Brunner-Munzel test~\cite{Brunner.2000}) to compare two distributions given by samples, which are each based on different assumptions 
regarding the sample distribution and the (in)equality of the two variances. A visualization of the calibrated marginal distributions implies that we cannot consistently assume normal-distributed samples or equal variances. Therefore, we use the Brunner-Munzel test in Python (\texttt{scipy.stats.brunnermunzel}) to compare marginal distributions. However, this test (as well as the others) is not designed for very large samples sizes. From a mathematical perspective, 
\NEW{it is nearly impossible to achieve analytically equal marginal distributions due to the approximative and non-deterministic nature of the calibration.} 
Hence, we will always encounter mathematically different marginal posteriors in our setting. The more accurate the approximation of these distributions, i.e. the larger the sample size, the more obvious the differences between them show up for the statistical tests. In fact, trying to use the whole particle sets as samples for statistical testing appeared to be ineffective since this yields vanishing~$p$-values. Therefore, we need to compare the distributions based on a considerably smaller sample size (note that before testing we resample the particles for uniform weights). To get an impression of the role of the sample size on the significance results, we perform the test with different sample sizes which should work well for this testing method~\cite{Brunner.2000}:~500, 100, 50 and 30. Furthermore, we repeat each test 1000 times to consider numerical variation of the test results. Eventually, the visualization of the obtained $p$-values help to finally judge statistical significance (see Figure~\ref{Fig:sigCheck}). This testing approach will be referred to as \emph{significance check}. We use this method in Section~\ref{sec:results} to compare particular marginal posteriors \NEW{as well as for comparing obtained fitted
estimates with the given uncertainty of the fit. Note that the latter is given by a confidence interval of a normal distribution. Hence, we use the two-tailed Welch's test~(Python:~\texttt{scipy.stats.ttest\_ind(..., equal\_var=False)}) for checking the significance of observed differences between fitted estimates.} 
\noindent However, this approach can only discern clear statistical significance or provide an overview over the degree of similarity between distributions. As mentioned, not observing obvious statistical significance does not imply irrelevance or non-existence of differences between estimates.

\subsubsection{Adaption of the kernel density estimation}
\label{SUP-sec:CAL-KDE}
In this section, we explain how we truncate a kernel density estimation~(KDE), i.e. a smoothed approximation of the PDF, based on the particle representation of a marginal posterior distribution with bounded support. The boundedness of the posterior support is carried over from the prior (recall that we used either uniform or triangular priors), as the posterior measure~${\mu^\mathcal{I}}$ is absolutely continuous with respect to the prior~$\mu_0$~\cite{Bulte.2020}, i.e.~${\supp(\pi^\mathcal{I})\subseteq\supp(\pi_0)}$.
\paragraph*{Native KDE.} Suppose we want to approximate the PDF of the marginal posterior of the \mbox{$j$-th} parameter (${j=1\Ldots d\in\mathbb{N}}$), for which we assumed a prior distribution whose support is bounded on an interval~$[a,b]$, ${a<b}$. We use a KDE with Gaussian kernels and Scott's bandwidth selection method~\cite{Scott.2015}. In practice, this can be calculated by applying the Python function 
\begin{equation*}
	\texttt{scipy.stats.gaussian\_kde(}\underbracket[0.187ex]{\,\texttt{particles\_j}\,}_{\{(\theta_p)_j\}_{p=1}^P}\texttt{, weights\,=}\underbracket[0.187ex]{\,\texttt{w\_particles\,}}_{\{W_p\}_{p=1}^P}\texttt{)}\,,
\end{equation*}
where the arguments are the \mbox{$j$-th} component of all particles with the corresponding weights. If~${a\ll(\theta_p)_j\ll b~\forall p\in\{1\Ldots P\}}$, i.e. the collection of marginalized particles is situated \qm{far enough} (which depends on the bandwidth) away from the bounds, the support of this KDE is virtually included\footnote{To be precise, the actual support of the KDE is unbounded due to the Gaussian kernels. However, suppose~${\theta_{\text{min},j}}$ resp.~${\theta_{\text{max},j}}$ being the smallest/largest value of~${\{(\theta_p)_j\}_{p=1}^P}$, then the value of the KDE will approach zero if evaluated at points sufficiently smaller/larger than~${\theta_{\text{min},j}}$ resp.~${\theta_{\text{max},j}}$.} in~$[a,b]$. Therefore, in this case, there is no need to truncate the KDE to obtain the desired boundedness.

However, the above scenario might not always apply in practice, especially if the bounds of the prior are motivated by mathematical modeling. For instance, if a parameter~${c\in[0,1]}$ quantifies some kind of response, the mathematical bounds can be interpreted as \qm{no response}~(${c\equiv0}$) and \qm{maximal response}~(${c\equiv1}$). Then, depending on the underlying data, it can actually happen that the corresponding marginal particles accumulate at one of these bounds. In general, if a considerable part of the particles is close to a border, the support of the KDE will overshoot this border due to its construction by~\cite{Parzen.1962}, that is
\begin{equation*}
	\text{KDE}_{h}(\theta)=\frac{1}{Ph}\sum_{p=1}^{P}\kappa\left(\frac{\theta-(\theta_{p})_j}{h}\right),
\end{equation*}
where~${h>0}$ is the bandwidth,~$P$ the sample size and~$\kappa$ the kernel density, i.e. in our case the PDF of a standard normal distribution. We see that with the latter, a particle close to the border~$b$ (w.l.o.g.), i.e.~${(\theta_{p})_j\lesssim b}$, contributes to the estimator in such a way that~${\text{KDE}_{h}(\theta)>0}$ also for~${\theta\gtrsim b}$. This contribution accumulates with the number of particles which are situated close to~$b$. Hence, in such a scenario, the above KDE is not feasible to obtain an approximation of the marginal PDF with bounded support. To preserve the boundedness of the marginal posterior particles, we present an adapted version of the native KDE in the following.

\begin{figure}[H]
	\centering
	\includegraphics[width=\textwidth]{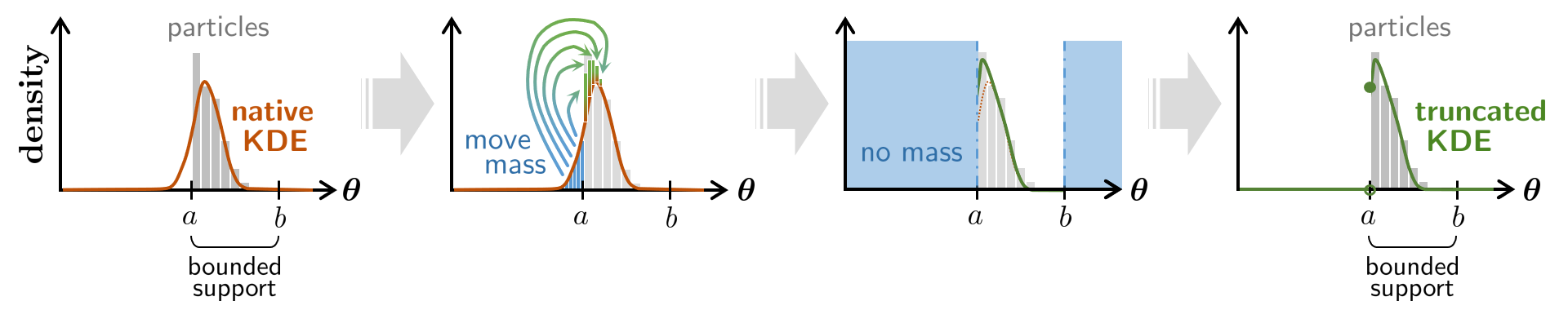}
	\caption[How the truncated KDE is obtained from the native KDE]{Outline of how the truncated KDE is obtained from the native KDE, based on the weighted particle approximation of a marginal posterior.}
	\label{Fig:KDE}
\end{figure}

\paragraph*{Truncated KDE.} Figure~\ref{Fig:KDE} illustrates how we appropriately limit the native KDE to the bounded support of the prior distribution -- we will refer to this as the \emph{truncated KDE} (trKDE). Note that the latter needs to sustain the features of a PDF, especially having an integral of one. 
\noindent Without the truncation, the probability mass of the particle approximation close to a boundary gets spread outside of the bounded support of the prior (Figure~\ref{Fig:KDE}: first plot). In practice, we observed that the distance of any external (i.e. outside of~$[a,b]$), significantly non-zero mass of the KDE has a distance from the boundaries which is not larger than the interval length~${b-a}$. Hence, we can move all external mass into the interval~$[a,b]$ by defining the truncated KDE as
\begin{equation*}
	\text{trKDE}_{h}(\theta)=\Big(\text{KDE}_h(a+\tau_a)+\underbracket[0.187ex]{~\text{KDE}_h(a-\tau_a)\,}_{\text{external mass}}+\,\text{KDE}_h(b-\tau_b)+\underbracket[0.187ex]{~\text{KDE}_h(b+\tau_b)\,}_{\text{external mass}}\Big)\cdot \mathbbm{1}_{[a,b]}(\theta)\,,
\end{equation*}
with~${\tau_a,\tau_b\in[0,b-a]}$ being the distance of~${\theta\in[a,b]}$ from~$a$ resp.~$b$ and~${\mathbbm{1}_{[a,b]}}$ denoting the indicator function on~$[a,b]$. In particular, we basically \qm{shift} external mass with distance~${\tau_{*}}$ (${*\in\{a,b\}}$) from the respective boundary to the point with the same distance, but inside the interval~$[a,b]$ (Figure~\ref{Fig:KDE}: second plot). Therefore, we can set~${\text{trKDE}_{h}(\theta)=0~\forall \theta\notin[a,b]}$ without losing any mass (Figure~\ref{Fig:KDE}: third plot), i.e. preserving an integral of one. This results in an appropriate approximation of the marginal posterior PDF given by the particles and truncated on~$[a,b]$ (Figure~\ref{Fig:KDE}: fourth plot).

\section{Complementary calibration results}\label{sec:othercal}
Here we present further investigations of the results regarding the model calibrations for the two cell lines Hep3B2 and HepG2.

\subsection{Chemoresistance of Hep3B2}\label{APP-Res2.1}
We present the marginal estimates, investigate the parameter correlations and compare the marginal distributions of specific parameters.

\subsubsection{\NEW{Marginal estimates}} \label{APP-Res2.1-EST}
\begin{figure}[H]
	\centering
	\includegraphics[width=\linewidth]{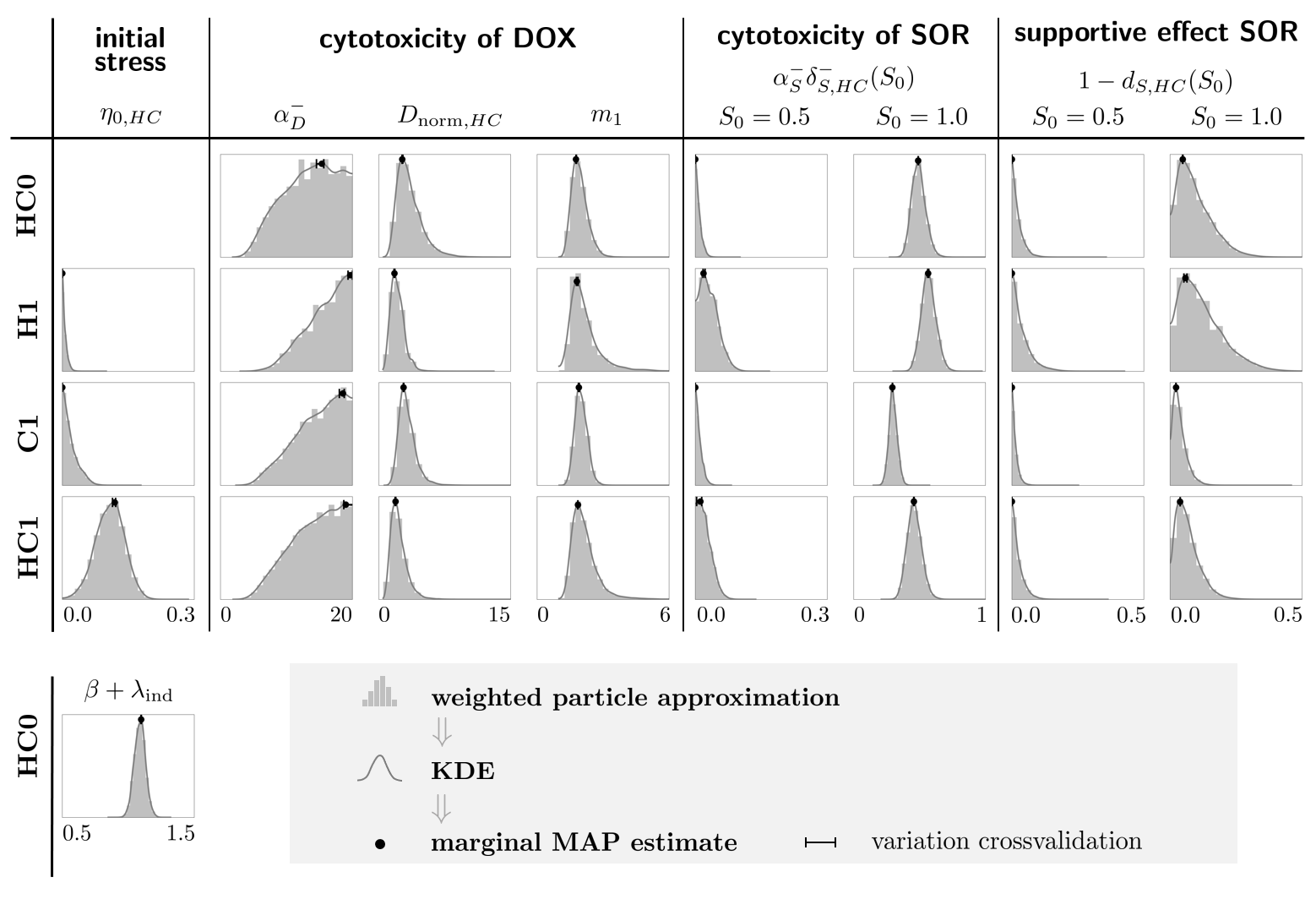}
	\caption[Hep3B2: Marginal posteriors]{\NEW{Marginal posteriors of Hep3B2 calibrations. The numerical deviations for cross-validating the MAP estimates according to Subsection~\ref{sub:RES-ana-var} are shown as error bars, which might not be visible due to high robustness of the respective estimate. Recall that~${\prol+\indDeath}$ resp.~$\initStCH$ are only calibrated for HC0 resp. H1/C1/HC1 (${\prol+\indDeath}$ is fixed to its MAP for H1/C1/HC1 and ${\initStCH=0}$ for HC0 per definition).}}
	\label{A-Fig:CL1-marg}
\end{figure}

\NEW{Figure~\ref{A-Fig:CL1-marg} compares the marginal distributions of the model parameters based on the particle approximation collecting all SMC runs. 
Note that the parameters~$\DnormCH$ and~${m_1}$ were calibrated in log scale to improve the coverage of the high probability region with the prior sample, but their marginal distributions and corresponding MAPs are determined by calculating the KDE in linear scale.}
\subsubsection{Parameter correlations} \label{APP-Res2.1-CORR}
We check the (linear) parameter correlations of the particle approximation collecting all SMC runs. To capture numerical variations, we repeated the calculation of the correlation coefficients 1000 times for respectively different set of 5000 samples. Since this resulted in robust coefficients\footnote{\NEW{The standard deviation of the numerical dispersion is in the magnitude of~$10^{-2}$\,, i.e. the deviations are of minor interest.}}, we give the average values of $r$ in the following Figures~\ref{A-Fig:CL1-corr2} and~\ref{A-Fig:CL1-corr13}\,. Note that in all subplots of both figures the axis limits are not consistent and especially not labeled as this is not the focus of this investigation. Figure~\ref{A-Fig:CL1-corr2} depicts the correlations between the DOX susceptibility threshold~$\DnormCH$ and the remaining parameters of the ODE for the ESL~$\stress$.
\begin{figure}[H]
	\centering
	\includegraphics[width=\linewidth]{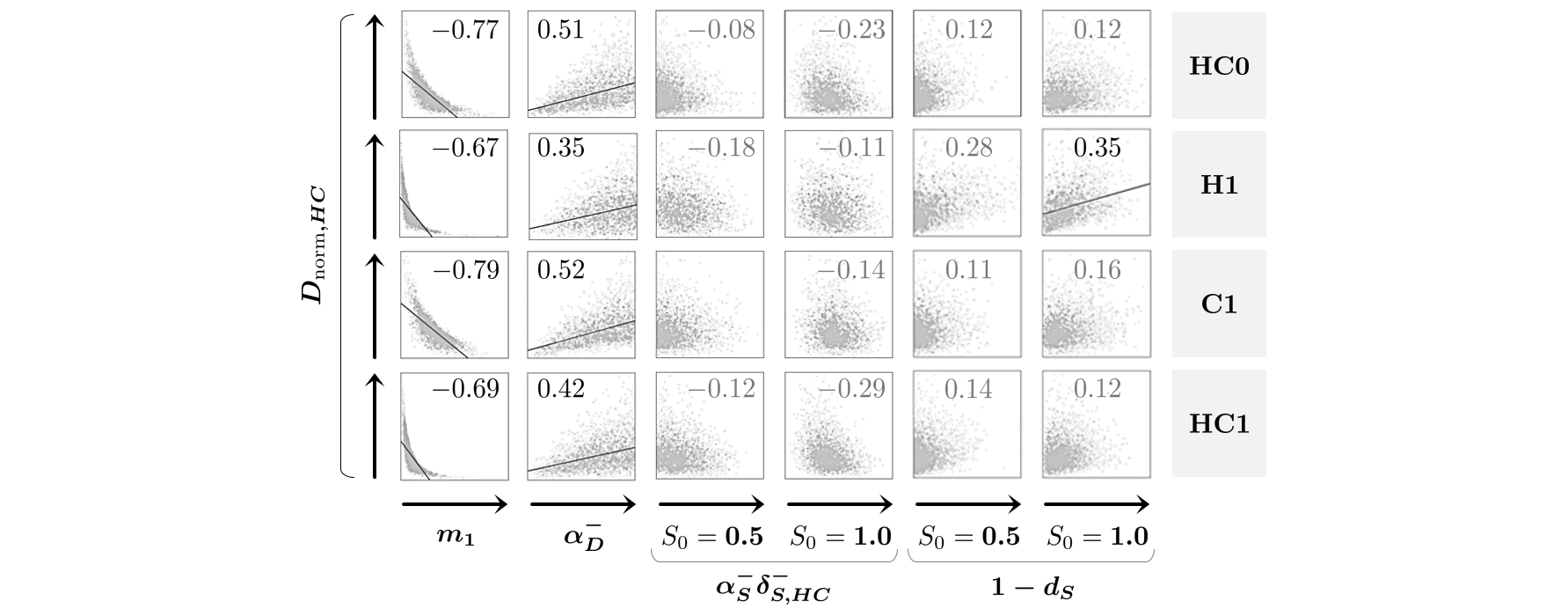}
	\caption[Hep3B2: Correlations between~$\DnormCH$ and remaining calibrated parameters]{Scatter plots with $5000$~samples drawn from the 2D distributions of pairwise parameter combination of~$\DnormCH$ with the remaining calibrated parameters (columns) resulting from model calibrations with  Hep3B2 data. Only statistically significant (${\pval<0.05}$) correlation coefficients~$r$ are given and a regression line is depicted if at least a moderate linear correlation (${|r|>0.3}$) is observable.}
	\label{A-Fig:CL1-corr2}
\end{figure}
\noindent We observe a strong negative correlation with~$m_1$ (potentially nonlinear as the L-shaped point cloud suggests) and a moderate positive one with~$\sensRate{D}^-$ for all environmental conditions. Furthermore, there are less obvious indications from the shape of the point clouds that there is a relevant positive correlation with~${1-d_S}$. Note that for~${1-d_S}$ the correlation results could be distorted due to the proximity of the respective marginal posteriors to the lower bound of zero (see Figure~\ref{Fig:CL1-dS}). In summary, all parameters involved in the DOX stress response~${\adDCH}$ show considerable correlations.

For the sake of completeness, Figure~\ref{A-Fig:CL1-corr13} shows analogous plots for the remaining parameter estimations. Note that we do not show the correlations of pairwise combinations of~$\sensRate{D}^-$,~${\sensRate{S}^-\deact_{S,HC}}$ and~${1-d_{S}}$ since they are mostly found to be statistically insignificant and/or very weak. The only exception is a moderate positive correlation between~${1-d_{S}(0.5)}$ and~${1-d_{S}(1.0)}$ which is an expected result as we used the parametrization~${d_S(1.0)=c_d\cdot d_S(0.5)}$ and calibrated~${c_d\in(0,1]}$ instead of~${d_{S}(1.0)}$.
\begin{figure}[H]
	\centering
	\begin{subfigure}{\textwidth} \label{A-Fig:CL1-corr1}
		\includegraphics[width=\linewidth]{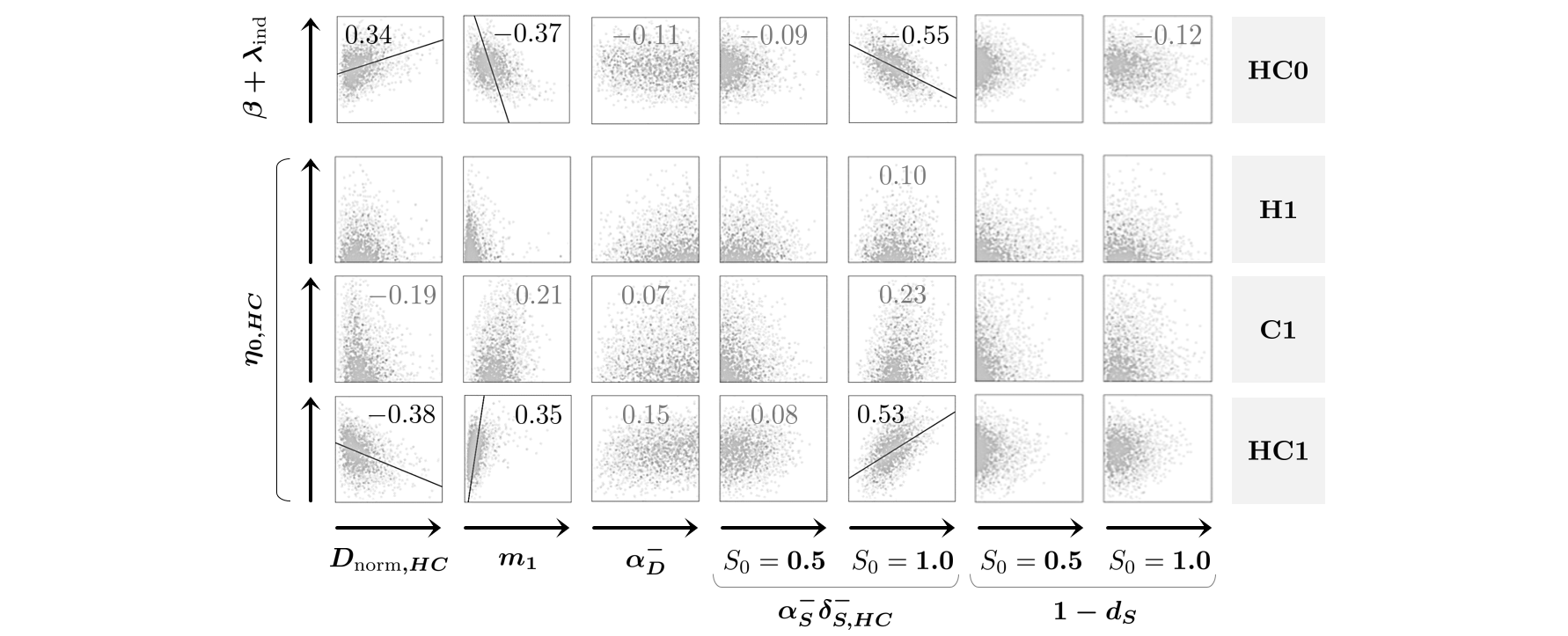}
		\caption{Illustration of the correlations between~${\prol+\indDeath}$ (top row) resp.~$\initStCH$ (rows 2--4) with the remaining calibrated parameters (columns).}
	\end{subfigure}\\[10pt]
	\begin{subfigure}{\textwidth} \label{A-Fig:CL1-corr3}
		\includegraphics[width=\linewidth]{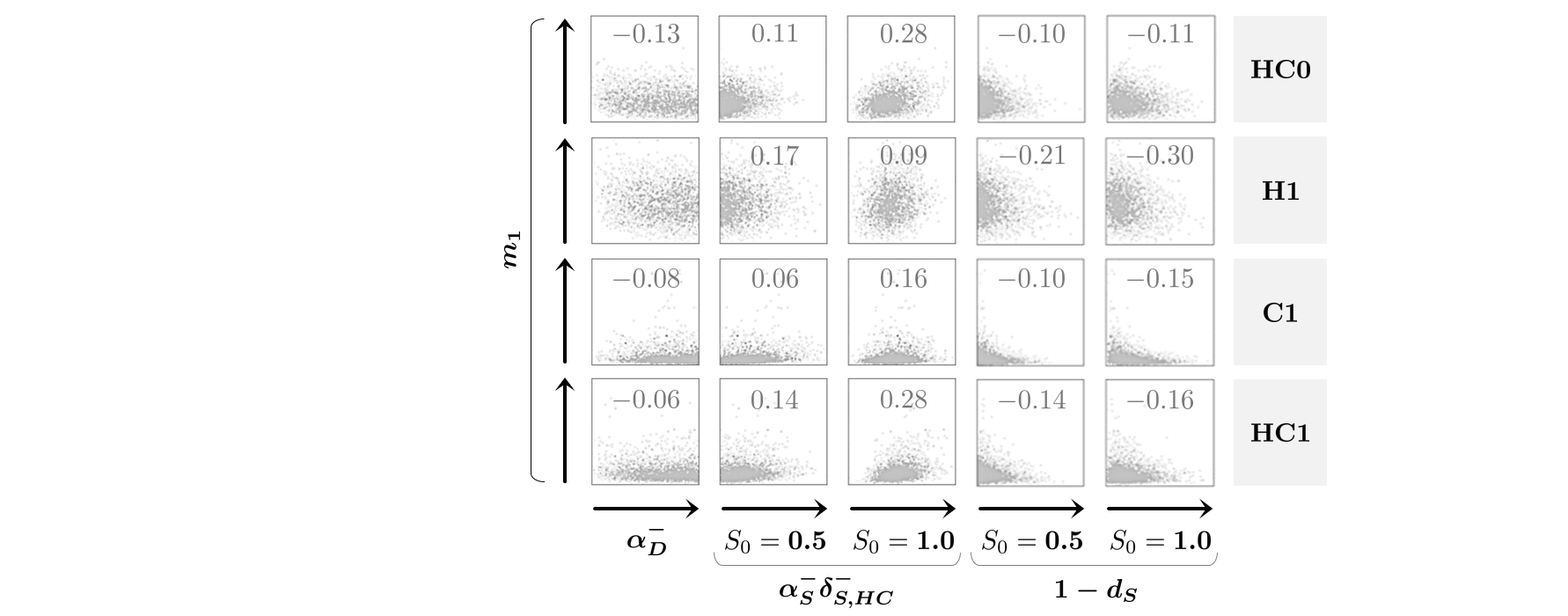}
		\caption{Illustration of the correlations between~$m_1$ with the remaining calibrated parameters (columns).}
	\end{subfigure}
	\caption[Hep3B2: Further parameter correlations]{Scatter plots with $5000$~samples drawn from the 2D distributions of pairwise parameter combinations resulting from model calibrations with  Hep3B2 data. Only statistically significant (${\pval<0.05}$) correlation coefficients~$r$ are given and a regression line is depicted if at least a moderate linear correlation (${|r|>0.3}$) is observable.}
	\label{A-Fig:CL1-corr13}
\end{figure}
%
\subsubsection{Comparison of the marginal parameter estimates} 
\label{APP-Res2.1-MARG}
\begin{figure}[H]
	\centering
	\includegraphics[width=\linewidth]{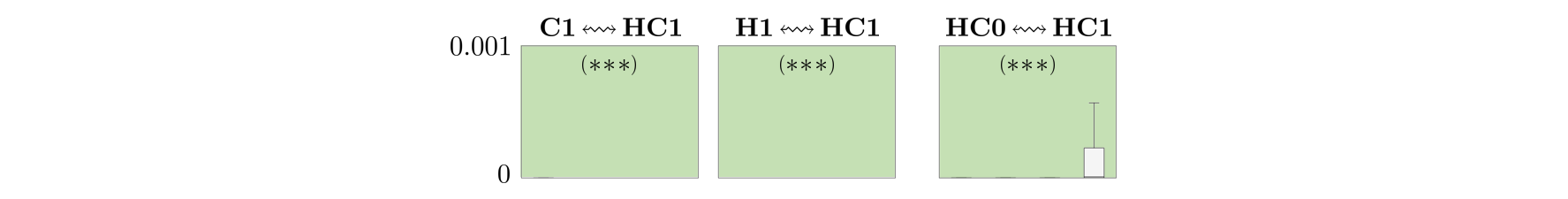}
	\caption[Hep3B2: Significance check for marginal distributions of~$\initStCH$]{Resulting $p$-values of the significance check (sample sizes per subplot from left to right: 500, 100, 50, 30) comparing the marginal distributions of~$\initStCH$ for Hep3B2 in different environmental settings. Vertical axis limits:~0--0.001. Note that some boxplots are not visible due to vanishing~$p$-values.}
	\label{A-Fig:CL1-P-stress0}
\end{figure}

\begin{figure}[H]
	\centering
	\includegraphics[width=\linewidth]{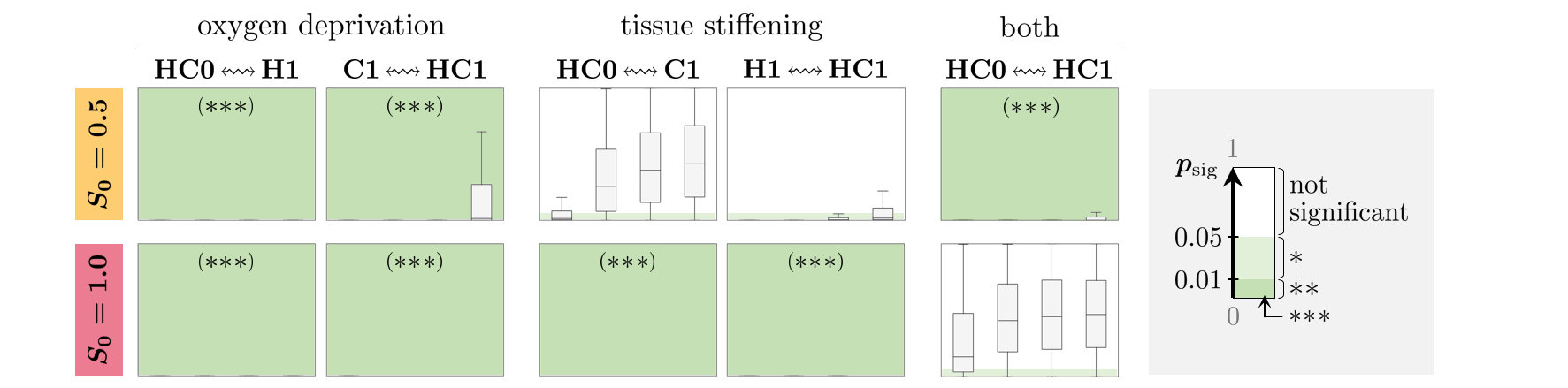}
	\caption[Hep3B2: Significance check for marginal distributions of~${\adSCH(\initS)}$]{Resulting $p$-values of the significance check (sample sizes per subplot from left to right: 500, 100, 50, 30) comparing the marginal distributions of~${\adSCH(\initS)}$ for Hep3B2 in different environmental settings. Vertical axis limits are given by the receptive mark in the subplot:~\mbox{0\,--\,1/0.001}~(no mark\,/\,$\AAAst$). Note that some boxplots are not visible due to vanishing~$p$-values.}
	\label{A-Fig:CL1-P-aS}
\end{figure}

\begin{figure}[H]
	\centering
	\includegraphics[width=\linewidth]{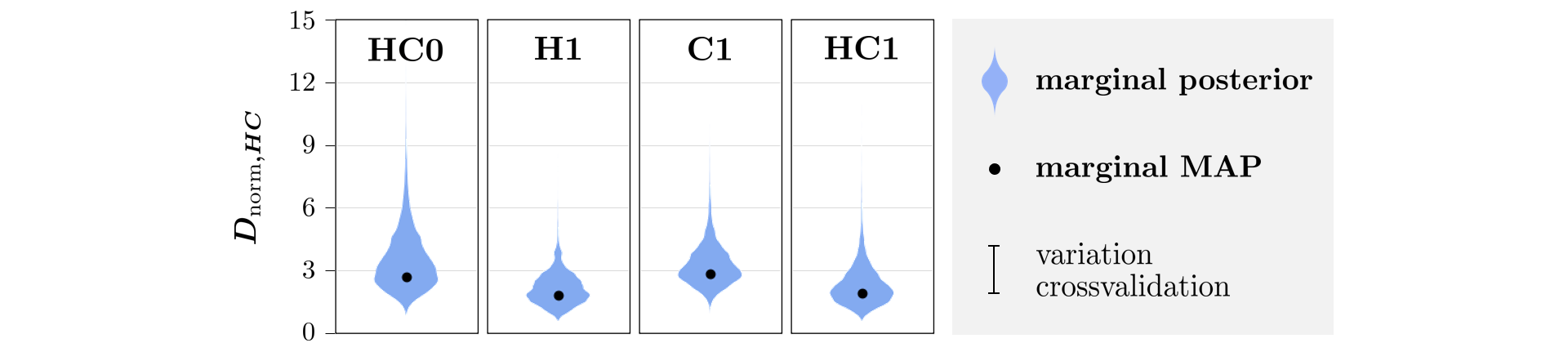}
	\caption[Hep3B2: Marginal posteriors and MAPs of~$\DnormCH$]{Comparison of marginal posteriors (violin plots) and corresponding MAPs of~$\DnormCH$ for Hep3B2. Due to minor numerical variations, the cross-validated 95\% confidence intervals (error bars) are not visible.}
	\label{Fig:CL1-Dnorm}
\end{figure}

\begin{figure}[H] 
	\centering
	\includegraphics[width=\linewidth]{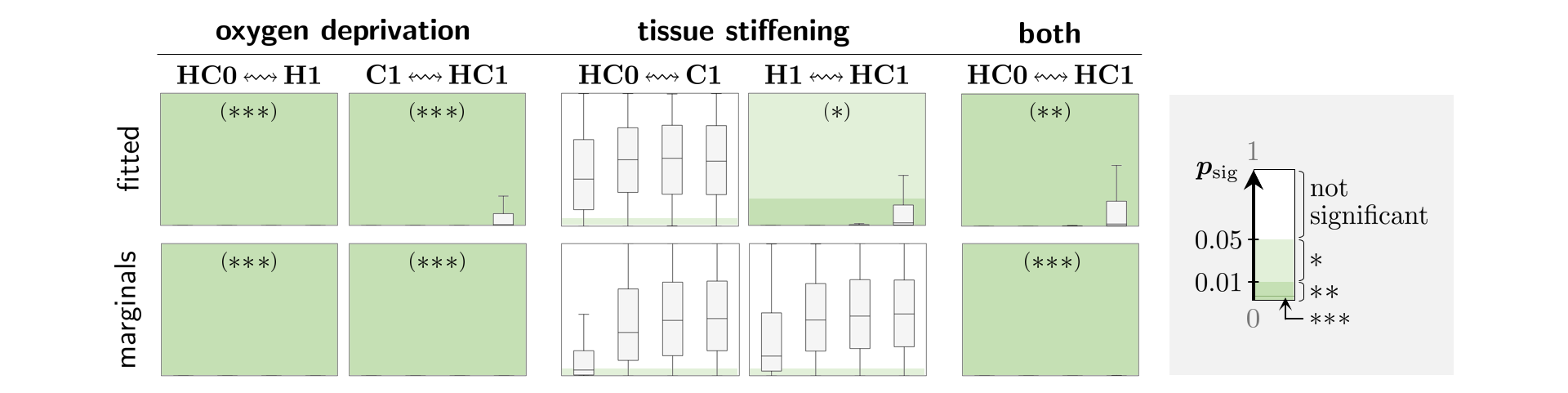}
	\caption[Hep3B2: Significance check for marginal/fitted estimates of~$\DnormCH$]{Resulting $p$-values of the significance check comparing the marginal distributions resp. fitted estimates of~$\DnormCH$ (in Figure~\ref{Fig:CL1-Dnorm}) for Hep3B2 in different environmental settings (sample sizes per subplot from left to right: 500, 100, 50, 30). Vertical axis limits are given by the receptive mark in the subplot:~\mbox{0\,--\,1/0.05/0.01/0.001}~(no mark\,/\,$\Ast$\,/\,$\AAst$\,/\,$\AAAst$). Note that some boxplots are not visible due to vanishing~$p$-values.}
	\label{A-Fig:CL1-P-Dnrom}
\end{figure}

\begin{figure}[H]
	\centering
	\includegraphics[width=\linewidth]{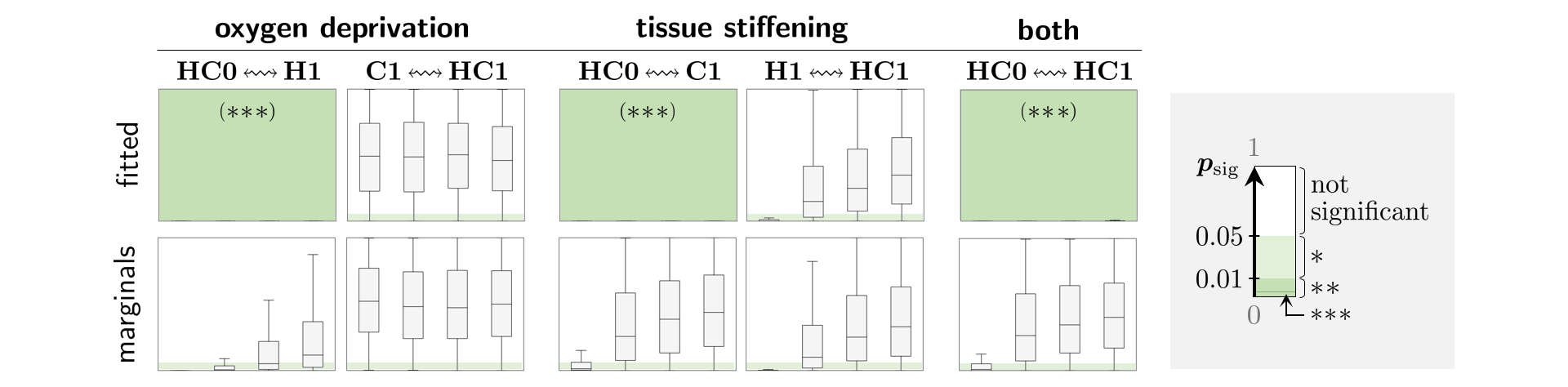}
	\caption[Hep3B2: Significance check for marginal/fitted estimates of~$\sensRate{D}^-$]{Resulting $p$-values of the significance check comparing the marginal distributions resp. fitted estimates of~$\sensRate{D}^-$ for Hep3B2 in different environmental settings (sample sizes per subplot from left to right: 500, 100, 50, 30). Vertical axis limits are given by the receptive mark in the subplot:~\mbox{0\,--\,1/0.001}~(no mark\,/\,$\AAAst$). Note that some boxplots are not visible due to vanishing~$p$-values.}
	\label{A-Fig:CL1-P-aD1000}
\end{figure}

\begin{figure}[H]
	\centering
	\includegraphics[width=\linewidth]{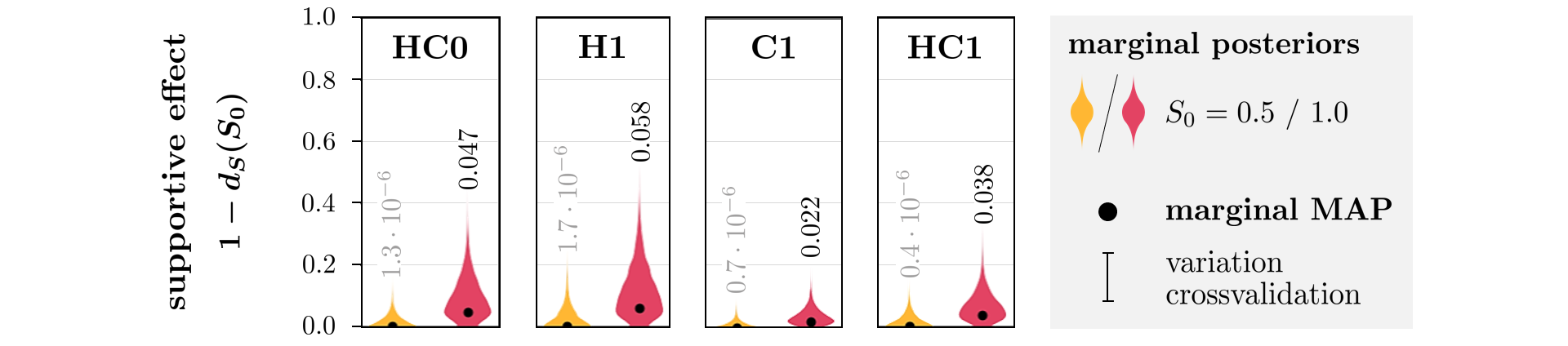}
	\caption[Hep3B2: Marginal posteriors and MAPs of~${1-d_{S}(\initS)}$]{Comparison of marginal posteriors (violin plots) and corresponding MAPs of~${1-d_{S}(\initS)}$ for Hep3B2. Note that for~${\initS=0.5}$ we directly calibrated~$d_S(\initS)$, while for~${\initS=1.0}$ the estimates are obtained by calibration of~${c_d=d_S(1.0)\,/\,d_S(0.5)}$\,. Due to minor numerical variations, the cross-validated 95\% confidence intervals (error bars) are not visible.}
\label{Fig:CL1-dS}
\end{figure}

\begin{figure}[H]
	\centering
	\includegraphics[width=\linewidth]{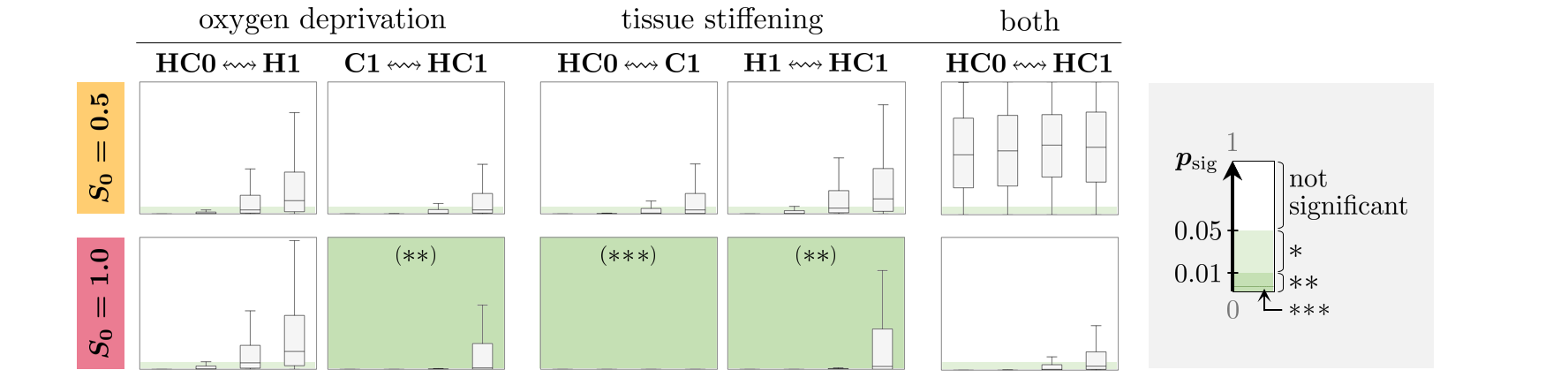}
	\caption[Hep3B2: Significance check for marginal distributions of~${1-d_S(\initS)}$]{Resulting $p$-values of the significance check comparing the marginal distributions of~${1-d_S(\initS)}$ for Hep3B2 in different environmental settings (sample sizes per subplot from left to right: 500, 100, 50, 30). Vertical axis limits are given by the receptive mark in the subplot:~\mbox{0\,--\,1/0.01/0.001}~(no mark\,/$\AAst$\,/\,$\AAAst$). Note that some boxplots are not visible due to vanishing~$p$-values.}
	\label{A-Fig:CL1-P-dS}
\end{figure}

We provide further results from investigating the marginalized estimates (which neglects correlations) of particular parameters. We focus on comparing the respective marginal distributions in different environments with the significance check as proposed in Subsection~\ref{sub:RES-ana-var} and Figure~\ref{Fig:sigCheck} of our article.
\paragraph*{Marginal posterior comparison: Initial stress level.} Figure~\ref{A-Fig:CL1-P-stress0} illustrates the statistical significance of the differences between the marginal distributions of~$\initStCH$ from Figure~\ref{Fig:CL1-stress0} in the paper. We see clear significance for all cases.
\paragraph*{Marginal posterior comparison: Cytotoxic efficacy of SOR.} Figure~\ref{A-Fig:CL1-P-aS} statistically compares the marginal distributions of~${\adSCH(\initS)}$ from Figure~\ref{Fig:CL1-aS} in the paper. \NEW{We see clear significance for the marginals for almost all scenarios, except for the cases of tissue stiffening with~${\initS=0.5}$ (potentially insignificant for sole cirrhosis resp. potentially significant for cirrhosis in hypoxic conditions) and combined hypoxia and cirrhosis with~${\initS=1.0}$ (potentially insignificant).}
\paragraph*{Marginalized investigation: Estimates of DOX susceptibility.}
We investigate the marginals of the DOX susceptibility threshold~${\DnormCH(\initH,\initC)=\Dnorm\cdot d_H(\initH)\cdot d_C(\initC)}$ for~${\initH,\initC\in\{0,1\}}$, i.e. without considering correlations to other parameters. Figure~\ref{Fig:CL1-Dnorm} compares the respective marginal posteriors and MAPs of~$\DnormCH$ for different environments.
\noindent The marginal MAPs of~$\DnormCH$ show the following environmental effects:
\begin{align*}
	\text{sole hypoxia (HC0${\,\leadsto\,}$H1):}\quad2.659\searrow1.778~~(-33.1\%)\,,\\
	\text{hypoxia in cirrhosis (C1${\,\leadsto\,}$HC1):}\quad2.805\searrow1.903~~(-32.2\%)\,,\\[4pt]
	\text{sole cirrhosis (HC0${\,\leadsto\,}$C1):}\quad2.659\nearrow2.805~~\hphantom{0}(+5.5\%)\,,\\
	\text{cirrhosis in hypoxia (H1${\,\leadsto\,}$HC1):}\quad1.778\nearrow1.903~~\hphantom{0}(+7.0\%)\,.
\end{align*}
Since a smaller value for~$\DnormCH$ translates to an enhanced susceptibility, we observe hypoxia resp. cirrhosis to increase/decrease the DOX susceptibility. Furthermore, we do not see a large difference between the effect of sole hypoxia/cirrhosis and a respective combination of both. This suggest no synergistic effect between hypoxia and stiffness on the susceptibility. In general, the relative decrease of~$\Dnorm$ due to hypoxia~(approx.~${-31\%}$) appears to outweigh the increase by cirrhosis~(approx.~${+6\%}$). This is consistent with the observation
\begin{equation*}
	\text{HC0}{\,\leadsto\,}\text{HC1}:\quad2.659\searrow 1.903~~(-28.4\%)\,.
\end{equation*}
Comparing the marginal posteriors of~$\DnormCH$ for different environmental settings shows a clear statistical significance of the increasing effect of hypoxia on the DOX susceptibility, in contrast to the influence of ECM stiffness (see Figure~\ref{A-Fig:CL1-P-Dnrom}). This could indicate that the DOX susceptibility is actually not effected by cirrhosis, however this cannot be concluded distinctively. \NEW{Figure~\ref{A-Fig:CL1-P-Dnrom} shows that the significance results for observed differences between the fitted estimates of~$\DnormCH$ are mostly very similar to the ones for the marginals. The only exception is visible for comparing H1 with HC1, which appears to be clearly significant for the fitted estimates in contrast to the marginals. This weakens the indications obtained with the marginals that the DOX susceptibility is generally not influenced by cirrhosis.}
\paragraph*{Marginal posterior comparison: DOX impact.} Figure~\ref{A-Fig:CL1-P-aD1000} depicts the statistical significance of the differences between the marginal distributions of~$\sensRate{D}^-$ from Figure~\ref{Fig:CL1-aD1} in the paper. \NEW{We observe potentially insignificant differences between the marginals for all cases. For the fitted values however, the differences appear to be clearly significant for sole hypoxia resp. cirrhosis (columns~1 and~3) and a combination of both (column~5). In general, this could hint on a significant influence of hypoxia/stiffening on the DOX impact rate.}
\paragraph*{Marginal posterior comparison: Supportive effect of SOR.} Although the estimates for the supportive effect~${1-d_S(\initS)}$ do not yield a noticeable treatment support, it might be interesting to take a closer look at its marginals. Figure~\ref{Fig:CL1-dS} shows the corresponding marginal posteriors and the estimated marginal MAPs.
\noindent As expected, the MAPs for a standard SOR dosage~(${\initS =0.5}$) show no significant support on the DOX treatment under all environmental conditions. For a high dosage~(${\initS =1.0}$) we see non-zero MAPs indicating that hypoxia ({HC0\,$\leadsto$\,H1} and {C1\,$\leadsto$\,HC1}) resp. high ECM stiffness ({HC0\,$\leadsto$\,C1} and {H1\,$\leadsto$\,HC1}) might increase resp. decrease the supportive effect of SOR, if there is any. When comparing the marginal distributions in different environmental settings for given~$\initS$\,, clear statistically significant differences are only visible for a high SOR dosage: {HC0~vs.~C1}, {H1~vs.~HC1} (influence of stiffness) and {C1~vs.~HC1} (influence of hypoxia in stiff environment), see Figure~\ref{A-Fig:CL1-P-dS}. However, it is hard to judge whether these are relevant observations in the context of the negligible support that we observed.

Figure~\ref{A-Fig:CL1-P-dS} shows the statistical comparison of the marginal distributions for~${1-d_S(\initS)}$ from Figure~\ref{Fig:CL1-dS}. \NEW{For~${\initS=0.5}$\,, we see potentially insignificant differences between the marginals when comparing HC0 with HC1 (column~5) and an unclear significance for the remaining cases (columns~{1--4}). For~${\initS=1.0}$\,, the observed differences appear to be clearly (columns~{2--4}) or potentially (columns~1 and~5) significant.}
\subsection{Chemoresistance of HepG2}\label{APP-Res2.2}

We start with some additional information regarding the obtained parameter estimates. Next, we check the statistical significance of observed differences between parameter estimates for HC0 and H1. Eventually, we take a look at the parameter correlations in the particle sample.
\subsubsection{Marginal estimates} \label{APP-Res2.2-EST}
\begin{figure}[H]
	\centering
	\includegraphics[width=\linewidth]{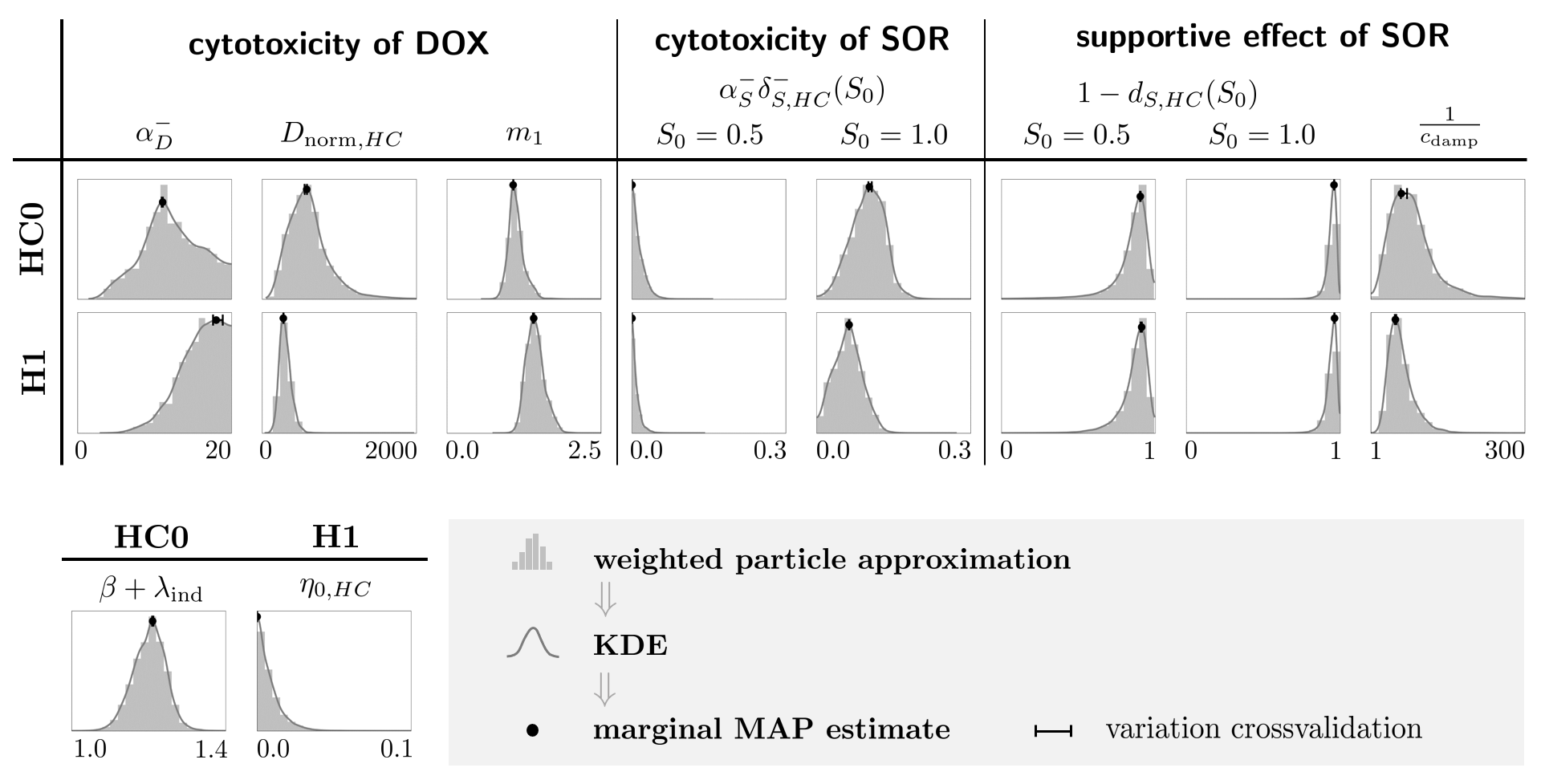}
	\caption[HepG2: Marginal posteriors]{Marginal posteriors of HepG2 calibrations. The numerical deviations for cross-validating the MAP estimates according to Subsection~\ref{sub:RES-ana-var} are shown as error bars, which might not be visible due to high robustness of the respective estimate. Recall that~${\prol+\indDeath}$ resp.~$\initStCH$ are only calibrated for HC0 resp. H1 (${\prol+\indDeath}$ is fixed to its MAP for H1 and ${\initStCH=0}$ for HC0 per definition).}
	\label{A-Fig:CL2-marg}
\end{figure}

\begin{figure}[H] 
	\centering
	\includegraphics[width=\linewidth]{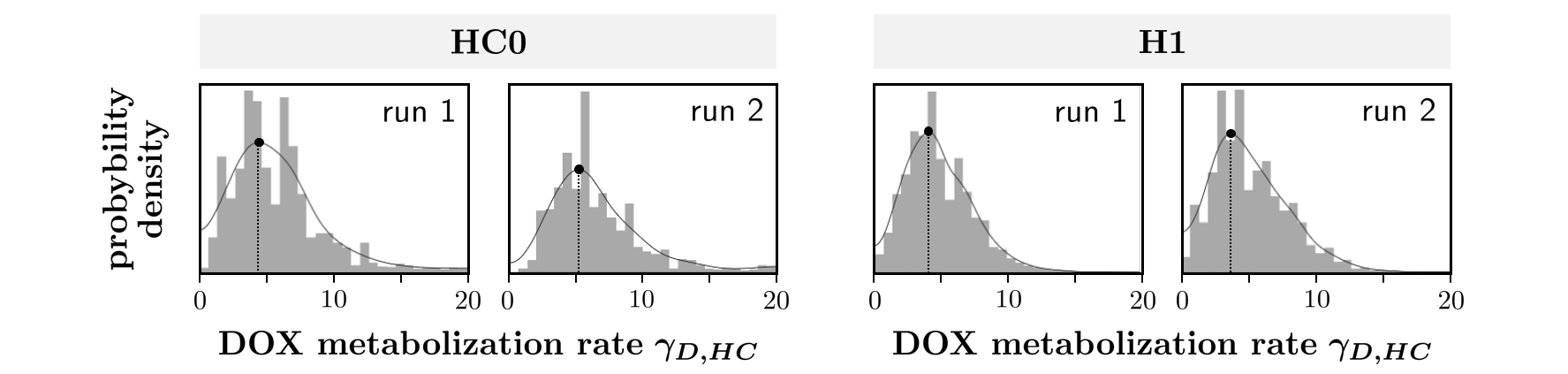}
	\caption[HepG2 -- test runs calibrating~\ref{eq:M-chemo}: Marginal posteriors and MAPs of~$\metDCH$]{Marginal distributions and corresponding MAPs estimates (markers and dotted line) of the DOX metabolization rate~$\metDCH$ from two test runs for calibrating model~\ref{eq:M-chemo} to HepG2 (HC0/H1) data. The solid line shows the KDEs and the scaling of vertical axis is not consistent for all subplots.}
	\label{A-Fig:CL2-metRate}
\end{figure}

Figure~\ref{A-Fig:CL2-marg} compares the marginal distributions of the model parameters based on the particle approximation collecting all SMC runs.
Note that the parameters~$\DnormCH$ and~${1/\cdamp}$ were calibrated in log scale to improve the coverage of the high probability region with the prior sample, but their marginal distributions and corresponding MAPs are determined by calculating the KDE in linear scale.
\noindent Furthermore, we obtained alternative estimates for parameters regarding the stress response to DOX by a weighted least square fit to take the corresponding parameter correlations (see later Subsection~\ref{APP-Res2.2-CORR}) into account. Table~\ref{A-Tab:CL2-fitsSTD} provides the corresponding 95\% confidence intervals of these fits.
\begin{table}[H]
	\caption[HepG2: 95\%~confidence intervals resulting from the least square fits]{95\%~confidence intervals~(${\pm\,1.96\,\cdot\,}$standard deviation) resulting from the least square fits in Figure~\ref{Fig:CL2-aD} of the paper and Table~\ref{Tab:CL2-parsDOX}.}\label{A-Tab:CL2-fitsSTD}
	\centering
		\begin{tabular}{rrrr} \toprule
			~&&\textbf{HC0}&\textbf{H1}\\\midrule
			{$\boldsymbol{\sensRate{D}^-}$}&&$10.694\pm\hphantom{00}4.589$&$17.413\pm\hphantom{00}3.391$\\
			{$\boldsymbol{\DnormCH}$}&&$501.122\pm305.410$&$318.290\pm103.267$\\
			{$\boldsymbol{m_1}$}&&$1.128\pm\hphantom{00}0.129$&$1.433\pm\hphantom{00}0.147$\\[4pt]
			{$\boldsymbol{1-d_S(0.5)}$}&&$0.879\pm\hphantom{00}0.069$&$0.873\pm\hphantom{00}0.084$\\
			{$\boldsymbol{1-d_S(1.0)}$}&&$0.948\pm\hphantom{00}0.032$&$0.942\pm\hphantom{00}0.043$\\
			\multirow{2}{*}{$\boldsymbol{\frac{1}{\cdamp}}$}&&$22.542\pm\hphantom{0}12.446$&$18.864\pm\hphantom{00}8.756$\\[-2pt]
			&&$22.467\pm\hphantom{00}9.174$&$19.128\pm\hphantom{00}6.296$\\
			\bottomrule
		\end{tabular}		
\end{table}
\noindent Recall that it was not possible to achieve reasonable parameter estimates with the complete model~\ref{eq:M-chemo} to appropriately reconstruct the HepG2 (HC0/H1) data from~\cite{Ozkan.2021}. Nevertheless, for some parameters, these calibrations yield fairly concentrated marginal distributions. One of them is the DOX metabolization rate~$\metDCH$, which results in estimates~${\metDCH\approx 5}$ for both HC0 and H1 (see Figure~\ref{A-Fig:CL2-metRate}). This value is close to the estimates constructed in the paper. However, we cannot be entirely sure that these marginals are reliable since the underlying model/algorithmic assumptions are currently not at a stage to get reasonable results as a whole. Hence, there is no interest in presenting further results from these calibrations at the moment.
\subsubsection{Significance checks} \label{APP-Res2.2-SIG}
\begin{figure}[H]
	\centering
	\includegraphics[width=\linewidth]{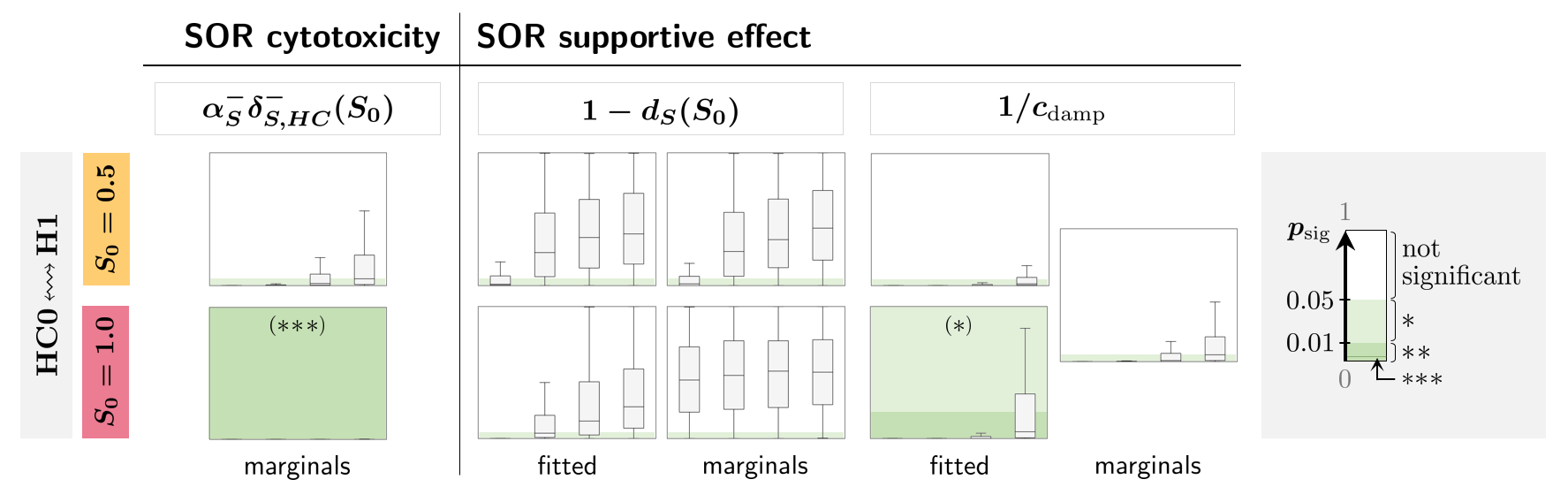}
	\caption[HepG2: Significance check for marginal/fitted estimates of SOR-related parameters]{Resulting $p$-values of the significance check comparing the marginal distributions of the SOR-related parameters (HepG2) for HC0 and H1 (sample sizes per subplot from left to right: 500, 100, 50, 30). Vertical axis limits are given by the receptive mark in the subplot:~\mbox{0\,--\,1/0.05/0.001}~(no mark\,/$\Ast$\,/\,$\AAAst$). Note that some boxplots are not visible due to vanishing~$p$-values.}
	\label{A-Fig:CL2-P-SOR}
\end{figure}

\begin{figure}[H]
	\centering
	\includegraphics[width=\linewidth]{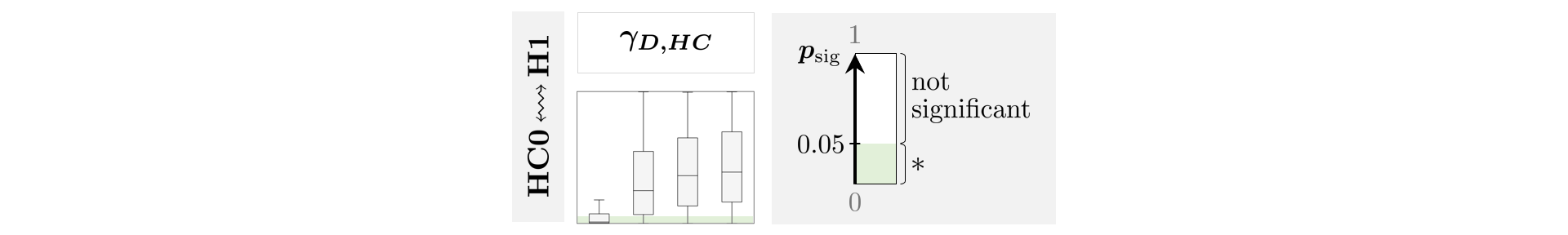}
	\caption[Significance check for estimates of~$\metDCH$]{Resulting $p$-values of the significance check comparing estimates of~$\metDCH$ (HepG2) for HC0 and H1 (sample sizes per subplot from left to right: 500, 100, 50, 30). Vertical axis limits:~\mbox{0\,--\,1}\,.}
	\label{A-Fig:CL2-P-CYP}
\end{figure}

This subsection compares parameter estimates (HC0~vs.~H1) according to Subsection~\ref{sub:RES-ana-var} and Figure~\ref{Fig:sigCheck} in the paper. Figure~\ref{A-Fig:CL2-P-SOR} shows the statistical comparison of the marginal distributions of~${\adSCH(\initS)}$\,, ${1-d_S(\initS)}$ and~$1/\cdamp$ from Figure~\ref{A-Fig:CL2-marg} as well as of their least square fits from~Table~\ref{A-Tab:CL2-fitsSTD}. We see clear significance for the marginals of~${\adSCH(\initS)}$ for~${\initS=1.0}$ in contrast to~${\initS=0.5}$\,. The differences between corresponding estimates for~${1-d_S(\initS)}$ (fit and marginals) are potentially statistically insignificant. For~${1/\cdamp}$\,, significance is only clear for the fit with~${\initS=1.0}$\,, however the remaining plots indicate a similar tendency.
\noindent Analogously, Figure~\ref{A-Fig:CL2-P-SOR} statistically compares the marginal distributions of~$\sensRate{D}^-$\,,~$\DnormCH$ and~$m_1$ from Figure~\ref{A-Fig:CL2-marg} as well as their least square fits from~Table~\ref{A-Tab:CL2-fitsSTD}. For all parameter estimates (fitted and marginals), we observe a clear statistical significance of observed differences. 
\begin{figure}[H]
	\centering
	\includegraphics[width=\linewidth]{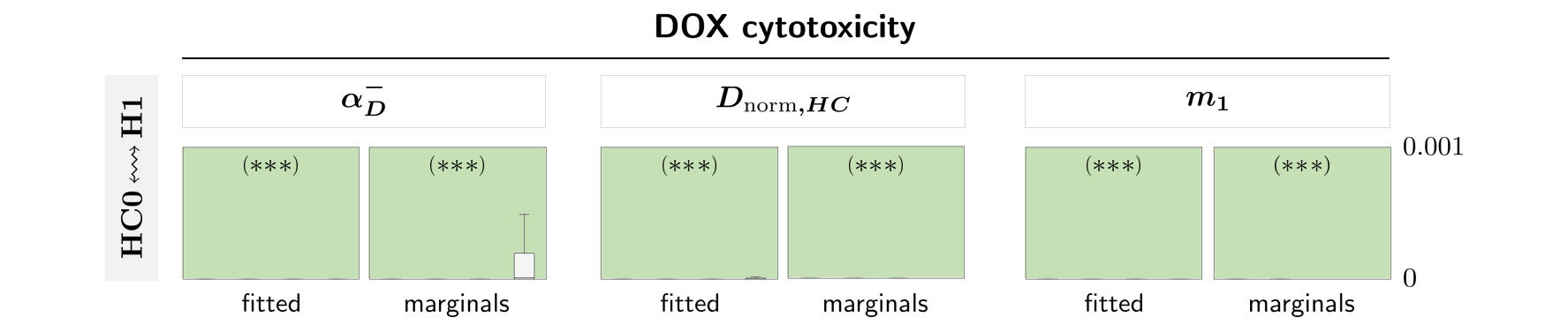}
	\caption[HepG2: Significance check for marginal/fitted estimates of DOX-related parameters]{Resulting $p$-values of the significance check comparing the fitted resp. marginal estimates of the parameters regarding the unsupported stress response to DOX (HepG2) for HC0 and H1 (sample sizes per subplot from left to right: 500, 100, 50, 30). Vertical axis limits:~\mbox{0\,--\,0.001}\,. Note that most of the boxplots are not visible due to vanishing~$p$-values.}
	\label{A-Fig:CL2-P-DOX}
\end{figure}
\noindent Lastly, Figure~\ref{A-Fig:CL2-P-CYP} shows the potential statistical insignificance of the observed differences for the DOX metabolization rate~$\metDCH$\,, which were reconstructed with the approach in~\eqref{eq:CYP-assump}.
\subsubsection{Parameter correlations} \label{APP-Res2.2-CORR}
\begin{figure}[H]
	\centering
	\begin{subfigure}{\textwidth} 
		\includegraphics[width=\linewidth]{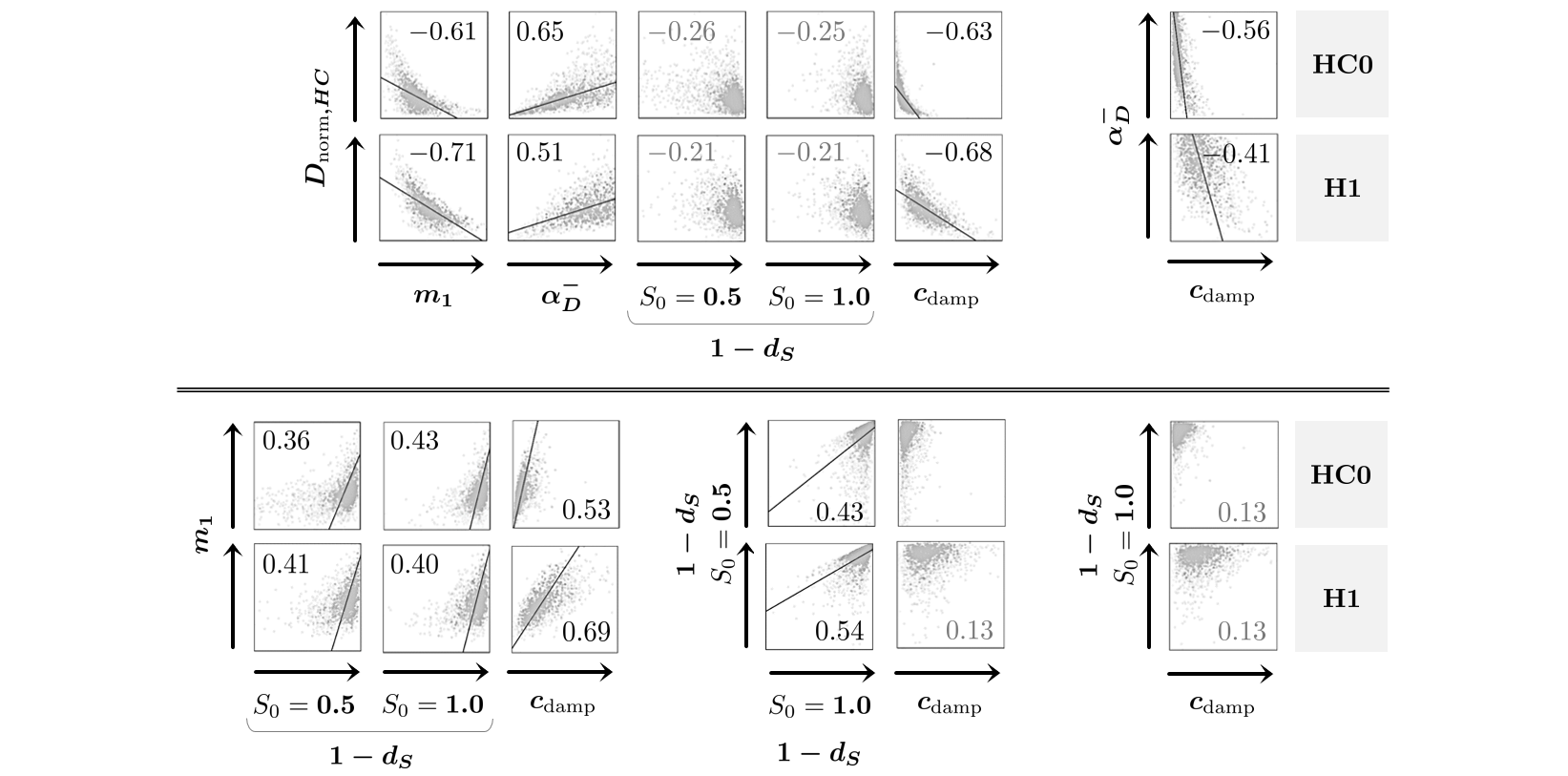}
		\caption{Illustration of considerable correlations between parameters regarding the stress response to DOX.}\label{A-Fig:CL2-corr1}
	\end{subfigure}\\[10pt]
	\begin{subfigure}{\textwidth} 
		\includegraphics[width=\linewidth]{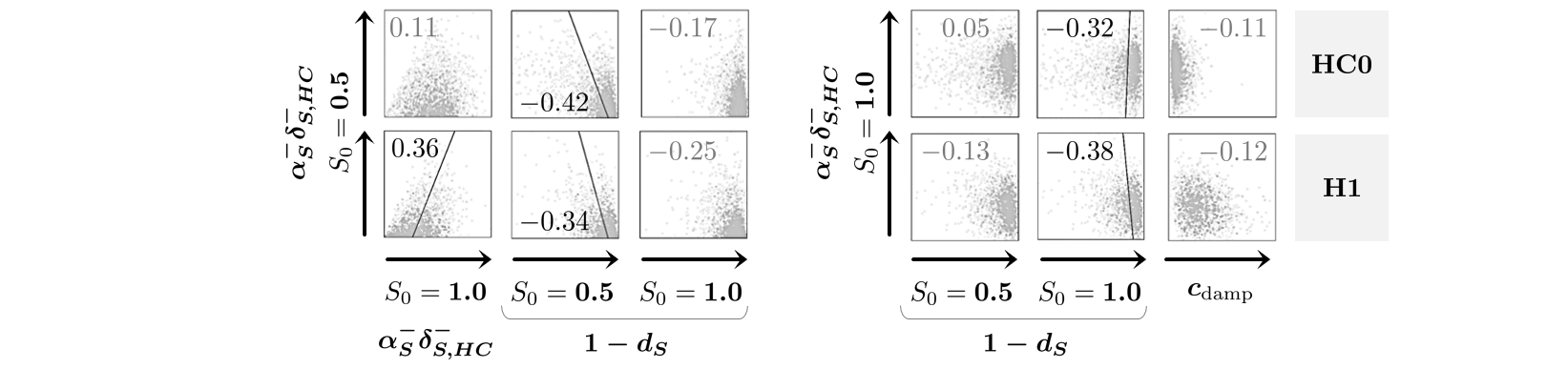}
		\caption{Illustration of further considerable correlations (for the sake of completeness).}\label{A-Fig:CL2-corr2}
	\end{subfigure}
	\caption[HepG2: Parameter correlations]{Scatter plots with $5000$~samples drawn from the 2D distributions of pairwise parameter combinations resulting from model calibrations with  HepG2 data. Only statistically significant~(${\pval<0.05}$) correlation coefficients~$r$ are given and a regression line is depicted if at least a moderate linear correlation (${|r|>0.3}$) is observable.}
	\label{A-Fig:CL2-corr}
\end{figure}

We check the (linear) parameter correlations of the particle approximation collecting all SMC runs analogously to the previous Subsection~\ref{APP-Res2.1-CORR}. Again, we get sufficiently robust coefficients (standard deviation in the magnitude of~$10^{-2}$\,). Hence, we present the average values of~$r$ in the following. Figure~\ref{A-Fig:CL2-corr} depicts the selection of considerable correlations (i.e.~$\pval<0.05$ and~$|r|>0.1$) between the parameters.
\noindent As expected, we see relevant correlations between all parameters regarding the stress response to DOX~(Subfigure~\ref{A-Fig:CL2-corr1}), which is similar to Hep3B2~(recall Figure~\ref{A-Fig:CL1-corr2}). As HepG2 exhibits a considerable supportive influence of SOR contrary to Hep3B2, the correlations of~$\DnormCH$\,,~$\sensRate{D}^-$ and~$m_1$ to the supportive parameters~(${1-d_S}$ and~$\cdamp$) are more distinct. We observe L-shaped point clouds indicating non-linear correlations between~$\DnormCH$ and~$m_1$ resp.~$\cdamp$\,.